\documentclass{aa}  
\usepackage[version=4]{mhchem}
\usepackage{graphicx}
\usepackage{mathtools}
\usepackage{natbib}
\usepackage{txfonts}
\bibpunct{(}{)}{;}{a}{}{,}             %% natbib format for A&A and ApJ

\usepackage[breaklinks=true]{hyperref}
\hypersetup{linkcolor=red,citecolor=blue,filecolor=cyan,urlcolor=magenta}%% to avoid 

\newcommand{\cii}{[C\,{\sc ii}] }
\newcommand{\ci}{[C\,{\sc i}] } 
\newcommand{\ohp}{\ce{OH^+} }
\newcommand{\water}{\ce{H_2O } }
\newcommand{\molpop}{{\sc MOLPOP-CEP} }
\begin{document}

   \title{Probing the interstellar medium of the quasar BRI\,0952$-$0115}

   \subtitle{An analysis of \cii, \ci, CO, OH, and \water}

   \author{K. Kade
          \inst{1}
          \and
          K.K. Knudsen
          \inst{1}
          \and 
          A. Bewketu Belete
          \inst{1}
          \and
          C. Yang
          \inst{1}
          \and
          S. K\"onig 
          \inst{1}
          \and 
          F. Stanley 
          \inst{2}
          \and
          J. Scholtz
          \inst{3}
        }

   \institute{Department of Space, Earth \& Environment, Chalmers University of Technology, SE-412 96
              Gothenburg,  Sweden  \\ \email{kiana.kade@chalmers.se}
    \and 
        Institut de Radioastronomie Millimétrique (IRAM), 300 Rue de la Piscine, 38400 Saint-Martin-d’Hères, France
        \and 
        Kavli Institute for Cosmology, University of Cambridge, Madingley Road, Cambridge, CB30HA, UK
        }

   \date{Received September 15, 1996; accepted March 16, 1997}

% \abstract{}{}{}{}{} 
% 5 {} token are mandatory
 
  \abstract
   {The extent of the effect of active galactic nuclei (AGN) on their host galaxies at high-redshift is not apparent. The processes governing the co-eval evolution of the stellar mass and the mass of the central supermassive black hole, along with the effects of the supermassive black hole on the host galaxy remain unclear. Studying this effect in the distant universe is a difficult process as the mechanisms of tracing AGN activity can often be inaccurately associated with intense star formation and vice versa.}
   {Our aim is to better understand the processes governing the interstellar medium (ISM) of the quasar BRI\,0952$-$0952 at $z = 4.432$, specifically with regard to the individual heating processes at work and to place the quasar in an evolutionary context. 
   }
   {We analyzed ALMA archival bands 3, 4, and 6 data and combined the results with high-resolution band-7 ALMA observations of the quasar. We detect \ci(2--1), \cii(\ce{{}^{2}P_{3/2}-{}^{2}P_{1/2}}), CO(5--4), CO(7--6), CO(12--11), OH $^{2}\Pi_{1/2} (3/2-1/2)$, \water($2_{11} - 2_{02}$), and we report a tentative detection of OH$^+$. We update the lensing model from \citet{Kade23} and we use the radiative transfer code MOLPOP-CEP to produce line emission models which we compare with our observations. }
   {We use the \ci line emission to estimate the total molecular gas mass in the quasar. We present results from the radiative transfer code MOLPOP-CEP constraining the properties of the CO emission and suggest different possible scenarios for heating mechanisms within the quasar. We extend our results from MOLPOP-CEP to the additional line species detected in the quasar to place stronger constraints on the ISM properties.}
   {Modeling from the CO SLED suggests that there are extreme heating mechanisms operating within the quasar in the form of star formation or AGN activity; however, with the current data it remains unclear which of the two is the preferred mechanism as both models reasonably reproduce the observed CO line fluxes. The updated lensing model suggests a velocity gradient across the \cii line, suggestive of on-going kinematical processes within the quasar. We find that the \water emission in BRI\,0952 is likely correlated with star-forming regions of the ISM. We use the molecular gas mass from \ci to calculate a depletion time for the quasar. We conclude that BRI\,0952$-$0952 is a quasar with a significant AGN contribution while also showing signs of extreme starburst activity, indicating that the quasar could be in a transitional phase between a starburst-dominated stage and an AGN-dominated stage.}
   

   \maketitle
%-------------------------------------------------------------------
%%%%%%%%%%%%%%%%%%%%%%%%%%%%%%%%%%%%%%%%%%%%%%%%%%%%%%%%%%%%%%%%%%%%%%%%%%
\section{Introduction}
%%%%%%%%%%%%%%%%%%%%%%%%%%%%%%%%%%%%%%%%%%%%%%%%%%%%%%%%%%%%%%%%%%%%%%%%%%
The formation and evolution of massive galaxies in the high-redshift universe remains to be understood in detail. The correlation between the mass of the central supermassive black hole (SMBHs) and the velocity dispersion of the galaxy suggests that there is a co-evolution of these two properties \citep[e.g.,][]{Kormendy13}. This correlation has been confirmed by several studies \citep[e.g.,][]{Magorrian98, Ferrarese00, Gebhardt00, Haring04, Gultekin09, Beifiori12, Kormendy13, Bennert15, Reines15}. A common method of studying the interplay between the central active galactic nuclei and the ISM is by studying the host galaxies of active galactic nuclei (AGN) across cosmic time and, in particular, studying quasars and their host galaxies. High-redshift quasar host galaxies are often found to be dust-rich and star forming, possibly undergoing a starburst phase \citep[e.g.,][]{Venemans18, Venemans20}. Thus, detailed analyses of the interstellar medium (ISM) properties of AGN host galaxies and other dust-rich starburst galaxies in the high-redshift universe have become increasingly common to understand the early phases of the evolution of massive galaxies. Several studies have been conducted in this area, such as those by \citet{Chen17}, \citet{Circosta19}, \citet{Venemans19}, \citet{Li20}, \citet{Jarugula21}, and \citet{Hagimoto}.

The ISM consists of multiple gas phases, including molecular, neutral, and ionized gas phases. The heating and cooling mechanisms of these phases, as well as the gas and dust content, imply that studies considering a range of properties are required in order to establish a complete understanding of the physical conditions in a galaxy. At high redshift, the CO molecule is often employed as a tool for investigating the ISM due to the variation in density and temperature necessary to excite different population levels of CO \citep[e.g.,][]{Carilli13}. The low-$J$ ($J$ < 3) CO lines trace regions of the ISM with lower densities and temperatures, and are commonly used to trace the bulk molecular gas mass \citep[e.g.,][]{Ivison11}. Mid-$J$ ($4 \leq J \leq 8$) CO lines trace warmer and denser regions of the ISM. These lines have also been shown to linearly correlate with infrared luminosity and are thought to trace star-formation activity \citep[e.g.,][]{Greve14, Lu15, Liu15, Yang17}. High-$J$ CO lines ($J \geq 9$) require both high density and high temperature to excite them and are therefore associated with more extreme ISM conditions. Studies have shown that high-$J$ lines can be enhanced by extreme heating mechanism such as X-ray dominated regions (XDRs) from AGN or could be produced by turbulence, shocks, and cosmic rays \citep[e.g.,][]{Weiss07, vanderWerf10, Bradford11, Gallerani14, Mashian15}. In general, CO emission has been used extensively to study the ISM of local and high-redshift galaxies \citep[e.g.,][]{Weiss07, Daddi15, Rosenberg15, Yang17, Canameras18, Boogaard20, Valentino20, Stanley23}.

A powerful tool to investigate the heating mechanisms present in the ISM is the use of the CO spectral line energy distribution (SLED). This allows for the study of the temperature and density in the ISM as well as providing an estimate of the incident radiation field strength \citep[e.g.,][]{Vallini19}; the CO SLED can be used as a method to study the ISM of individual galaxies or as a means to compare a broad sample of galaxies. CO SLED shape comparisons have been performed in a number of studies and used as a means of classifying, for instance, AGN contribution \citep[e.g.,][]{Rosenberg15, Kirkpatrick19}. The turnover point, that is the CO $J$-transition at which the SLED peaks has been shown to be higher for AGN than, for instance, `normal' star-forming galaxies, submillimeter galaxies (SMGs), starbursts, or ultra-luminous infrared galaxies \citep[ULIRGs;][]{Daddi15, Rosenberg15, Kirkpatrick19}. In the high-redshift Universe, high-$J$ CO lines have been detected in a very limited number of sources, including the quasar SDSS J231038.88+185519.7 at $z = 6.003$ \citep{Carniana19, Li20} and the quasar J1148+5251 at $z = 6.4$ \citep{Gallerani14}. 

Despite its usefulness, CO SLED analysis may not always provide an accurate classification of heating mechanisms in the ISM. For example, \citet{Vallini19} have shown that more compact galaxies can exhibit CO SLED signatures similar to those of AGN-host galaxies. Indeed, \citet{Mashian15} studied high-$J$ CO lines out to $J = 30$ in a diverse sample of local sources and found that the detection of high-$J$ CO lines alone does not provide sufficient information for a determination of the necessity of an AGN. Therefore, combining an analysis of the CO SLED with other tracers can provide a more accurate means of determining the specific heating mechanisms in the ISM. One such tracer is H$_2$O, which is abundant in regions of the ISM that host star formation or are affected by shocks \citep[e.g.,][]{Bergin03, Cernicharo06, GA10}. A detailed study by \citet{Liu17} of water lines in galaxies at $z > 1$ found that different water transitions can be correlated with different regions of the ISM, corresponding to different excitation processes. The analysis of a variety of water lines in the quasars Mrk231 and APM08279+5255 has demonstrated the effect of AGN on water excitation \citep{GA10, vanderWerf11}. Furthermore, studies have found a linear correlation between far-infrared luminosity and water line luminosity for various water lines \citep[e.g.,][]{Yang13, Yang16}. Water lines are increasingly becoming a common tool for ISM analysis in the high-redshift Universe \citep[e.g.,][]{Yang19, Apostolovski19, Yang20, Stanley21, Jarugula21, Pensabene21, Pensabene22, Decarli23}.

The observation of a variety of atomic and molecular line species has become available in high-resolution to the high-redshift Universe with the advent of the Atacama Large Millimeter/submillimeter Array (ALMA). Studies leveraging ALMA data such as \citet{Yang19}, \citet{Rybak20}, \citet{Jarugula21}, \citet{Gururajan22}, \citet{Dye22}, and \citet{Spilker22} have begun leveraging lines such as H$_2$O, CO, \cii, and \ci in order to perform detailed studies of the ISM of high-redshift galaxies. Although these studies are still limited in many aspects, especially considering the angular resolution, direct conclusions about the kinematics, morphology, and gas properties are now able to be drawn. 

In order to better understand the ISM properties of high-redshift AGN, we present line results of the quasar BRI\,0952-0115 (hereafter BRI\,0952), a doubly-imaged gravitationally lensed quasar at $z = 4.432$. BRI\,0952 is a radio-quiet quasar first detected by \citet{Omont96}. The quasar has been studied in a variety of lensing studies \citep[e.g.,][]{Lehar00, Eigenbrod07, Momcheva15}, a detection of CO(5--4) was reported in \citet{Guilloteau99}, and \cii was first detected in the quasar by \citet{Maiolino09}. A follow-up study of \cii was conducted by \citet{Gallerani12} where, in addition to the two images of the quasar, a companion located to the south-west of the quasar was detected. \citet{Kade23} used high-resolution band-7 ALMA observations to perform an in-depth study of the \cii properties of the quasar; here we summarize that \citet{Kade23} detected strong \cii emission from both northern and southern images (Img-N and Img-S) of the quasar. Fitting the line with two Gaussian profiles, the authors found evidence for a broad, complex \cii line profile in BRI\,0952, suggesting the possible presence of out-flowing material from the quasar. Additionally, the analysis revealed the presence of two additional companion sources, Comp-N and Comp-SW, that exhibited emission in \cii. Here we present a follow-up study of the quasar host galaxy using multiple CO transitions, \ci, \cii, \water, and OH, therefore throughout this analysis we will commonly refer to methodology and results from \citet{Kade23}.

In Section \ref{sec:Observations} we describe the ALMA observations and data reduction methods used in this paper. In Section \ref{sec:results} we describe the line emission results, provide a detailed method for correcting the observed spectra for gravitational lensing, calculate the bulk molecular gas mass and investigate line ratios. In Section \ref{sec:Molpop} we describe and present the results of radiative transfer modelling of the detected line species. Section \ref{sec:discussion} discusses the ISM properties of BRI\,0952 in the context of local and high-redshift galaxies. We provide our conclusions in Section \ref{sec:conclusions}. Throughout this paper, we adopt a flat $\Lambda$CDM cosmology with ${H_{0}}$ = 70\,km $\mathrm{s^{-1}\,Mpc^{-1}}$ and $\mathrm{\Omega_{m}}$ = 0.3. 

\begin{table*}[t]
    \centering
    \caption{Details of Observations}
    \begin{tabular}{l c c c c c c} \hline \hline
        Band$^a$ & Date of Obs.$^b$ & $\mathrm{N_{ant}}^c$ & $\nu_{\mathrm{spw, central}}^d$ & Channel Width$^e$ & Synthesized Beam$^f$ & RMS$^g$\\
         & [yyyy mm dd] & & [GHz] & [$\mathrm{km\,s^{-1}} $] & [$'' \times ''$] & [mJy/beam]\\ \hline 
        $3_{\rm CO(5-4)}$ & 2017 12 04 & 47 & 105 & 44 & [$0.50 \times 0.38$] & 0.2 \\ 
        $4_{\rm [CI] \& CO(7-6)}$ & 2016 02 04 & 42 & 147 & 31 & [$1.44 \times 1.04$] & 0.4 \\
        $4_{\rm H_{2}O}$ & 2016 02 04 & 42 & 139 & 31 & [$1.55 \times 1.12$] & 0.4  \\
        $6_{\rm CO(12-11)}$ & 2016 01 06 & 45 & 255 & 18 & [$1.51 \times 0.93$] & 0.3 \\
        $7_{\rm [CII]}$ & 2019 04 06 & 41 & 348 & 13 & [$0.48 \times 0.35$] & 0.2 \\
        $7_{\rm [OH]}$ & 2019 04 06 & 41 & 338 & 28 & [$0.50 \times 0.37$] & 0.2 \\
        $7_{\rm [OH^{+}]}$ & 2019 04 06 & 41 & 347 & 28 & [$0.48 \times 0.36$] & 0.2 \\ \hline
    \end{tabular}
    \tablefoot{
        \tablefoottext{a}{ALMA band for each line observation.}
        \tablefoottext{b}{Date of ALMA observation.}
        \tablefoottext{c}{Number of antennas used in observation.}
        \tablefoottext{d}{Central frequency of the spectral window containing the specific line.}
        \tablefoottext{e}{Channel width of the calibrated and imaged data for the specific line.}
        \tablefoottext{f}{Synthesized beam of the spectral window.}
        \tablefoottext{g}{Per-channel RMS of the spectral window.}
        }
    \label{tab:obs_details}
\end{table*}

%%%%%%%%%%%%%%%%%%%%%%%%%%%%%%%%%%%%%%%%%%%%%%%%%%%%%%%%%%%%%%%%%%%%%%%%%%
\section{Observations} \label{sec:Observations}
%%%%%%%%%%%%%%%%%%%%%%%%%%%%%%%%%%%%%%%%%%%%%%%%%%%%%%%%%%%%%%%%%%%%%%%%%%
This paper makes use of high-resolution band 7 ALMA data for which the full description of the data reduction process is described in \citet{Kade23} as well as ALMA archival bands 3 (2017.1.01081.S, P.I. Leung), 4 (2015.1.00388.S, P.I. Lu) and 6 (2015.1.00388.S, P.I. Lu) data. The scientific goals of each of these datasets vary, and therefore the setup and details of the data reduction are different for each. The original purpose of the band 3 data was to map CO(5--4) line emission at sub-kpc resolution in BRI\,0952 and a sample of other high-$z$ quasar host galaxies and thus the spectral and angular resolution of this data is high in comparison to that of the band 7 data. In contrast, the band 4 and 6 data were designed to observe the CO(7--6) and [N\,{\sc ii}]$205\,\mu$m lines in a sample of high-$z$ galaxies; as this was a detection study the angular resolution of the data is lower compared to both the band 3 and band 7 data. Due to these differences, sensitivity and resolution vary depending on the specific observation set. Additionally, all archival observations were performed in continuum mode, including those containing lines, which affects the resolution and noise properties between channels, but is negligible for the purposes of this paper.

\begin{figure}[h!]
\centering
\includegraphics[width=0.8\columnwidth]{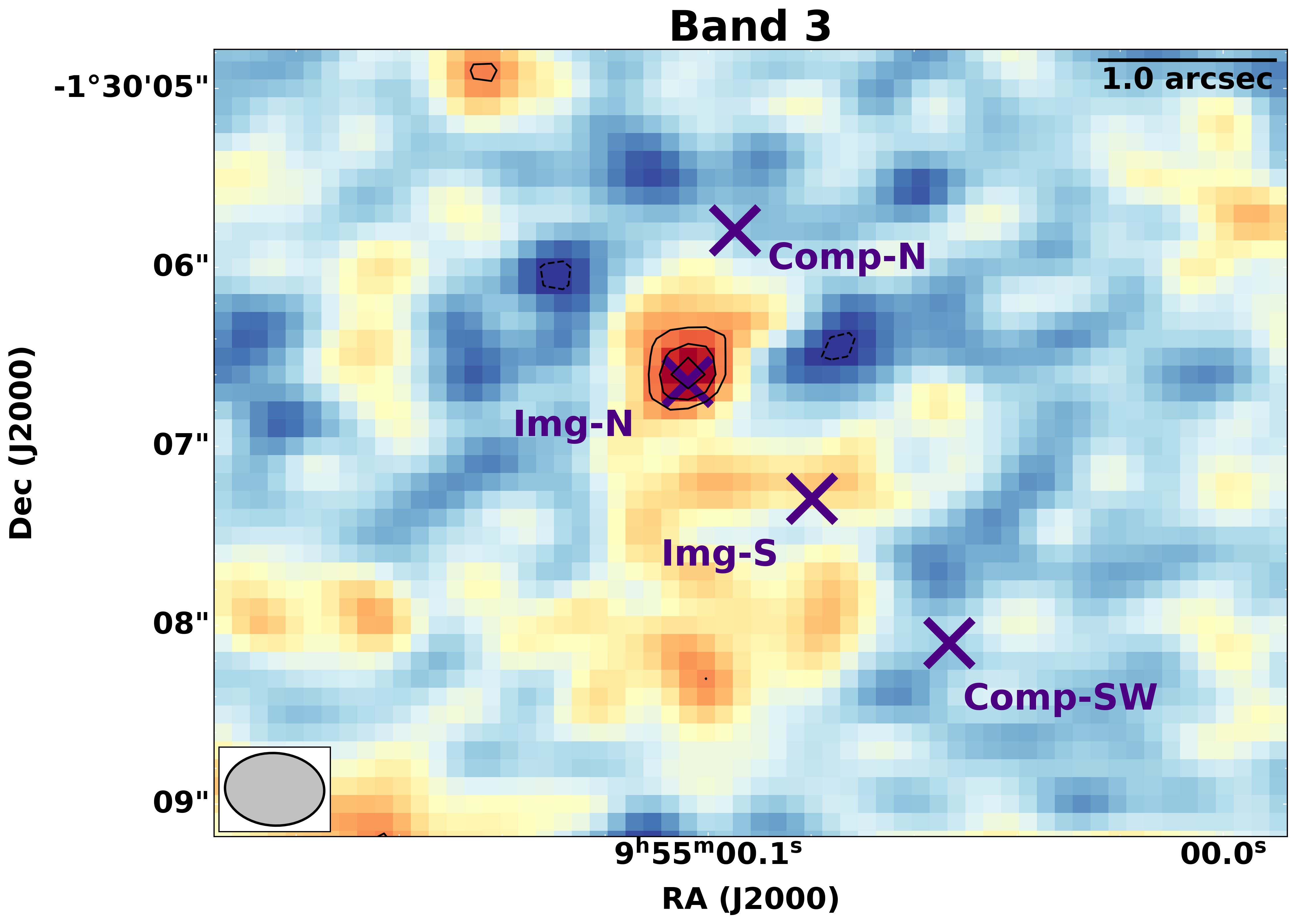}
\includegraphics[width=0.8\columnwidth]{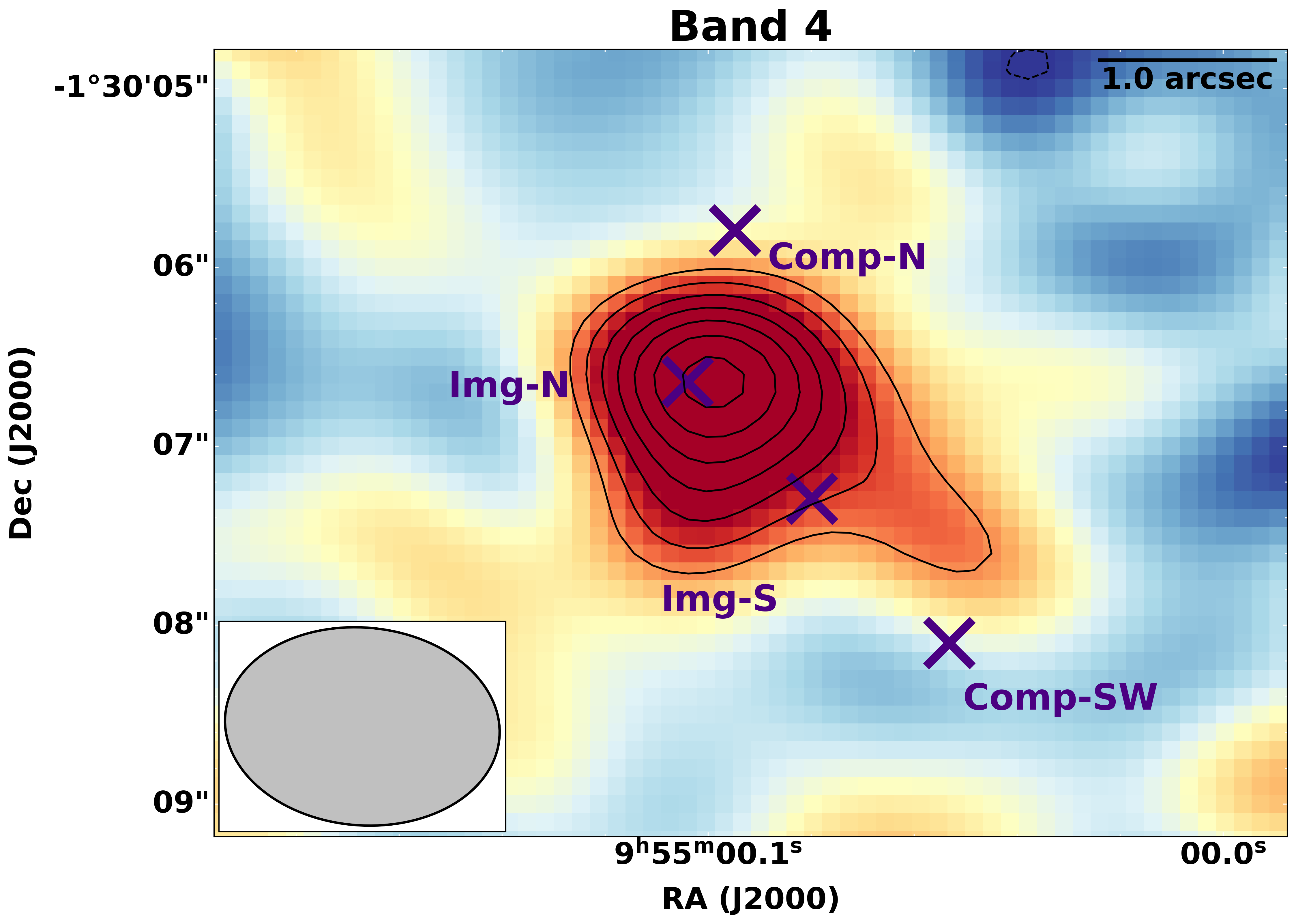}
\includegraphics[width=0.8\columnwidth]{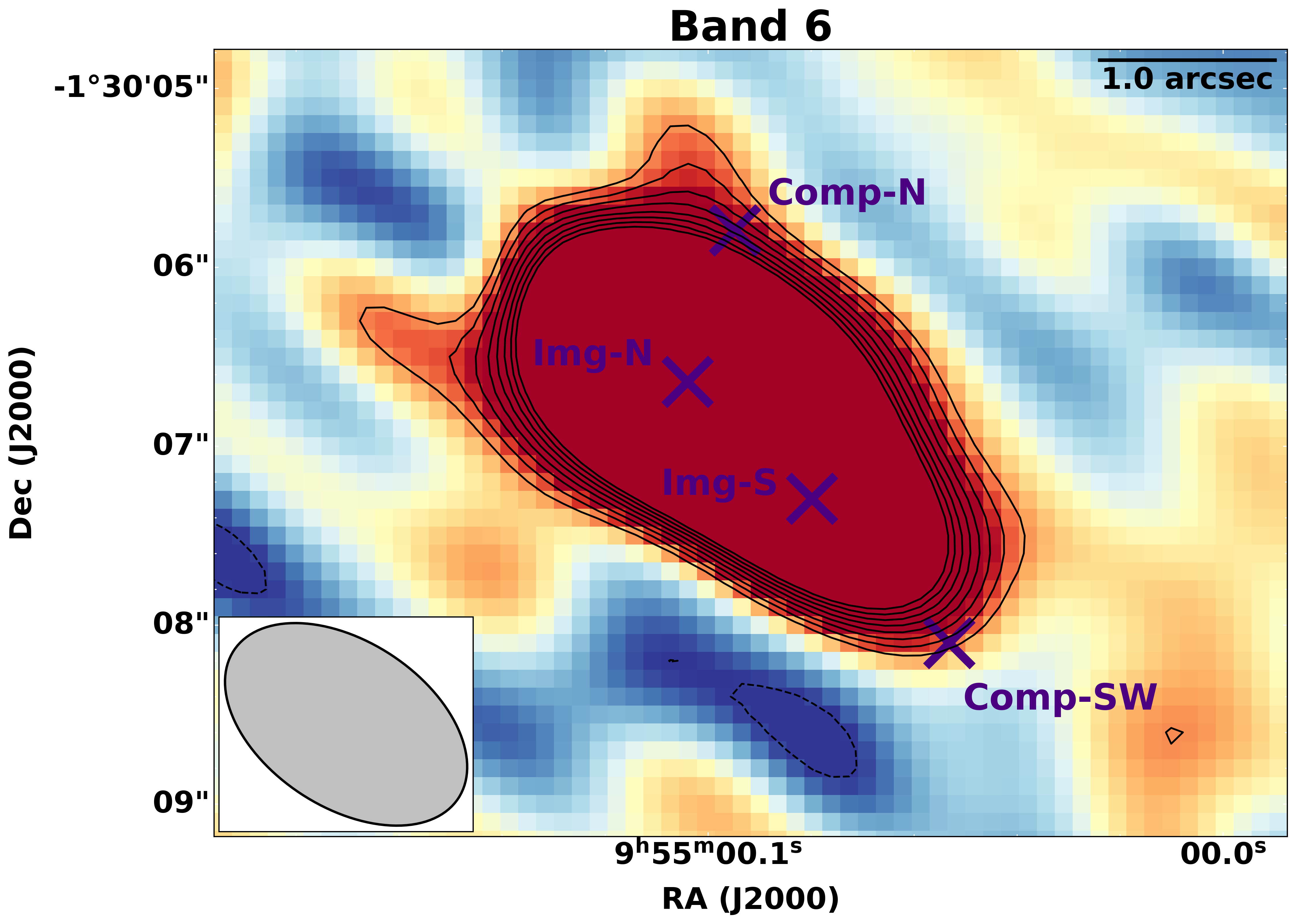}
\includegraphics[width=0.8\columnwidth]{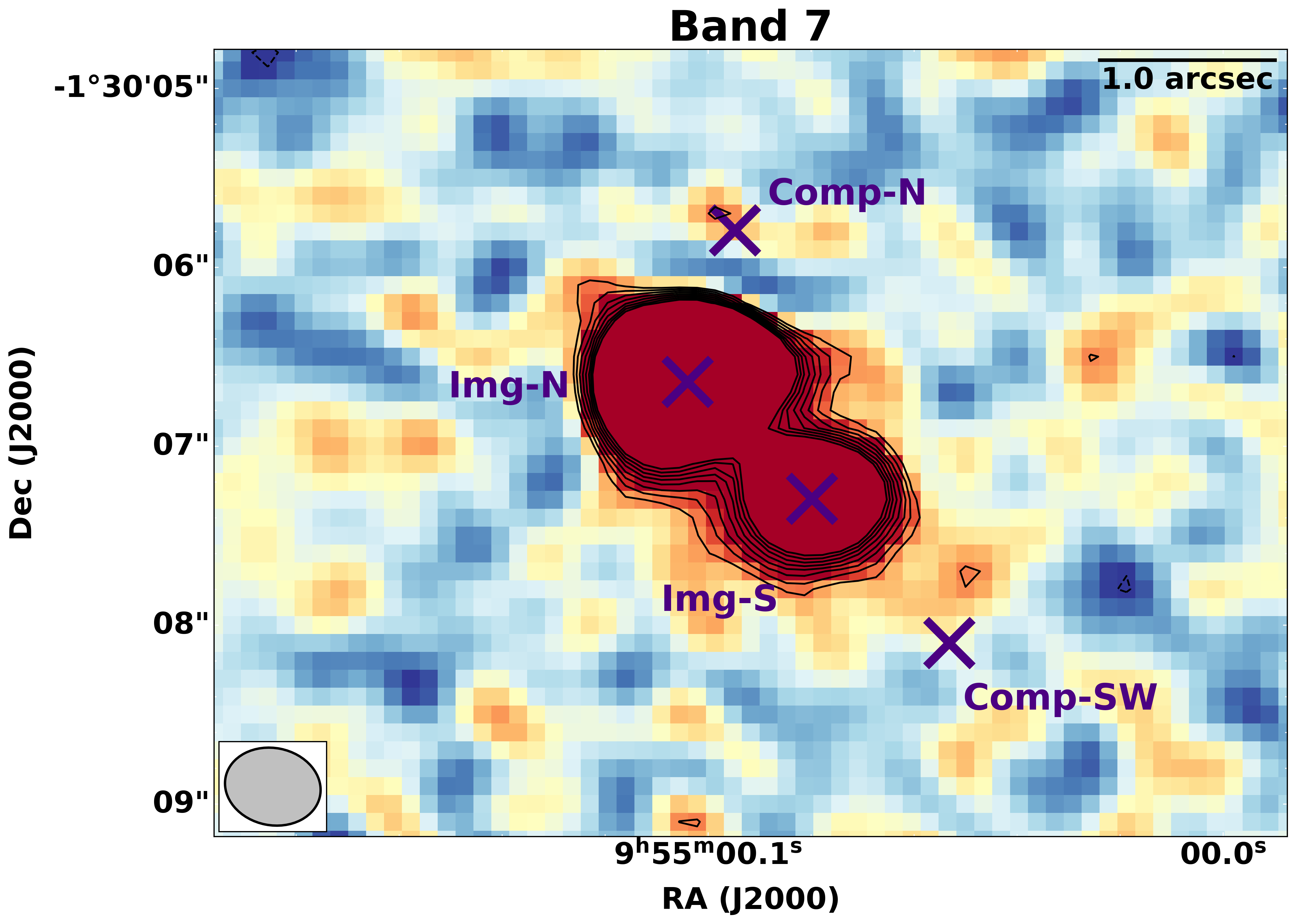}
\caption{ALMA bands 3, 4, 6, and 7 continuum images created using line free channels. The contours are shown at $-3, 3,4,5,6,7,8,9,$ and $10\sigma$ levels. The purple `X's show the position of the companion sources, Comp-N and Comp-SW, detected in \citet{Kade23} as well as the position of Img-N and Img-S, the two images of the quasar BRI\,0952. The synthesized beam is shown in the bottom left of each image.}
\label{fig:BRI_continuum}
\end{figure} 

Similar to the band 7 data, all calibration and imaging processes for the archival band 3, 4, and 6 data were performed in the Common Astronomy Software Application package \citep[{\sc CASA}][]{CASA}. The data in each band were calibrated using the CASA pipeline calibration scripts in version {\sc CASA} 4.5.1 for the band 4 and 6 data and {\sc CASA} version 5.1.1 for the band 6 data. The pipeline includes calibration of the phase, bandpass, flux, and gain. The following phase calibrators were used; band 3: J1058+0133, J0948+0022, J1000+0005; band 4: J1058+0133, Ganymede, J0948+0022; band 6: J0854+2006, J1037-2934, J0948+0022. The pipeline calibrated data were checked to ensure quality. Imaging was performed in {\sc CASA} using the {\sc TCLEAN} task, cleaning in the region of expected emission using a circular mask with a radius of $\sim 20\,''$ from the band 7 data including both BRI\,0952 and the companions detected in \citet{Kade23}. The images were cleaned down to a $2\sigma$ level for each band. Pipeline-provided images of the calibrated data from the ALMA data reduction pipeline were visually inspected to ensure data quality and identify line-free channels. Continuum subtraction was performed using {\sc CASA}'s task {\sc UVCONTSUB} using line-free channels and a polynomial fit order of 0 for all bands. Continuum images were created using the line-free channels with a Briggs weighting scheme with a robust parameter of 0.5. Spectral cubes were created employing the same weighting scheme using the continuum subtracted measurement set, with the exception of the band 3 data which was imaged using natural weighting. As mentioned above, the band 3 data were of significantly higher angular resolution than the other bands. In order to improve data continuity, a taper was performed to the \textit{uv}-data which effectively lowered the resolution of produced continuum images and line-emission cubes to approximately match that of the band 7 data. The band 3 data were also of significantly higher spectral resolution than the other bands ($\sim5.5$\,km\,s$^{-1}$); we binned the data by a factor of 8$\times$ in order to improve the signal-to-noise ratio in each channel. To account for the quasar not being centered in the observational field, primary beam correction was applied to the band 3 data in addition to the taper. However, primary beam correction was not necessary for the band 4 and 6 data since the observations were centered on the source. A conservative estimate of 10\% is used to represent the uncertainty in the absolute flux calibration (note that the fiducial value for bands 3 and 4 is $\sim5\%$ and for bands 6 and 7 is $\sim 10\%$)\footnote{https://almascience.eso.org/documents-and-tools/cycle10/alma-technical-handbook}.

Table \ref{tab:obs_details} provides details on the different bands and spectral windows in which lines were detected including synthesized beam sizes, observing time, spectral resolution, and final image rms'.

%%%%%%%%%%%%%%%%%%%%%%%%%%%%%%%%%%%%%%%%%%%%%%%%%%%%%%%%%%%%%%%%%%%%%%%%%%
\section{Results} \label{sec:results}
%%%%%%%%%%%%%%%%%%%%%%%%%%%%%%%%%%%%%%%%%%%%%%%%%%%%%%%%%%%%%%%%%%%%%%%%%%
\begin{figure*}[h!]
    \centering
    \includegraphics[width = 0.9\linewidth]{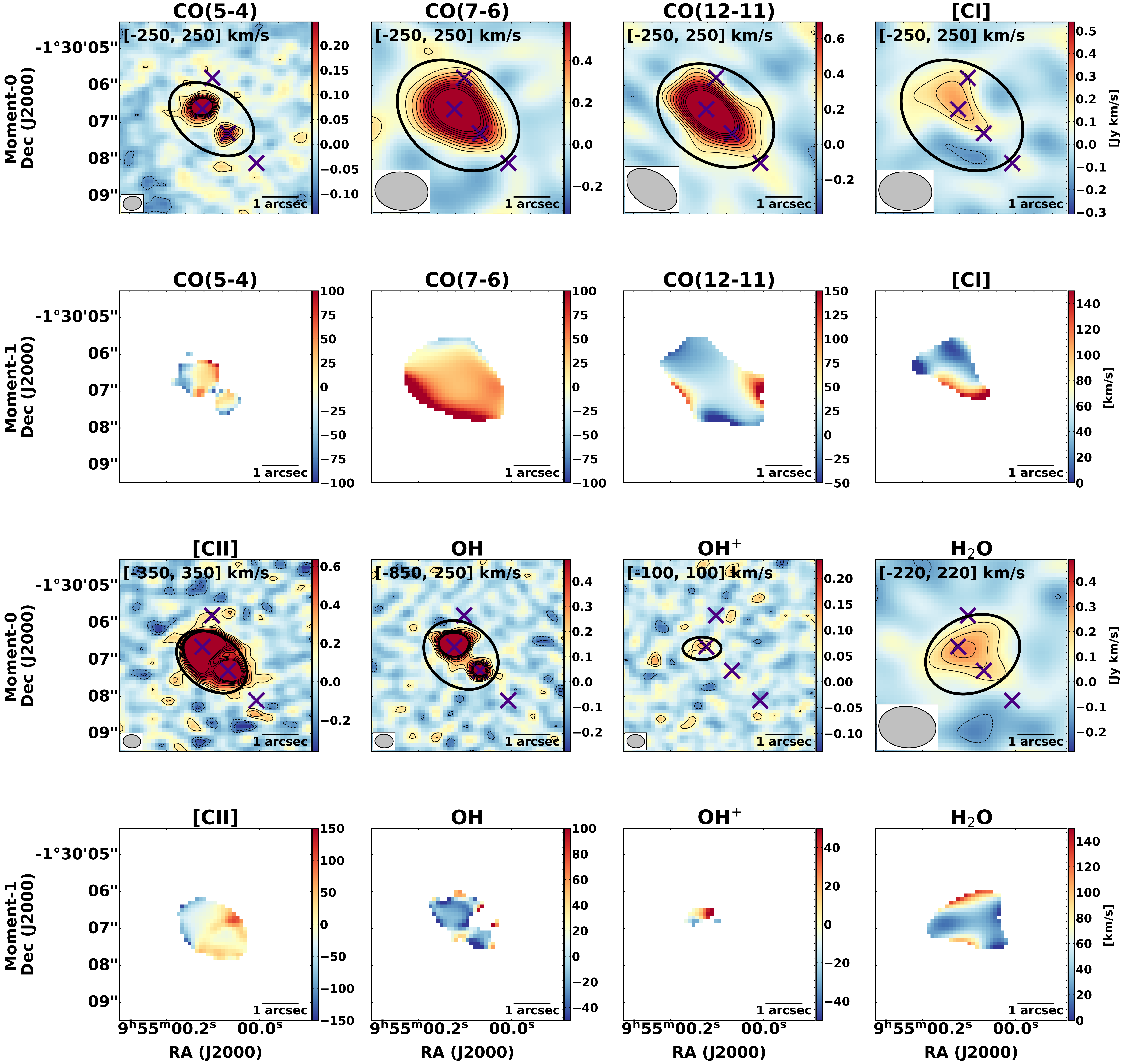}
    \caption{Moment-0 and moment-1 maps of atomic and molecular species detected in BRI\,0952. Rows one and three show the moment-0 maps for the entire field. The purple `X's show the position of Img-N and Img-S of the quasar as well as the location of the two companions, Comp-N and Comp-SW, detected in \citet{Kade23}. Contours are at $-3, -2, 2, 3, 4, 5, 6, 7, 8, 9,$ and $10\sigma$ levels. The black ellipses show the region of spectral extraction and the velocity range used for the extraction is show at the top of the image. Rows two and four show the moment-1 maps within the region of spectral extraction where the flux is limited to only that above a $2\sigma$ level. The synthesized beam is shown in the bottom left of each image.}
    \label{fig:all_mom0}
\end{figure*}
%%%%%%%%%%%%%%%%%%%%%%%%%%%%%%%%%%%%%%%%%%%%%%%%%%%%%%%%%%%%%%%%%%%%%%%%%%
\subsection{Continuum} \label{sec:continuum}
%%%%%%%%%%%%%%%%%%%%%%%%%%%%%%%%%%%%%%%%%%%%%%%%%%%%%%%%%%%%%%%%%%%%%%%%%%
We have detected continuum emission in bands 3, 4, 6, and 7. For details on how we extracted the band 7 continuum data, refer to \citet{Kade23}. For the archival data, we extracted continuum fluxes using {\sc CASA}'s {\sc IMFIT} routine following \citet{Kade23}. Continuum fluxes are provided in Table \ref{tab:cont_fluxes} and images are shown in Fig. \ref{fig:BRI_continuum}. As the spatial resolution of the band 4 and 6 data is lower than that of the band 3 and 7 data, the two images of the quasar are not resolved in the continuum images presented in Fig. \ref{fig:BRI_continuum}. While extended emission is visible in the band 4 continuum data, it is improbable that it arises from the south-western companion as the continuum sensitivity is poorer in the band 4 data than in the band 7 data, and the south-western companion was not detected in the latter \citep[see][]{Kade23}. Without higher resolution continuum data it is not possible to draw robust conclusions about this emission feature. 

\begin{table}[h]
    \centering
    \caption{Continuum properties for BRI\,0952, corrected for lensing.}
    \begin{tabular}{c c c c} \hline \hline
        Band & $\rm S_{\nu}$ & Final Image RMS\\ 
         & [mJy] & [mJy/beam] \\ \hline
        $3^{a}$ & N: $0.08 \pm 0.01$ & 0.01\\
         & S: $<0.03	\pm 0.01$ &  \\
        4 & $0.12 \pm 0.02$ &  0.03 \\
        6 & $0.92 \pm 0.04$ &  0.09 \\
        7 & N:$1.44 \pm 0.40$ &  0.05 \\
         & S:$0.52 \pm 0.14$ &  \\
        \hline
    \end{tabular}
     \tablefoot{
        \tablefoottext{a}{Continuum in band 3 is only detected in the northern image of the quasar, therefore we report a $3\sigma$ upper limit for the southern image of the quasar.}}
    \label{tab:cont_fluxes}
\end{table}

Although the angular resolution of both the band 3 and band 7 data are sufficient to distinguish between the two images (Img-S and Img-N; shown in Fig. \ref{fig:BRI_continuum}) of BRI\,0952, we do not detect Img-S in the band 3 data. This non-detection can be attributed to two factors. Firstly, the relatively faint continuum of Img-S is compounded with the high angular resolution of the observations, making it difficult to detect. Secondly, the frequencies covered by the band 3 data, which probe the long-wavelength end of the Rayleigh-Jeans tail, are intrinsically fainter than those covered by the band 7 data, which probe closer to the peak of the Rayleigh-Jeans regime. This difference in frequency coverage, combined with the native angular resolution of the band 3 observations, may have resulted in the fainter flux from Img-S to go undetected. We note that Img-S is also not visible in the non-tapered continuum image, shown in Fig. \ref{fig:b3_cont_untapered}. 

\begin{figure*}
\centering
\includegraphics[width=0.8\columnwidth]{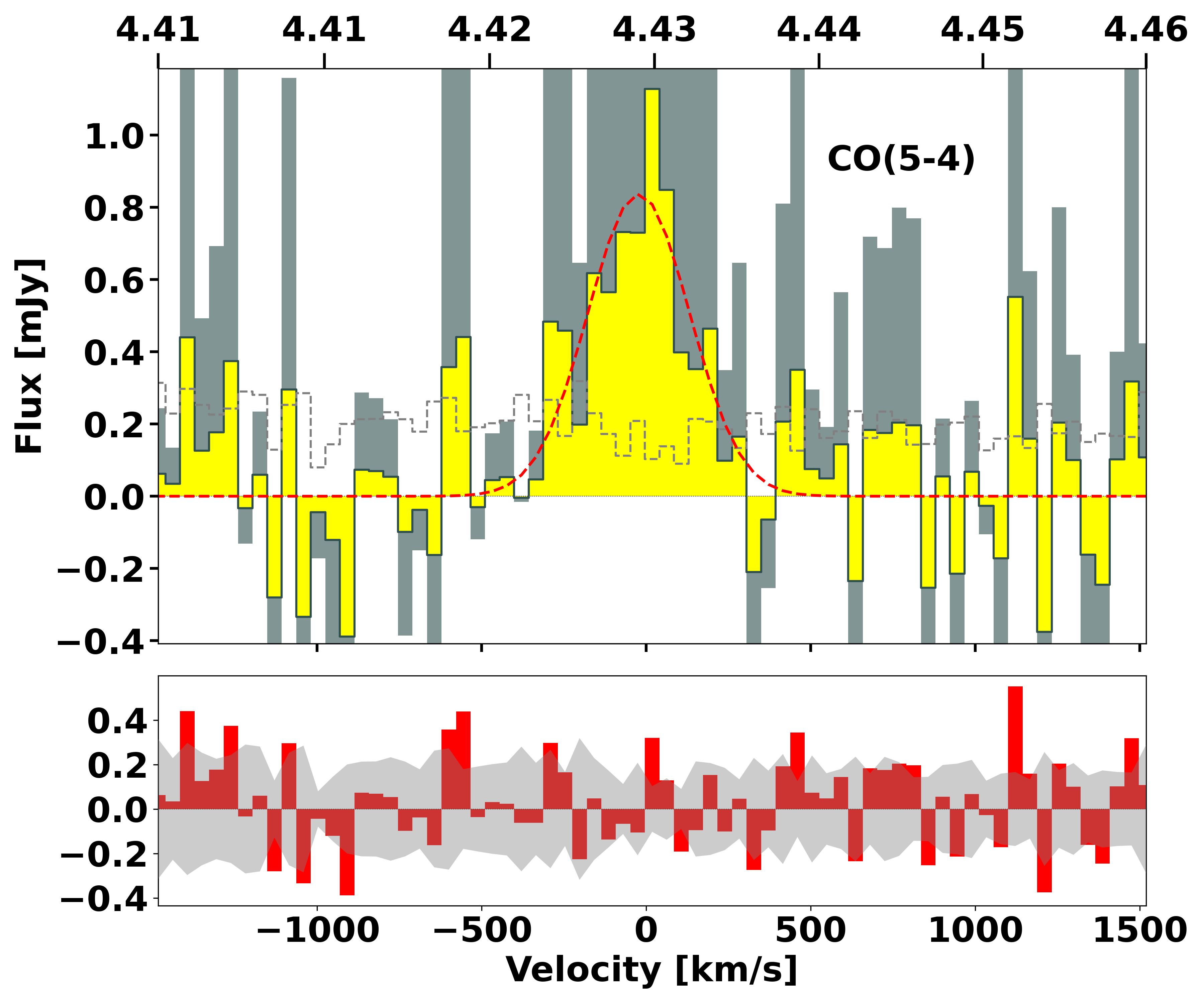}
\hspace{0.5cm}
\includegraphics[width=0.8\columnwidth]{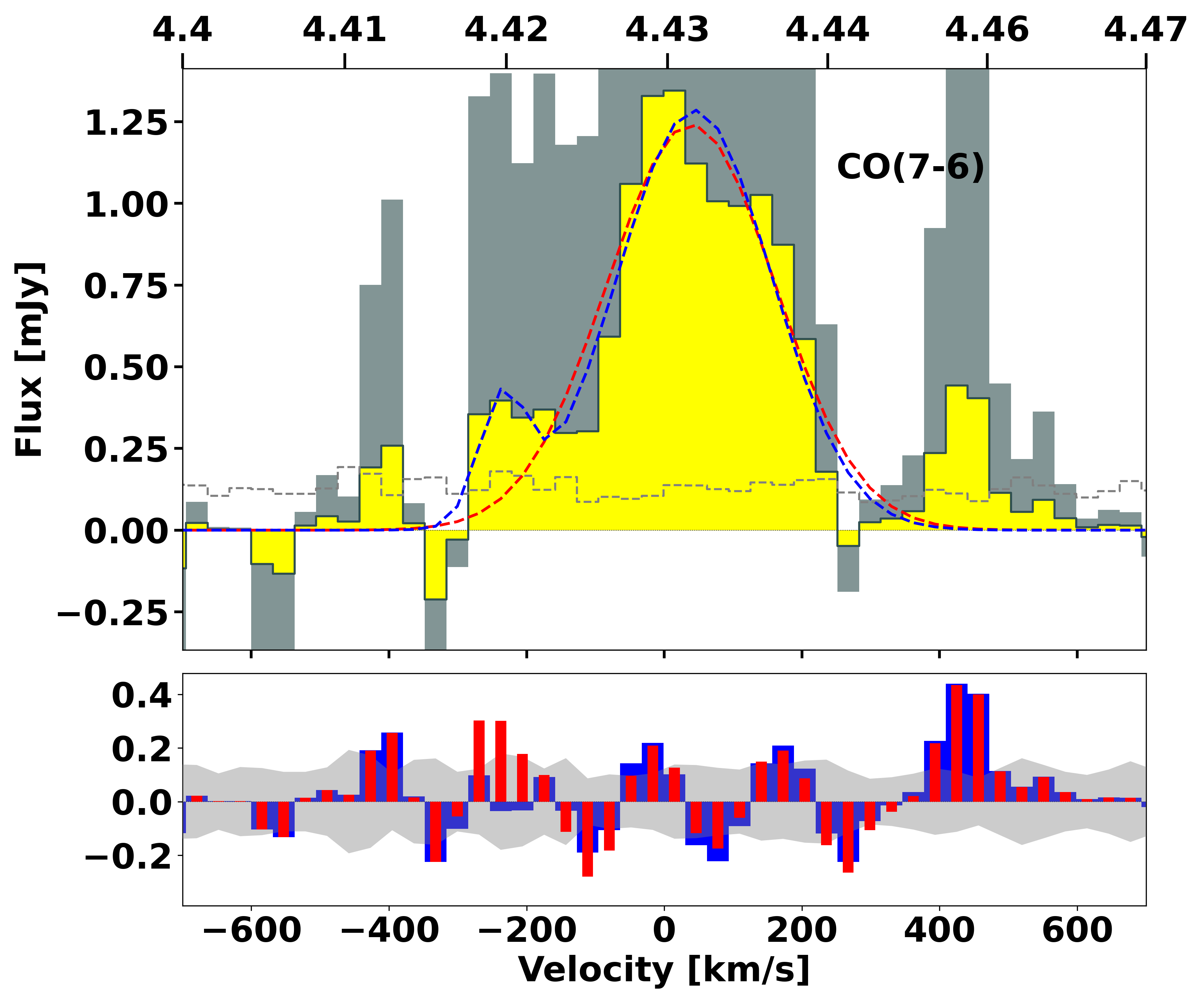}
\hspace{0.5cm}
\includegraphics[width=0.8\columnwidth]{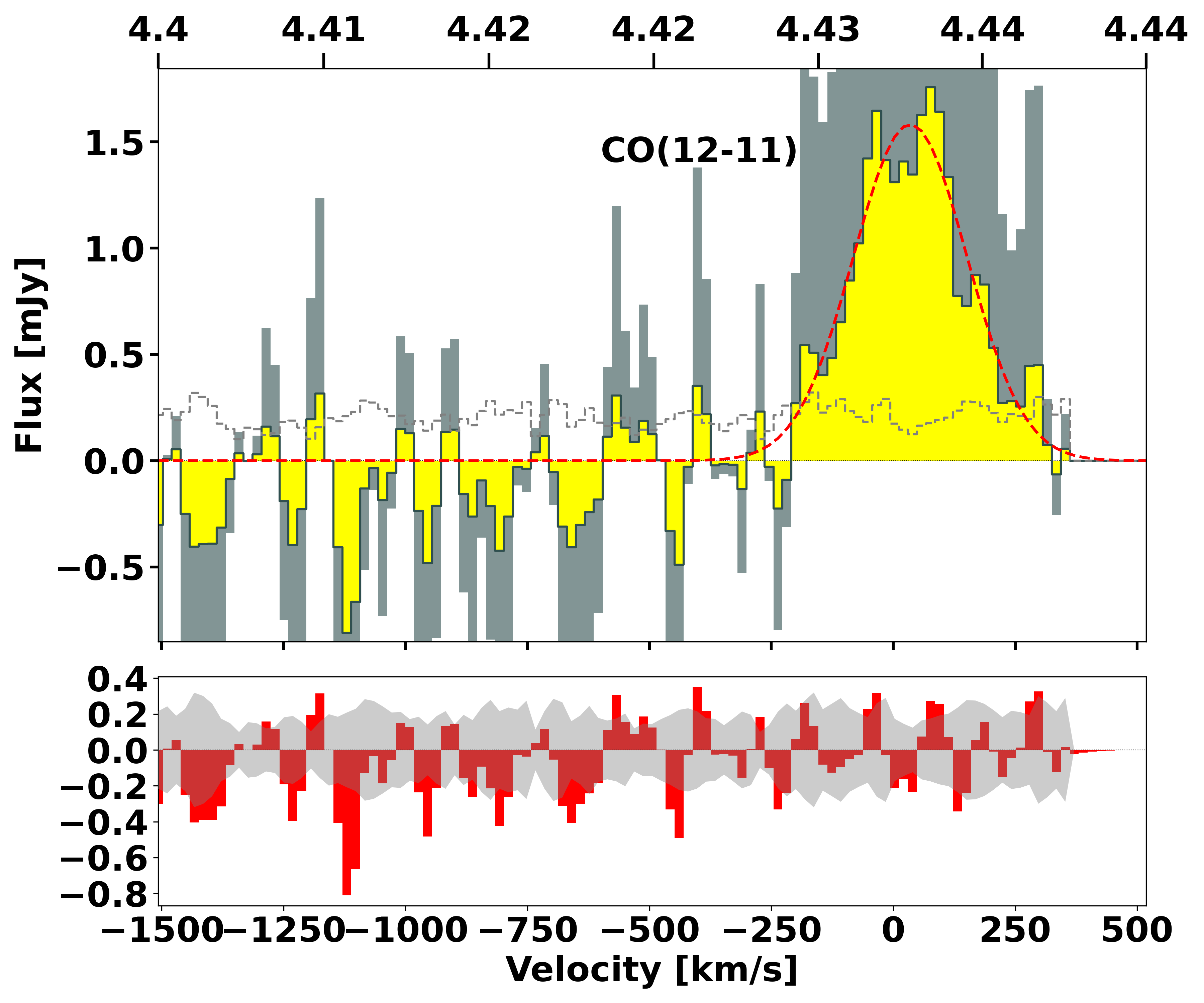}
\hspace{0.5cm}
\includegraphics[width=0.8\columnwidth]{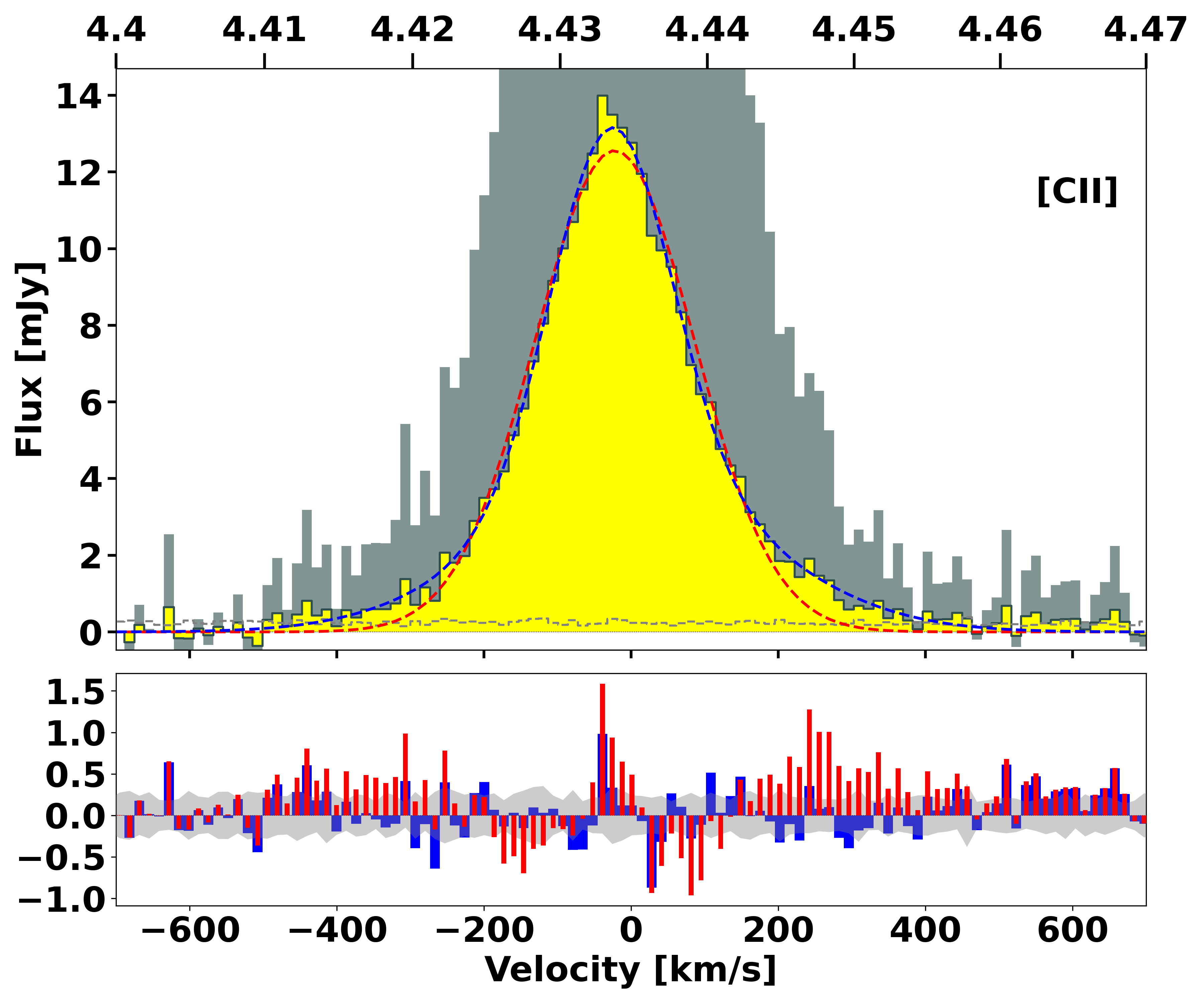}
\includegraphics[width=0.8\columnwidth]{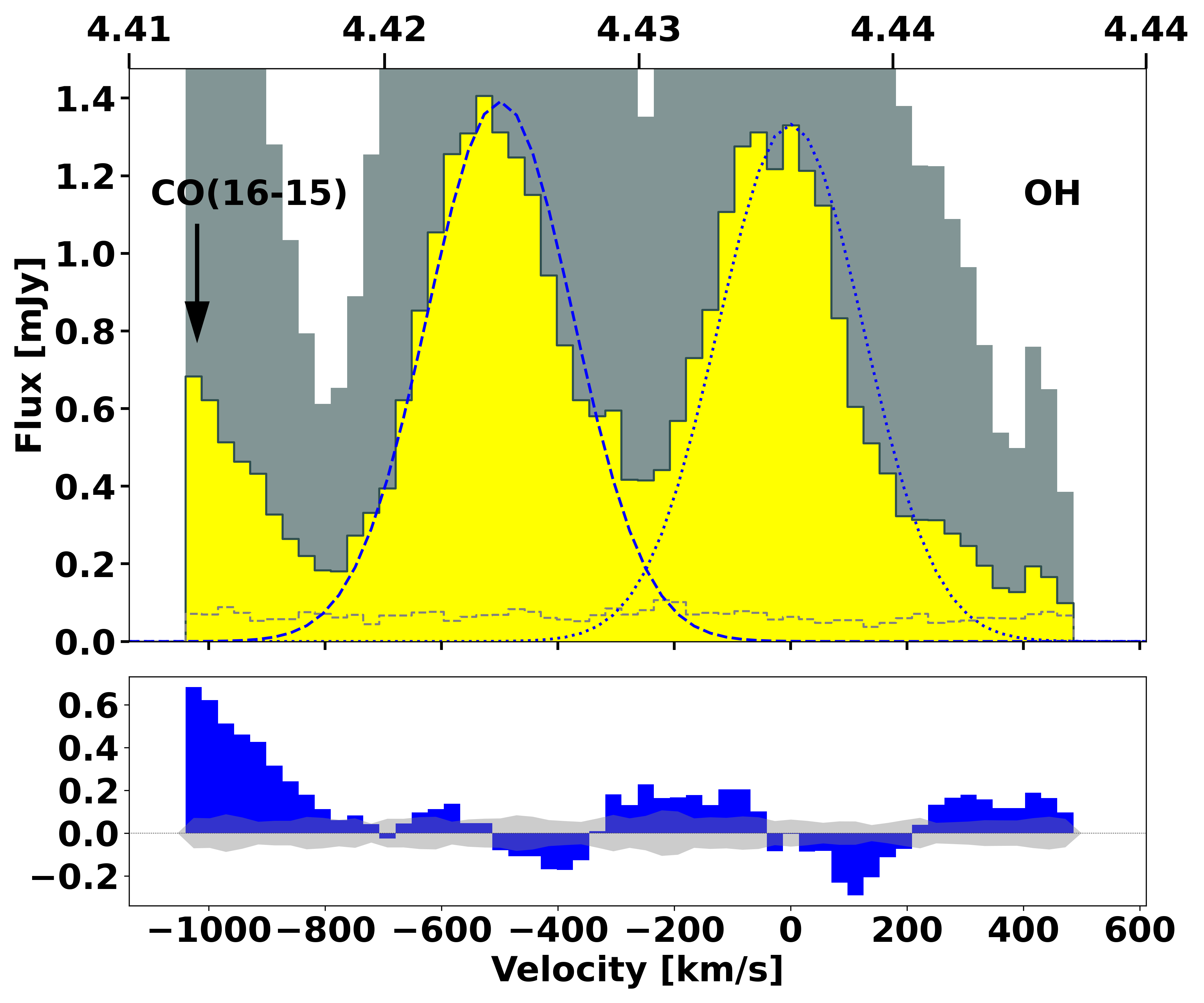}
\caption{Right: Lensing magnification corrected spectra, corrected using the process described in Section \ref{sec:lensing_reconstruction}. The gray background profile is the spectra before lensing correction. Similar to the left panel, single Gaussian fits are shown in red and double Gaussian fits are shown in blue. The dashed gray line represents the rms. The bottom panel shows the residuals from the Gaussian fits where the red corresponds to single fits and the blue to fits using two Gaussian profiles and the gray shaded region represents the rms. The OH redshift-axis is calibrated to the 1834.74735\,GHz line. Note that the red residuals in the bottom panel are shown to be narrower than the blue only the increase clarity.} 
\label{fig:BRI_spec_corrected}
\end{figure*} 

\begin{figure*}
\centering
\includegraphics[width=0.8\columnwidth]{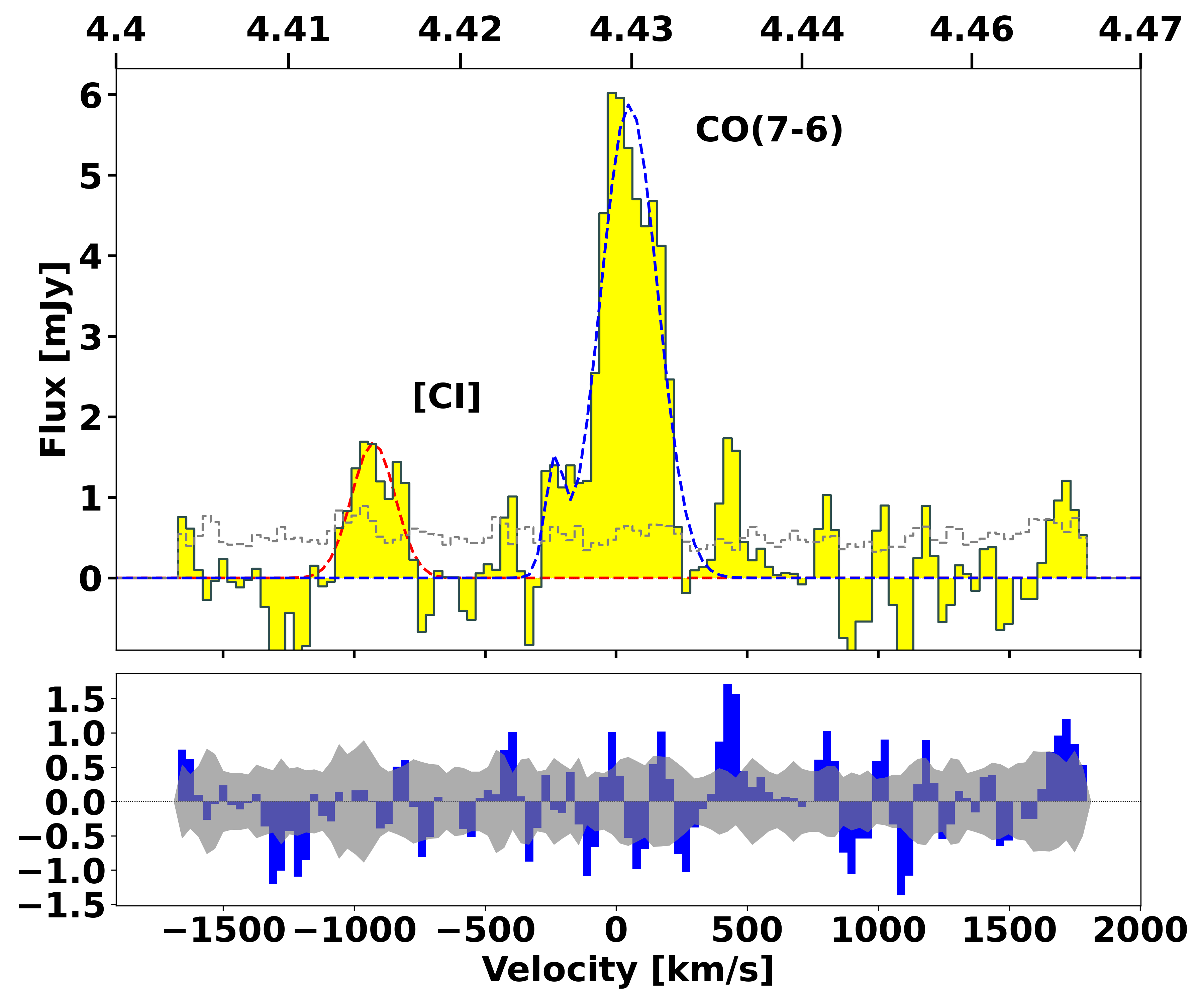}
\hspace{0.5cm}
\includegraphics[width=0.8\columnwidth]{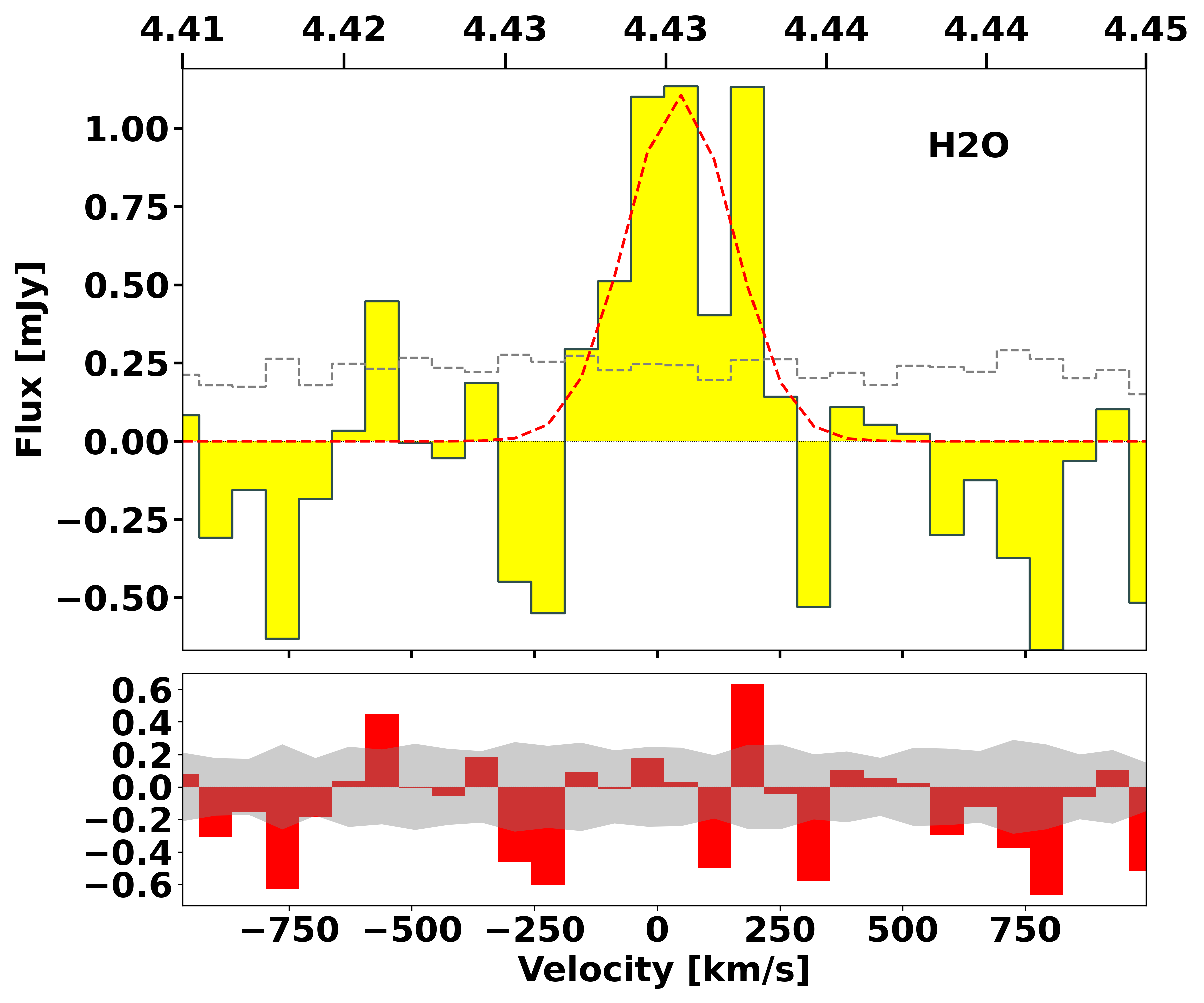}
\hspace{0.5cm}
\includegraphics[width=0.8\columnwidth]{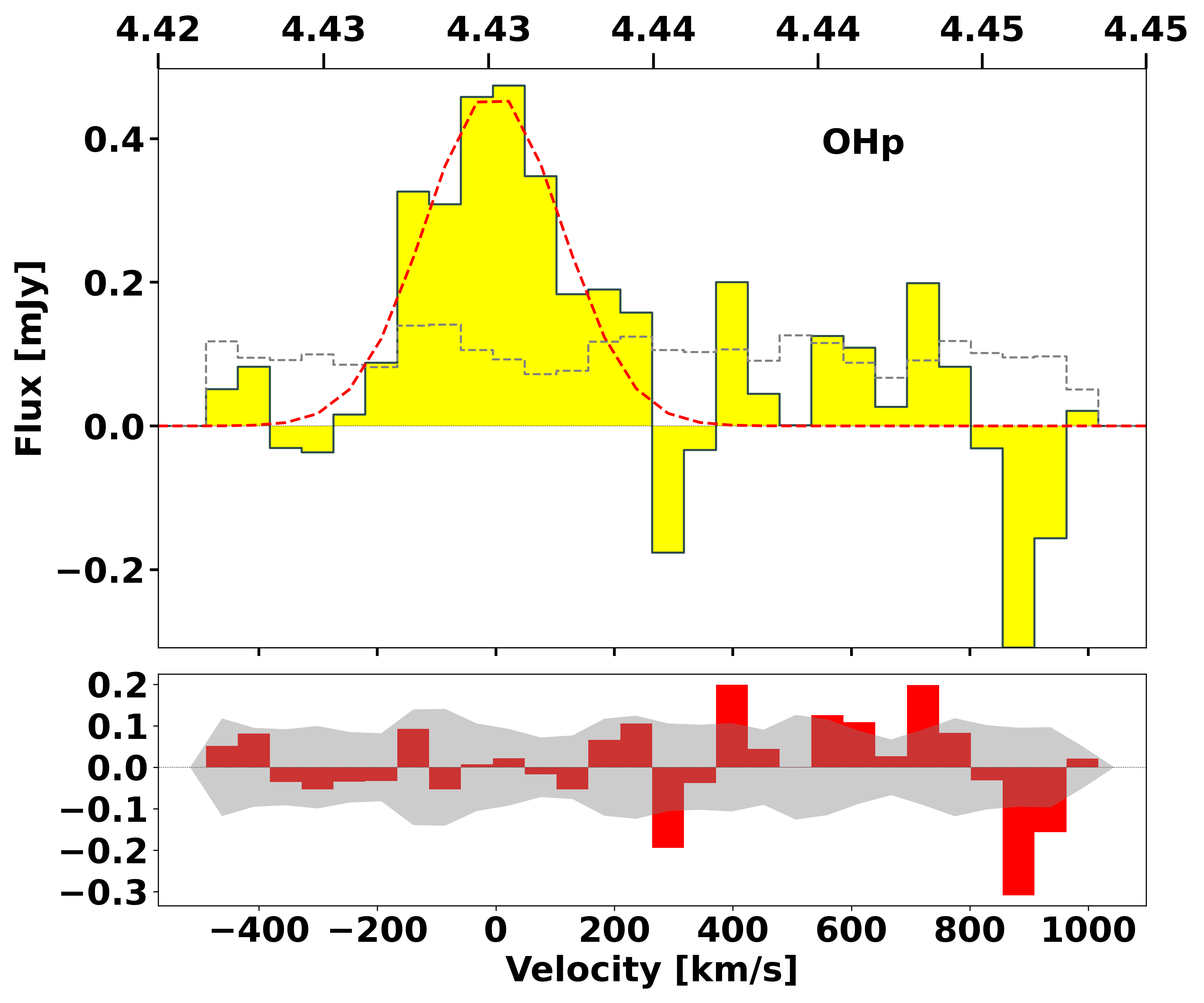}
\caption{Left: Spectra of molecular and atomic species detected in BRI\,0952, not corrected for gravitational lensing. The top panel shows the observed spectra. Single Gaussian fits are shown in red and double Gaussian fits are shown in blue. The dashed gray line represents the rms. The bottom panel shows the residuals from the Gaussian fits where the red corresponds to single fits and the blue to fits using two Gaussian profiles and the gray shaded region represents the rms. The \water and \ohp spectra have been binned by a factor two in frequency/velocity.}
\label{fig:BRI_spec}
\end{figure*}

%%%%%%%%%%%%%%%%%%%%%%%%%%%%%%%%%%%%%%%%%%%%%%%%%%%%%%%%%%%%%%%%%%%%%%%%%%
\subsection{Line Emission} \label{sec:line_emission}
%%%%%%%%%%%%%%%%%%%%%%%%%%%%%%%%%%%%%%%%%%%%%%%%%%%%%%%%%%%%%%%%%%%%%%%%%%

Previous studies of the quasar BRI\,0952 have detected CO(5--4) \citep{Guilloteau99} and \cii emission \citep{Maiolino09, Gallerani12, Kade23}. Here, using a combination of high-resolution ALMA band 7 and ancillary archival data, we report detections of CO(5--4) (rest frequency 576.26\,GHz), CO(7--6) (rest frequency 806.65\,GHz), CO(12--11) (rest frequency 1381.995\,GHz), \ci(2--1) (rest frequency 809.34197\,GHz), \cii(rest frequency 1900.5369\,GHz) \water($2_{11} - 2_{02}$) (rest frequency 752.033\,GHz), OH $^{2}\Pi_{1/2} (3/2-1/2)$ (rest frequencies 1834.749927\,GHz and 1837.746108\,GHz; hereafter OH), and a tentative detection of \ohp ($2_1-1_1$) (rest frequency 1892.23\,GHz). We also report a partial tentative detection of CO(16--15) at the edge of the band 7 spectral window containing the OH emission. 

We do not detect the companion sources, Comp-N or Comp-SW, reported in \cii emission in \citet{Kade23} in any of the additional line species reported in this paper. This is not surprising as the companion sources are very faint in comparison to the quasar ($<10\%$ the flux of the quasar in \cii), and the band 7 \cii data are the data of highest quality among the observations used in this paper. Although the band 3 data has higher native spectral and angular resolution than the band 7 data, it has a shorter integration time and lower sensitivity. This, combined with the significantly lower observed brightness of CO(5--4) in relation to \cii limits the possibility of detecting emission from Comp-N and Comp-SW. Higher angular resolution data with longer integration times than that of the band 4 and 6 data would be required to detect these sources. In general, the observations reveal compact unresolved emission.

We perform a regional spectral extraction for each species detected, meaning we extract the spectrum for a given species from a specific region, shown by the black ellipses in Fig. \ref{fig:all_mom0}, at all $\sigma$ levels. We opt to extract the spectra from these regions rather than performing a $\sigma$ cut as this method has the benefit of 
preserving emission at $\sigma$ levels lower than the chosen cut level; this is particularly important towards the edges of emission regions where the lower-brightness surface emission is distributed \footnote{We note that this is only one method for extracting the spectra. However, the results of this analysis do not change in a meaningful way when instead extracting the spectra above 2 or $3\sigma$ levels.}. This is crucial for fainter the lines, specifically [C\,{\sc i}], H$_2$O, and \ohp lines. We choose the region of extraction to cover the extent of the emission and, where possible, exclude the location of the companions detected in \citet{Kade23}. In some cases it was not possible to entirely exclude the spatial regions in which the companions manifest in [C\,{\sc ii}], however, given the relative faintness of the companions in \cii compared to BRI\,0952, if there was any contribution from the companions to the flux of the line, it would likely be negligible. Although it would be beneficial to use the same region of extraction for all species, specifically the region used in \citet{Kade23}, the differing resolutions results in non-resolved observations in the band 4 and 6 data and therefore we opt to use individual extraction regions for each band in order to ensure we extract all emission and to reduce noise in the cases of \water and \ohp.  

We calculate the rms of the spectra individually for each species as follows. We obtained the per-channel rms of the spectral cube by sampling twelve distinct emission-free regions chosen to be the same size as those utilized for extracting the spectra of the specific species. Fig. \ref{fig:BRI_spec_corrected}, and \ref{fig:BRI_spec} present the spectra of each species detected in BRI\,0952, centered on the redshifted systemic velocity of the line with the exception of the \ci line which is offset from the CO(7--6) line by its relative velocity difference. The moment-0 and moment-1 maps were created from the same regions as used to extract the spectra, however, here we remove flux below a $2\sigma$ level in the moment-1 maps for clarity, shown in Fig. \ref{fig:all_mom0} along with the velocity range used for extraction. Table \ref{tab:BRI_line_fluxes} provides details on the detected emission lines. 

We calculate integrated flux densities and subsequently line luminosities using the following equations from \citet{Solomon97}:

\begin{equation}
    L_{\rm line} = (1.04 \times 10^{-3})\,I_{\rm obs}\, \nu_{\rm rest}\,D^{2}_{L}\,(1+\textit{z})^{-1} [L_{\odot}]
\end{equation}

\begin{equation}
    L^{'}_{\rm line} = (3.25 \times 10^7)\,I_{\rm obs}\,D^{2}_{L}\,(1+\textit{z})^{-3}\, \nu_{\rm obs}^{-2} [\rm K \, km\, s^{-1} pc^{-2}]
\end{equation} 

%%%%%%%%%%%%%%%%%%%%%%%%%%%%%%%%%%%%%%%%%%%%%%%%%%%%%%%%%%%%%%%%%%%%%%%%%%
\subsubsection{CO emission} \label{sec:co_emission}
%%%%%%%%%%%%%%%%%%%%%%%%%%%%%%%%%%%%%%%%%%%%%%%%%%%%%%%%%%%%%%%%%%%%%%%%%%
The CO(5--4) observations were carried out in an extended configuration with a maximum recoverable scale of $\sim 23$\,kpc. We detect CO(5--4) emission at $1.50 \pm 0.32$\,Jy\,km\,s$^{-1}$, which is within 1.5$\sigma$ of the previously reported value of $0.91 \pm 0.11$\,Jy\,km\,s$^{-1}$ from \citet{Guilloteau99}. This suggests that any filtered out diffuse emission is negligible\footnote{Note that neither of these values are corrected for lensing magnification.}. Therefore, it is unlikely that we are missing significant flux. 

The spectrum of the CO(7--6) emission exhibits a `bump' in the blue part of its spectrum at velocities $<-100$\,km\,s$^{-1}$. We fit this line with two Gaussian profiles to account for this. Since the current data is limited by the angular resolution, a spatial correlation between the bump and Img-N or Img-S is not feasible. We must therefore explore other methods of investigation. One possible explanation for the bump is that it is an effect of the gravitational lensing. One simple method of investigation is to extract the spectra from only the very central region of Img-N of the quasar where the flux of the CO line is highest. Hence, we extract the spectra from Img-N at levels >10$\sigma$; however, in this analysis the bump remains. A second investigation would be to use the magnification factor corrected spectra; the process for this is described in Section \ref{sec:lensing_reconstruction}; similarly, in this investigation, the bump remains. It is possible that this feature originates from the gravitational lensing given that the lensing corrected spectra is still biased by the extrapolation of the \cii magnification factor per channel to the CO(7--6) spectra (Section \ref{sec:lensing_reconstruction}). Higher angular resolution data would be necessary to further investigate this possibility. An additional explanation is that this emission originates from one of the companion galaxies detected in \citet{Kade23}, however we discard this possibility as the \cii emission from Comp-N and Comp-SW is located at very similar systemic velocities as that of BRI\,0952. We suggest that this bump provides an indication of the gas kinematics operating in BRI\,0952. The slight offset or asymmetry in the line profile of the CO(7--6) line is reminiscent of the line profile of CO(6--5) detected in the strongly lensed starburst galaxy G09v1.97 \citep{Yang19} in which the authors argued that the complex line profile was an indication of two interacting galaxies. It remains to be seen if this is the precise case for BRI\,0952 but it is suggestive of an on-going kinematical process, be it rotation or interaction. 

There appears to be an elongation of the CO(12--11) emission with respect to the semi-major axis of the emission compared to other emission lines, however the large and extended beam with a similar position angle to the extension of the CO(12--11) data suggests that this is merely an affect of the beam.
%%%%%%%%%%%%%%%%%%%%%%%%%%%%%%%%%%%%%%%%%%%%%%%%%%%%%%%%%%%%%%%%%%%%%%%%%%
\subsubsection{\water ($2_{11} - 2_{02}$)} \label{sec:H2O}
%%%%%%%%%%%%%%%%%%%%%%%%%%%%%%%%%%%%%%%%%%%%%%%%%%%%%%%%%%%%%%%%%%%%%%%%%%
We bin the \water spectra by a factor two to improve the per channel signal-to-noise ratio. We note that the \water line follows the trend observed by \citet{Omont13, Yang17, Stanley21, Li20} of having a similar line profile to that of the CO(7--6) line with a full width half max (FWHM) of $257 \pm 74$\,km\,s$^{-1}$ to the $259 \pm 20$\,km\,s$^{-1}$ of the CO(7--6) line. The CO(5--4) line is significantly broader than the CO(7--6) line, indicating the possibility that the CO(5--4) line originates from a different region than that of the \water and CO(7--6) emission. We show the profile of the normalized \water, CO(5--4), CO(7--6), and \ci spectra in Fig. \ref{fig:h2o_co_spec}. We include the \ci emission as it is likely that these lines trace similar regions of the ISM, see Sections \ref{sec:CI_vs_CO} and \ref{sec:h2o_lir}. We note that the line profile of the \water emission does not exhibit the so-called `bumps' that we find in the CO spectra. One possible explanation for this is the relative faintness of the \water line or the possibility that the `bumps' arise from, for example, an interaction with a companion source. Alternatively, if the \water originates from a different region, specifically a smaller and more compact region, of the galaxy than the CO(5--4) and CO(7--6) lines perhaps this feature would simply not be present in the spectra regardless of data quality. Indeed, Quinatoa et al. (in prep) find different \water and CO line profiles and postulate that the difference is due to a discrepancy in spatial distribution of the emitting regions. 

%%%%%%%%%%%%%%%%%%%%%%%%%%%%%%%%%%%%%%%%%%%%%%%%%%%%%%%%%%%%%%%%%%%%%%%%%%
\subsubsection{OH} \label{sec:OH}
%%%%%%%%%%%%%%%%%%%%%%%%%%%%%%%%%%%%%%%%%%%%%%%%%%%%%%%%%%%%%%%%%%%%%%%%%%
We detect OH emission towards BRI\,0952. As this is a doublet we fit the spectra in a slightly different manner than described for the other lines. We fit an individual Gaussian for both lines of the doublet, but we fix the peak velocity to be the systemic velocity of each line in the doublet. We do not fix the FWHM of the Gaussian to be the same value, but both have the same boundary conditions. We do this in order to determine if there are outflow signatures in the spectra in the form of red- or blue-shifted emission from the systemic velocity. We do not find that the Gaussian fits are offset from the observed emission, see the bottom left image in Fig. \ref{fig:BRI_spec}, providing no indication of in- or out-flowing gas. We find that the individual lines composing the doublet are nearly symmetrical with very similar FWHMs and line fluxes, see Table \ref{tab:BRI_line_fluxes}, in agreement with the detected CO lines. 

%%%%%%%%%%%%%%%%%%%%%%%%%%%%%%%%%%%%%%%%%%%%%%%%%%%%%%%%%%%%%%%%%%%%%%%%%%
\subsubsection{\ohp} \label{sec:OHp}
%%%%%%%%%%%%%%%%%%%%%%%%%%%%%%%%%%%%%%%%%%%%%%%%%%%%%%%%%%%%%%%%%%%%%%%%%%
We report a tentative detection of \ohp emission towards BRI\,0952. This emission arises solely from Img-N in the quasar, see Fig. \ref{fig:all_mom0}. We extract the spectra from a region smaller than that used for the \cii and OH emission to limit the effect of the noise and we bin the spectra by a factor of two to improve the per channel signal-to-noise ratio. The spectrum of the \ohp is shown in Fig. \ref{fig:BRI_spec}. Higher sensitivity data would be necessary to confirm this detection.

\begin{figure}[h]
\centering
\includegraphics[width=1.0\linewidth]{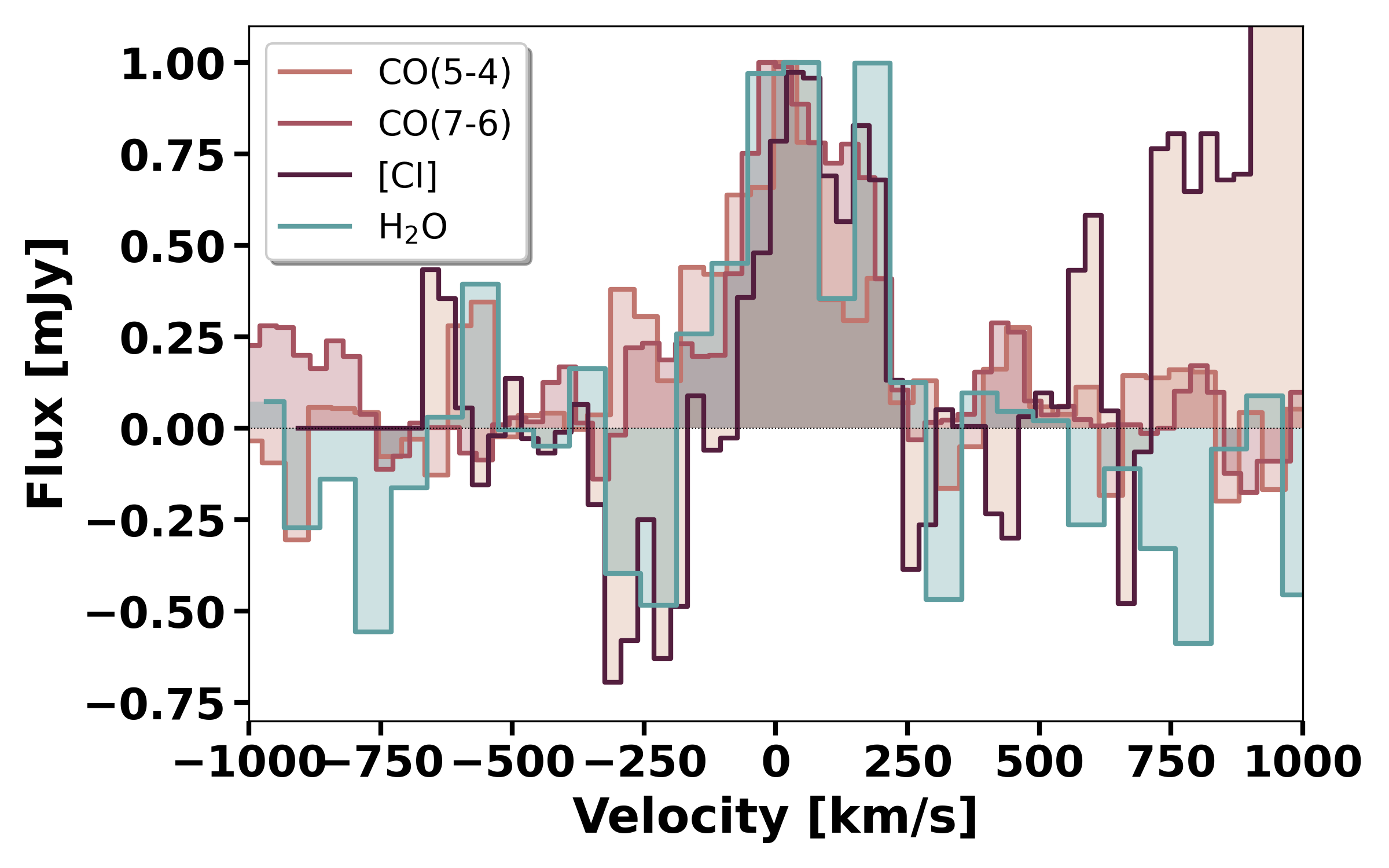}
\caption{Comparison of the normalized H$_2$O, CO(5--4) and CO(7--6) spectra. The CO lines are shown in pink and purple and the \water line is shown in blue. The \ci line appears at $\sim400$\,km\,s$^{-1}$ in the CO(7--6) spectra, and similarly the CO(7--6) line appears in the \ci line profile at $\sim>750$\,km\,s$^{-1}$. The line profile of the \water line broadly follows that of both CO lines as well as the \ci line.} 
\label{fig:h2o_co_spec}
\end{figure} 

%%%%%%%%%%%%%%%%%%%%%%%%%%%%%%%%%%%%%%%%%%%%%%%%%%%%%%%%%%%%%%%%%%%%%%%%%%
\subsection{Lensing Reconstruction} \label{sec:lensing_reconstruction}
%%%%%%%%%%%%%%%%%%%%%%%%%%%%%%%%%%%%%%%%%%%%%%%%%%%%%%%%%%%%%%%%%%%%%%%%%%
BRI\,0952 is a quasar lensed by a single galaxy \citep{Lehar00, Eigenbrod07, Gallerani12, Kade23}. Here we expand upon the lensing model described in \citet{Kade23}. While other papers, such as \citet{Apostolovski19}, have detected multiple atomic or molecular species and modelled their emission individually using gravitational lensing modelling codes like {\sc Visilens} \citep{Spilker16}, our paper focuses solely on modelling the \cii emission. This approach is necessary due to the varying data quality obtained from a combination of archival data and targeted high-resolution observations. 

The \cii line emission provides the highest quality observation of line emission in BRI\,0952 in terms of signal-to-noise, sensitivity, and angular resolution. Moreover, this emission is thought to originate from both colder neutral regions of the ISM and photo-dissociation regions (PDRs), thus likely tracing a large extent of the galaxy \citep[e.g.,][]{Carilli13, Pavesi18}. As a result, we use the \cii emission to determine the magnification factor across the \cii line, which is then extrapolated to the other atomic and molecular species detected in BRI\,0952 based on their systemic velocities. This approach is preferred because it enables us to leverage the high-quality \cii observations to derive magnification factors for other species without having to model each species individually using lower-quality data.

It is important to note that our approach of extrapolating magnification factors from the \cii emission to other species does not take into account that different emission lines likely originate from different locations within the galaxy. For instance, high-excitation CO lines may not necessarily originate in the same locations where \cii emission is present. Nevertheless, as long as the other line species detected follow the bulk motion and systematic kinematics of the entire galaxy, as traced by [C\,{\sc ii}], this method of extrapolating the magnification factor is valid. Biases in the magnification factor between species would arise in cases where individual species exhibit peculiar motions such as outflows or shocks which break from the bulk motion of the gas. While important to consider, with the current data available, specifically with regard to resolution limitations, the errors associated with performing a lensing analysis on lines with insufficient resolution to resolve the two images of the quasar would lead to an unnecessary increase in the systematic errors associated with those methods. 

%%%%%%%%%%%%%%%%%%%%%%%%%%%%%%%%%%%%%%%%%%%%%%%%%%%%%%%%%%%%%%%%%%%%%%%%%%
\subsubsection{{\sc Visilens} Modelling} \label{sec:visilens_modeling}
%%%%%%%%%%%%%%%%%%%%%%%%%%%%%%%%%%%%%%%%%%%%%%%%%%%%%%%%%%%%%%%%%%%%%%%%%%
We utilize the python-based code {\sc Visilens} \citep{Spilker16} to model each channel containing \cii emission. {\sc Visilens} is designed specifically for modeling radio observations of gravitationally lensed systems. The lens parameters used in our study are from \citet{Kade23}, where a full description of these parameters is provided. Although {\sc Visilens} does not generate a source plane reconstruction in the form of a spectral cube, we can perform a spectral reconstruction by modeling each channel in which \cii emission appears and adjusting for the best-fit magnification factor for that specific channel. Fig. \ref{fig:BRI_spec_corrected} and \ref{fig:BRI_spec} displays the magnification factor corrected spectra compared to non-corrected spectra. 

\begin{figure}[h]
    \centering
    \includegraphics[width=1.0\columnwidth]{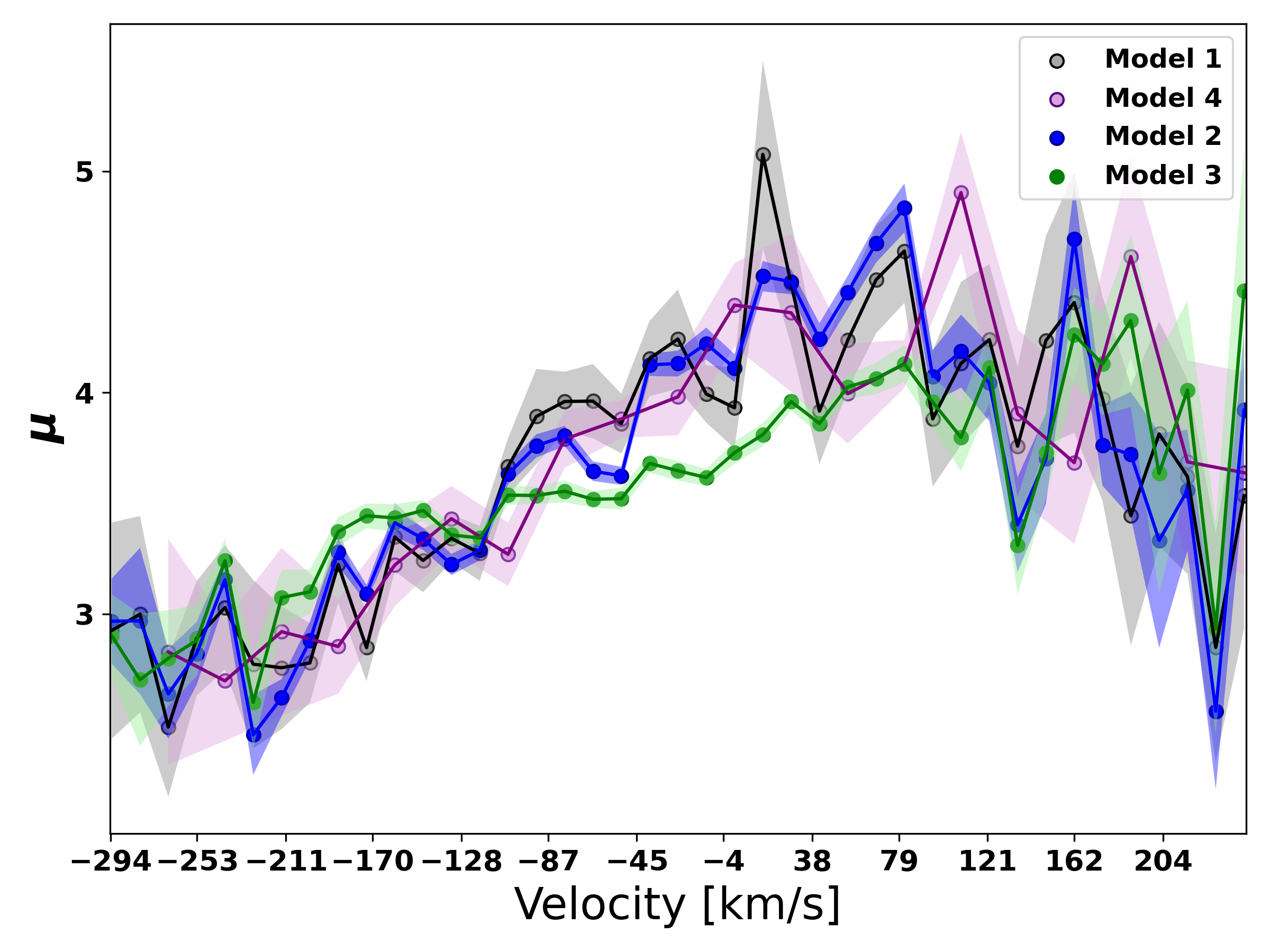}
    \caption{Magnification factor as a function of velocity across the \cii line. The different models are shown in the legend which corresponds to those described in Section \ref{sec:visilens_modeling}. It is clear that each model reproduces approximately the same results across the line where the magnification factor peaks at velocities redder than the systemic velocity of the \cii line.}
    \label{fig:mag_factor}
\end{figure}

We investigate four different methods to model the \cii emission, wherein we describe the source as having a S\'ersic profile parameterized by its position relative to the lens ($x_S$, $y_S$), flux density, S\'ersic index, half-light radius, axis-ratio, and position angle. Rather than optimizing each channel per model separately, we fit every channel using the same set of parameters. Although it is possible that individual channel optimization would marginally improve the accuracy of the model, we conclude that this would unnecessarily introduce user error without sufficient physical motivation. Below we describe the different models: 

\begin{enumerate}
    
    \item Model 1 (model 1 in Fig. \ref{fig:mag_factor} and \ref{fig:BRI_lensing}): We let the parameters vary within a reasonable range around best-fit values from \citet{Kade23}. 
    
    \item  Model 2(model 2 in Fig. \ref{fig:mag_factor} and \ref{fig:BRI_lensing}): The position of the source and S\'ersic index were fixed. We fixed the S\'ersic index to $n = 2.5$ (the average value of $n$ in model 1), and the source position to the best-fit source position in each channel from the first model. The remaining parameters are allowed to vary in the same range as in the first model. 
    
    \item Model 3 (model 3 in Fig. \ref{fig:mag_factor} and \ref{fig:BRI_lensing}): The position of the source and the S\'ersic index were fixed. The S\'ersic index was fixed to $n = 2.5$ as in model 2. In order to determine the source position in each channel we fit a linear regression of the best-fit source position from the first model and adopted those values as the positions in this model. The remaining parameters were allowed to vary in the same range as in the first model. 

    \item Model 4 (model 4 in Fig. \ref{fig:mag_factor} and \ref{fig:BRI_lensing}): The parameters were set to be the same as in model 1, however we bin the input data by a factor of 2 in order to increase data quality. This is particularly important in the high- and low-velocity wings of the \cii line not only due to the relatively lower signal to noise in these channels but it is also of special import due to the wing-like structures noted in \citet{Kade23} which may be indicative of differential magnification and could affect the lensing correction. 
\end{enumerate}

We note that we attempted to fit the \cii emission using a Gaussian source profile. This model is parameterized by its position relative to the lens ($x_S$, $y_S$), flux density, and Gaussian width. Although this model is more simplistic compared to the S\'ersic source parameterization, the physical motivation for one model over the other is not sufficiently clear so as to exclude a Gaussian source being a good description of the light profile of BRI\,0952. However, the fit from {\sc VISILENS} for this model was sufficiently poor that we disregarded it. 

We show the magnification factor as a function of velocity for the different methods described above in Fig. \ref{fig:mag_factor}. We show the change in individual parameters (source position, flux, position angle, etc.) in Fig. \ref{fig:BRI_lensing}. We note that there is no large discrepancy for in the output best-fit parameters between the different models. Most notably, the magnification factor across the velocity range does not change greatly depending on the model used. This is significant for models 1 and 4, specifically in the wings of the \cii line where the emission has lower signal-to-noise ratios. Despite the low emission levels, these models remain consistent with each other, indicating that the variation in magnification factor with velocity is not solely influenced by the lower flux in the wings. Due to the small discrepancies between different methodologies, we choose the most simple and reproducible to use throughout the remainder of this paper; namely, model 1 and model 4. We use the magnification factors obtained from the first model to reconstruct the spectra. An important conclusion from Fig. \ref{fig:mag_factor} is that the magnification factor is not symmetric or stable across the \cii line; therefore, conclusions drawn from the shape of line profiles should only be done using a spectra which has been corrected for magnification. The methodology used to correct the line profiles is described in Section \ref{sec:spec_reconstruction}. 

\begin{figure}
    \centering
    \includegraphics[width=1.0\columnwidth]{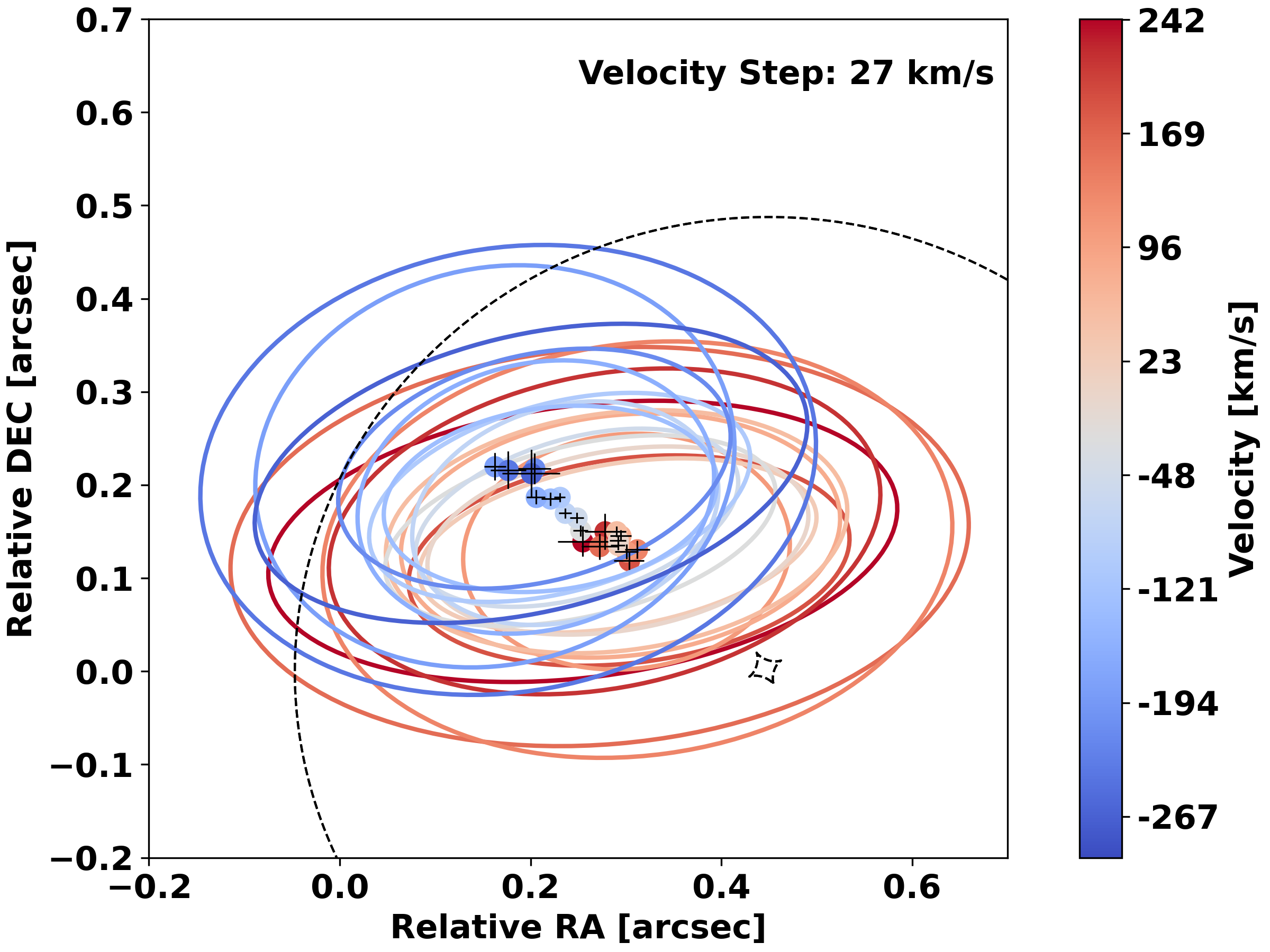}
    \caption{Best-fit position and half-light radius of BRI\,0952 across the \cii line in the image plane from model 4 (Section \ref{sec:visilens_modeling}). The black dashed lines show the critical and caustic lines.}
    \label{fig:Lensing_pv}
\end{figure}

We investigate the change in the position and half-light radius of the best fit S\'ersic source model across the \cii line. We plot the best-fit position and half-light radius as a function of velocity in Fig. \ref{fig:Lensing_pv} using model 4 to improve data quality through binning and ensure clarity. We find clear evidence of a velocity gradient across the line profile, suggesting that the galaxy may be rotating. This result is reminiscent of the results from \citet{Apostolovski19} and \citet{Yang19}.

%%%%%%%%%%%%%%%%%%%%%%%%%%%%%%%%%%%%%%%%%%%%%%%%%%%%%%%%%%%%%%%%%%%%%%%%%%
\subsubsection{Spectral Reconstruction} \label{sec:spec_reconstruction}
%%%%%%%%%%%%%%%%%%%%%%%%%%%%%%%%%%%%%%%%%%%%%%%%%%%%%%%%%%%%%%%%%%%%%%%%%%

We extrapolated the lensing factor based on the velocity relative to the center of the line. To correct the spectrum of the other line species, we used a closest value finding algorithm to search for the velocity bin in the \cii data which is nearest the velocity bin of the particular line and that specific channel was then corrected for the magnification factor in the closest \cii bin. We overplot the magnification corrected spectra in Fig. \ref{fig:BRI_spec_corrected}, where the corrected spectra are shown in yellow and the non-corrected spectra are shown in gray. 

We limit this correction to the lines in which the line profile provides relevant information and high signal-to-noise ratios; that is, the [C\,{\sc ii}], CO, and OH lines. We note a slight offset of the peak of both OH lines in the magnification corrected spectra. This could be a possible indication of gas motion, but given the errors associated with this approach and the limitations of the data, we do not investigate further. The reconstructed CO(7--6) line retains its `bump' in the blue part of the spectrum and it remains unclear what the source of this asymmetry could be. The CO(12--11) line seems to exhibit a double-horned shape, which is could be a feature indicative of rotation or absorption, however, given the low angular-resolution and the errors associated with this type of lensing magnification correction, we do not investigate this feature further. Additionally, the broad \cii wings reported in \citet{Kade23} remain, confirming that these are not due to differential lensing as hypothesized in \citet{Kade23}. Although the true origin of the wings remains unclear, this confirms that they do not arise from the lensing. We present the line luminosity values using $\mu = 3.92$ \citep{Kade23} and the per-channel $\mu$ correction in Table \ref{tab:BRI_line_fluxes}. It is worth noting that the magnification-corrected luminosities generally align within the margin of error comparing between the two methods. For the subsequent sections of this paper we discuss the fluxes and luminosities of the \cii and CO lines employing the magnification-corrected value obtained through the per-channel $\mu$ methodology. 

This approach suffers from the limitations of the method, as described in \ref{sec:visilens_modeling}. Given the constraints imposed by the differing qualities of the ALMA archival data we continue with this method for the remainder of the analysis. For a full analysis of the magnification factor of each individual line to be performed, a consistent survey would need to be done of the CO and \cii lines with the same sensitivity, angular, and spectral resolutions.

\begin{table*}[h]
    \centering
    \caption{Line Properties from our observations.}
    \begin{tabular}{l c c c c c} \hline \hline
        Line & $\nu_{\rm rest}$ & $I_{\mathrm{line}}$ & FWHM & $L_{\rm mol}^{a}$ & $L_{\rm mol}^{b}$ \\  
         & [GHz] & [Jy km $\rm s^{-1}$] & [km\,$\rm s^{-1}$] & [$10^{9}\,L_{\odot}$] & [$10^{9}\,L_{\odot}$]  \\ \hline
        
        CO(5--4) & 576.26 & $1.4 \pm 0.30$ & $278 \pm 46$ & $0.066 \pm 0.014$ & $0.063 \pm 0.02$ \\
        
        CO(7--6) & 806.65 & $1.7 \pm 0.18$  & $82 \pm 48$ & $0.12 \pm 0.012$ & $0.10 \pm 0.01$\\
        & & & $259 \pm 20$  &  & \\
        
        CO(12--11) & 1381.95 & $2.0 \pm 0.18$ &  $258 \pm 17$ &  $0.24 \pm 0.02$ &  $0.22 \pm 0.02$ \\ 
    
        \ci & 809.34 & $0.36 \pm 0.30$ & $195 \pm 123$ & $0.024 \pm 0.020$ & \\ 
        
        \cii & 1900.54 & $15.45 \pm 1.5$ &  $182 \pm 5$ & $2.5 \pm 0.23$ & $2.3 \pm 0.19$ \\
          & & & $410 \pm 21$ & & \\
        
        \water($2_{11}-2_{02}$) & 752.03& $0.31 \pm 0.12$ & $257 \pm 74$  & $0.020 \pm 0.0075$ & \\
        
        OH$^{c}$ & 1834.75 & $1.78 \pm 0.13$ & $270 \pm 15$ &  $1.1 \pm 0.080$ & \\
         
         OH$^{c}$ & 1837.75 & $1.90 \pm 0.14$ & $295 \pm 16$ & $1.1 \pm 0.08$  & \\
         
        \ce{OH^+} & 1892.23 & $0.14 \pm 0.04$ & $275 \pm 57$ & $0.02 \pm 0.006$ & \\ \\
        
        \hline
    \end{tabular}
        \tablefoot{
        \tablefoottext{a}{Corrected for lensing using $\mu = 3.92$ from \citet{Kade23}.}
        \tablefoottext{b}{Corrected for lensing using the per-channel magnification factor from the \cii emission as described in Section \ref{sec:lensing_reconstruction}.}
        \tablefoottext{c}{As the OH line is a doublet, we fit it with two single Gaussian components with a fixed distance between peaks, see Section \ref{sec:OH} for further discussion.}
        }
    \label{tab:BRI_line_fluxes}
\end{table*}

%%%%%%%%%%%%%%%%%%%%%%%%%%%%%%%%%%%%%%%%%%%%%%%%%%%%%%%%%%%%%%%%%%%%%%%%%%
\subsection{Gas Mass} \label{sec:gas_mass}
%%%%%%%%%%%%%%%%%%%%%%%%%%%%%%%%%%%%%%%%%%%%%%%%%%%%%%%%%%%%%%%%%%%%%%%%%%
A common procedure to estimate a galaxy's gas mass is to use CO(1--0) and assume an $\alpha_{\rm CO}$ to convert from CO to $\rm H_{2}$ whereby a total gas mass can be estimated. In cases where CO(1--0) is not observed, a conversion factor is used to translate to CO(1--0) luminosity which is subsequently converted to a total gas mass using the above procedure. However, there are a number of important caveats to these approaches. First, $\alpha_{\rm CO}$ has been shown to vary to a significant amount due to its dependencies on gas density and temperature as well as with galaxy type \citep[e.g.,][ and references therein]{Bolatto13}. In addition, conversions between any observed above-ground state CO line down to the ground state are also dependant on the conversion factor \citep[e.g.,][]{Boogaard20}. A substitute tracer for gas mass is the ground state \ci(1-0) line at 492\,GHz. Studies have shown that this line is linearly correlated with CO(1--0) \citep[e.g.,][and references therein]{Carilli13}. Hence, we can estimate the total gas mass using the equation from \citet{Weiss03}; 

\begin{align}
    M_{\rm [CI]} = 4.556 \times 10^{-4}\,Q(T_{\rm ex}) (\frac{1}{5}e^{\frac{62.5}{T_{\rm ex}}}) L^{'}_{\rm [CI](2-1)} [{\rm M_{\odot}}]
\end{align}
where $Q(T_{\rm ex})$ is the partition function defining the level difference between populations in the upper and ground state of \ci. This is given by the following:
\begin{equation}
    Q(T_{\rm ex}) = 1 + 3e^{\frac{-T_1}{T_{\rm ex}}} + 5e^{\frac{-T_2}{T_{\rm ex}}}    
\end{equation}

where $T_1 = 23.6$\,K and $T_2 = 62.5$\,K corresponding to the \ci(1-0) and \ci(2-1) transitions respectively. For the purposes of this paper we adopt a excitation temperature of $T_{\rm ex} = 30$\,K following \citet{Walter11} and \citet{Jarugula21}. Using this method we find a \ci gas mass of $(2.1 \pm 1.7) \times 10^6$\,M$_{\odot}$. 

Converting from the above calculation to an $\rm H_2$ gas mass can be done using a \ci/$\rm H_2$ ratio corresponding to an assumed abundance of \ci. The chosen abundance ratio value can affect the total molecular gas mass by a factor $\sim 10$ and may vary depending on the galaxy type or region from which the \ci emission arises. For this reason, we choose to use \ci/$\rm H_2 = (8.4 \pm 3.5) \times 10^{-5}$ from \citet{Walter11} which is calibrated from a sample of quasars and submillimeter galaxies at $z > 2$. However, it should be noted that this value may change in high extinction regions \citep[][and references therein]{Walter11}. 

From this result we obtain an $\rm H_{2}$ gas mass of $(6.2 \pm 4.9) \times 10^{9}\,M_{\odot}$. This result is similar to the $\rm H_{2}$ gas mass reported by \citet{Guilloteau99} where they find $\rm M_{H_{2}} \sim (2-3) \times 10^{9}\,M_{\odot}$. However, it is significantly lower than the $\rm H_{2}$ gas mass reported by \citet{Hughes97} where they find a value of $1.35 \times 10^{11} \rm\,M_{\odot}$. However, the methodology varies widely with the different values. \citet{Guilloteau99} use a conversion from the CO(5--4) to gas mass where \citet{Hughes97} calculate the dust mass through continuum measurements (which are likely contaminated by the AGN) and subsequently convert the obtained dust mass to an $\rm H_{2}$ mass. We note that the gas mass obtained through the \ci(2-1) emission is similar to the dynamical mass from the \cii emission reported in \citet{Kade23}. Given the previously mentioned caveats, this value should be interpreted as a proxy and not a robust measurement.

%%%%%%%%%%%%%%%%%%%%%%%%%%%%%%%%%%%%%%%%%%%%%%%%%%%%%%%%%%%%%%%%%%%%%%%%%%
\subsection{Atomic Carbon vs. CO(7--6)} \label{sec:CI_vs_CO}
%%%%%%%%%%%%%%%%%%%%%%%%%%%%%%%%%%%%%%%%%%%%%%%%%%%%%%%%%%%%%%%%%%%%%%%%%%
The ratio of mid-$J$ CO lines to \ci can also be conceptualized as a means of distinguishing the amount of bulk molecular mass (traced by \ci in lower density environments) versus that of molecular mass in dense environments (traced by mid-J CO lines). This was suggested in \citet{Papadopoulos12} for CO(4--3) and \ci(1--0), however studies have extended this method to CO(7--6) and \ci(2--1) \citep[e.g.,][]{Andreani18}. 

The CO(7--6) line is commonly taken to be a reliable tracer of star-formation due to its specific excitation conditions \citep[e.g.,][]{Greve14, Lu15, Yang17}; with excitation potential $T_{\rm ex} = 154$\,K and $n = 4.5 \times 10^5$\,cm$^{-3}$ \citep{Carilli13} this line originates from significantly denser and warmer regions of the ISM than the \ci(2-1) line which has an excitation potential of $T_{\rm ex} = 63$\,K and $n = 1.2 \times 10^3$\,cm$^{-3}$ \citep{Carilli13}. 

An additional powerful tracer of colder and more diffuse regions of the ISM are the \ci lines. Studies have shown that \ci line emission excitation temperature and carbon abundance with respect to hydrogen in low- and high-redshift sources does seem to exhibit a strong evolution with redshift, nor does there appear to be a strong discrepancy in \ci emission between galaxy types \citep[i.e., SMG or AGN;][]{Walter11}. \ci emission has the additional benefit of being less affected by radiation from the CMB and cosmic rays \citep[e.g.,][]{Zhang16, Bisbas15}, therefore the \ci lines have become a common proxy for calculations of the bulk molecular gas mass in galaxies at both low and high redshift \citep[e.g.,][]{} and the ratio of \ci to mid-$J$ CO lines has been used as a tracer for the amount of dense versus diffuse gas \citep[e.g.,][]{Andreani18}. 

The excitation conditions of \ci(2--1) and CO(7--6) are sufficiently different so as to make the ratio between the two a meaningful comparison regarding their excitation properties. Indeed, \citet{Andreani18} find this ratio to be approximately unity for disc-like galaxies such as the Milky Way whereas it may be up to a factor 10 higher in for example merger-driven galaxies. We find a ratio of CO(7--6)/\ci(2--1) $\sim 4.8$. This is significantly higher than the values found in \citet{Andreani18}, suggesting that a merger driven scenario could be occurring. \citet{Scholtz23} reported detections of \ci(2--1) and CO(7--6) in a sample of $z \sim 2.3$ extremely red quasars, from their reported values the average of their is $\sim 2$. The value we find for BRI\,0952 is significantly higher than the ratio found for the dusty star-forming galaxy (DSFG) SPT0311-58 \citep[$z = 6.9$;][]{Jarugula21} and the quasar J2348–3054 \citep[$z = 6.9$;][]{Venemans17}. 

%%%%%%%%%%%%%%%%%%%%%%%%%%%%%%%%%%%%%%%%%%%%%%%%%%%%%%%%%%
\section{Radiative Transfer Modeling} \label{sec:Molpop}
%%%%%%%%%%%%%%%%%%%%%%%%%%%%%%%%%%%%%%%%%%%%%%%%%%%%%%%%%%
We used the radiative transfer code \molpop \citep{Andres_2018} to model the line species detected in BRI\,0952. \molpop allows for exact calculations of multi-level line emission for any atomic or molecular line in the Leiden Atomic and Molecular Database database \citep{Elitzur06, Andres_2018}, and has the added advantage that the code is able to easily incorporate radiation fields from AGN, such as XDRs. \molpop utilizes a slab geometry where the line emitting region is divided into a number of zones as a function of optical depth where the input physical parameters can be either uniform or vary throughout the slab. The physical parameters of the slab are as follows: (1) zone width $\Delta L$; (2) gas volume density within the zone $n\rm (H_{2})$; (3) kinetic temperature $T_{\mathrm{kin}}$: (4) molecular abundance $X_{\mathrm{mol}}$; and (5) local line width (which corresponds to the line absorption/emission profile at each point in the geometry). Given these parameters, \molpop solves the level population problem between each zone in the slab and outputs an emergent flux which can be directly compared to observations. 

\citet{Andres_2018} suggested that even when the parameters are held constant across the entire slab, dividing the slabs into a number of zones helped achieve more accurate results. Hence, we divide the slab into 10 zones and employ a uniform setup in each zone throughout the slab following \citet{Andres_2018} and \citet{Li20}. This approach is particularly important for optically thick lines since the strength of the line emission depends on the distance into the cloud and therefore the the level populations of these lines is dependent on their position within the slab geometry. 

We generate a grid of models by varying the first three parameters, the zone width ($\Delta L$), the zone gas volume density $n(\rm H_2)$, and the zone kinetic temperature ($T_{\mathrm{kin}}$); these parameters provide the most direct means of identifying different regions of the ISM. The zone width was allowed to vary from 16\,cm to 19\,cm in steps of 0.25\,cm. The zone gas volume density was allowed to vary between 3\,cm$^{-3}$ to 6\,cm$^{-3}$ in steps of 0.25\,cm$^{-3}$. The zone kinetic temperature was allowed to vary between 50\,K to 200\,K in steps of 25\,K. The molecular abundance ($X_{\mathrm{species}}$ = [species]/[H$_{2}$]) was set to $10^{-4}$ for CO and $10^{-7}$ for H$_{2}$O, which are typical in molecular gas region \citep[e.g.,][]{Weiss03, Liu17}. The final grid consists of a total of 13$\times$13$\times$9 = 1521 models. 

It is important to note the interdependence of the zone width ($\Delta L$) and zone volume density $n(\rm H_2)$ and the column density within the entire slab ($N_{\mathrm{mol}}$). The total column density, defined as the sum of the column densities within each slab, is proportional to $\Delta L$ through the following: 
\begin{equation} \label{eq1}
N_{\rm{species}} = 10 \times n({\rm H_{2}}) \times \Delta L \times X_{\rm species},    
\end{equation}
where the subscript denotes the specific atom or molecule under consideration and the factor of 10 represents the number of zones. As $N_{\rm species}$ is in terms of both the zone column density and the zone width, there is a degeneracy between these two parameters which is important to consider when comparing different models and therefore throughout the remainder of the paper we use $N_{\rm species}$ when presenting our results. 

We also take into account the effect of the CMB at the source redshift ($z = 4.432$). At this redshift the CMB temperature is $T_{\rm CMB}$ = 14.742\,K using the relation $T_{\rm CMB}(z) = 2.725\,\rm K\times(1+z)$. Hence, the CMB is sufficiently hot to act as a heating source for the CO emission at the redshift of the quasar \citep[e.g.,][]{Andres_2018, Li20}. This is particularly important for line transitions with lower excitation temperature as those lines are more heavily affected by this phenomenon \citep[e.g.,][]{obreschkow_2009, cunha_2013}. 

Apart from allowing for variation of the molecular gas physical parameters within the slab, MOLPOP-CEP also offers the flexibility to incorporate various radiation fields. These radiation fields include the CMB (characterized by the CMB temperature at the specific redshift), blackbody radiation, dust heating, (characterized by the dust temperature and optical depth), and the inclusion of any other radiation field described by its SED. In regard to the CO SLED we consider heating from PDRs and XDRs while for the \water emission we consider heating via dust.

%%%%%%%%%%%%%%%%%%%%%%%%%%%%%%%%%%%%%%%%%%%%%%%%%%%%%%%%%%%%%%%%%%%%%%%%%%
\subsection{Heating via PDRs} \label{ext_heat}
%%%%%%%%%%%%%%%%%%%%%%%%%%%%%%%%%%%%%%%%%%%%%%%%%%%%%%%%%%%%%%%%%%%%%%%%%%
We investigate the contribution of the stellar heating via PDRs to the CO excitation in the quasar. Should the CO-emitting cloud be situated in a star-forming region, then the radiation from nearby young massive stars (mainly type O and B stars) should contribute to the excitation, namely to the mid-$J$ lines. To account for the contribution from a stellar radiation, we used the stellar population synthesis code {\sc Starburst99} \citep{Leitherer_1999} to generate an SED using the same assumptions as \citet{Vallini19} and assuming an SFR of $\sim 3000$\,M$_{\odot}$\,yr$^{-1}$. The stellar radiation field strength is generally given in terms of the interstellar FUV flux, $G_{0}$ \citep[e.g.,][]{Vallini19}:
\begin{equation} \label{eq2}
{G_{0} = \left(\frac{\rm SFR}{\rm SFR_{\rm MW}}\right)G_{\rm 0,MW}}
\end{equation}
where 
$G_{\rm 0,MW} = 1.6 \times 10^{-3}$\,erg\,cm$^{-2}$\,s$^{-1}$ \citep{Habing_1968}. We can express the strength of the radiation field from the SED in terms of $G_0$ and scale this value accordingly to increase the radiation field strength. For example, assuming an SFR of 3000 M$_{\odot}$\,yr$^{-1}$, and setting SFR$_{\rm MW}$ = 1\,M$_{\odot}$\,yr$^{-1}$ \citep[following][]{Vallini19}, the scale of the bolometric stellar radiation at the molecular source is set to be $4.8 \times 10^{-3}$\,W\,m$^{-2}$. In our analysis we consider the stellar radiation SED scaled as $\log(G_{0}) = 0.0$, 1.0, 2.0, 3.0, and 4.0.

%%%%%%%%%%%%%%%%%%%%%%%%%%%%%%%%%%%%%%%%%%%%%%%%%%%%%%%%%%%%%%%%%%%%%%%%%%
\subsection{Heating via X-rays from the AGN}\label{sec:agn_heating}
%%%%%%%%%%%%%%%%%%%%%%%%%%%%%%%%%%%%%%%%%%%%%%%%%%%%%%%%%%%%%%%%%%%%%%%%%%
An alternative heating scenario for the gas in BRI\,0952 is heating via the AGN. We model the AGN radiation in the form of XDRs using an SED for an extreme super-Eddington Type-1 quasar as described in \citet{2017MNRAS.471..706J} from {\sc CLOUDY} covering the frequency range $2.99\times10^{16}$ -- $5.98\times10^{19}$\,Hz \footnote{see Table \textit{$AGN\_Jin17\_Eddr\_highest.sed$} in the CLOUDY documentation: https://gitlab.nublado.org/cloudy/cloudy/-/tree/master/data/SED \citep{2017RMxAA..53..385F}.}. \molpop requires that the flux level of the SED (corresponding to the XDR flux) at the emitting region be specified in terms of both the overall luminosity and the distance to the emitting region. Here we assume a radiation source of luminosity 10$^{13}$\,L$_{\odot}$ at distances of 500\,pc, 1.0\,kpc, and 1.5\,kpc to the emitting cloud, normalized to bolometric energy density $1.45 \times 10^{-1}$\,W\,m$^{-2}$ \footnote{Note that the spatial extent of the major axis in \cii is $\sim 2.3$\,kpc, corrected for lensing \citep{Kade23}.}. 

%%%%%%%%%%%%%%%%%%%%%%%%%%%%%%%%%%%%%%%%%%%%%%%%%%%%%%%%%%%%%%%%%%%%%%%%%%
\subsection{One-Component Models of the ISM} \label{basic}
%%%%%%%%%%%%%%%%%%%%%%%%%%%%%%%%%%%%%%%%%%%%%%%%%%%%%%%%%%%%%%%%%%%%%%%%%%

We model the CO SLED of BRI\,0952 for each of these heating mechanisms, namely PDRs and XDRs, individually. To evaluate the goodness of the fit for each of the models we use a $\chi^{2}$ test, where the $\chi^{2}$ value is a weighted sum of squared deviations. In order to compare the accuracy of the different heating mechanisms we consider only those models with $\chi^{2} \leq 1\sigma$ where $1\sigma = 2.30$ for the degrees of freedom in our model.

We show the best fit and $1\sigma$ models in which $G_{0}$ is scaled to represent different stellar radiation field strengths and for AGN heating at 1.0\,kpc and 1.5\,kpc in \ref{fig:co_sleds_1per}. We discard models with AGN radiation modeled at a distance of 0.5\,kpc as we do not find a reasonable fit to the data in this case. Both heating scenarios fail to reproduce the observed CO fluxes; generally XDR-heating models are able to reproduce the CO(7--6) and CO(12--11) but fail to reproduce the CO(5--4) emission while PDR models reproduce only the CO(12--11) emission and struggle to reproduce the lower-$J$ CO line emission. In the stellar heating regime, the best-fit model corresponds to log($G_0$) = 3.0 and $\rm T_{kin} = 75.0\,K$ and log($N_{\rm CO}$/cm$^{-2}$) = 18.0. We note that there is no difference in the $\chi^2$ value between log($G_0$) = 0.0, 1.0, and 2.0. In the XDR regime, we find a preference for the XDR to be situated at a distance of 1.5\,kpc from the emitting region where the best-fit model has the same physical conditions as the  log($G_0$) = 3.0 model. Given the inability of these models to reproduce the observed CO SLED, we investigate additional heating scenarios below.

\begin{figure*}[h!]
    \includegraphics[width = 1.0\columnwidth]{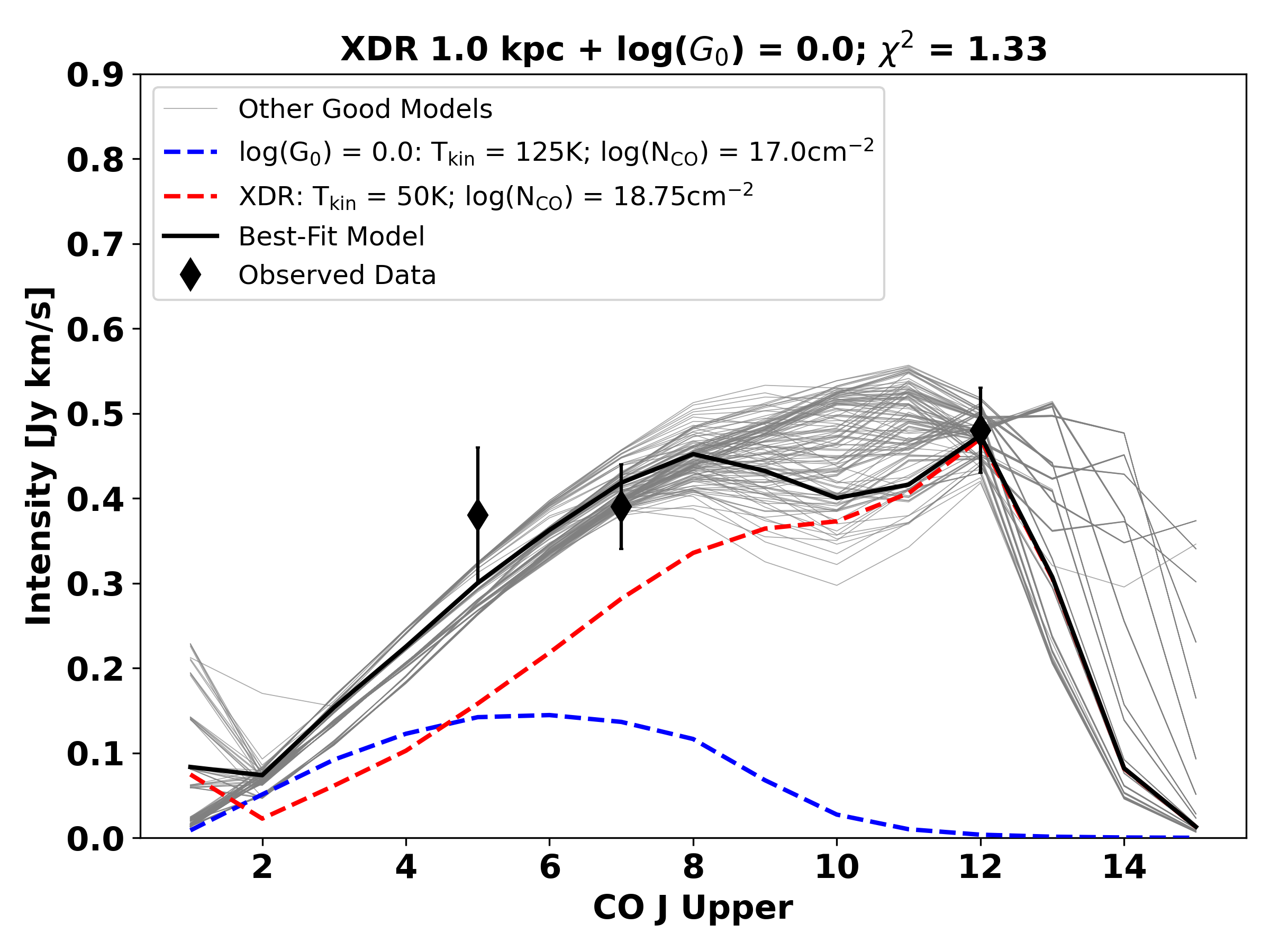}
    \includegraphics[width = 1.0\columnwidth]{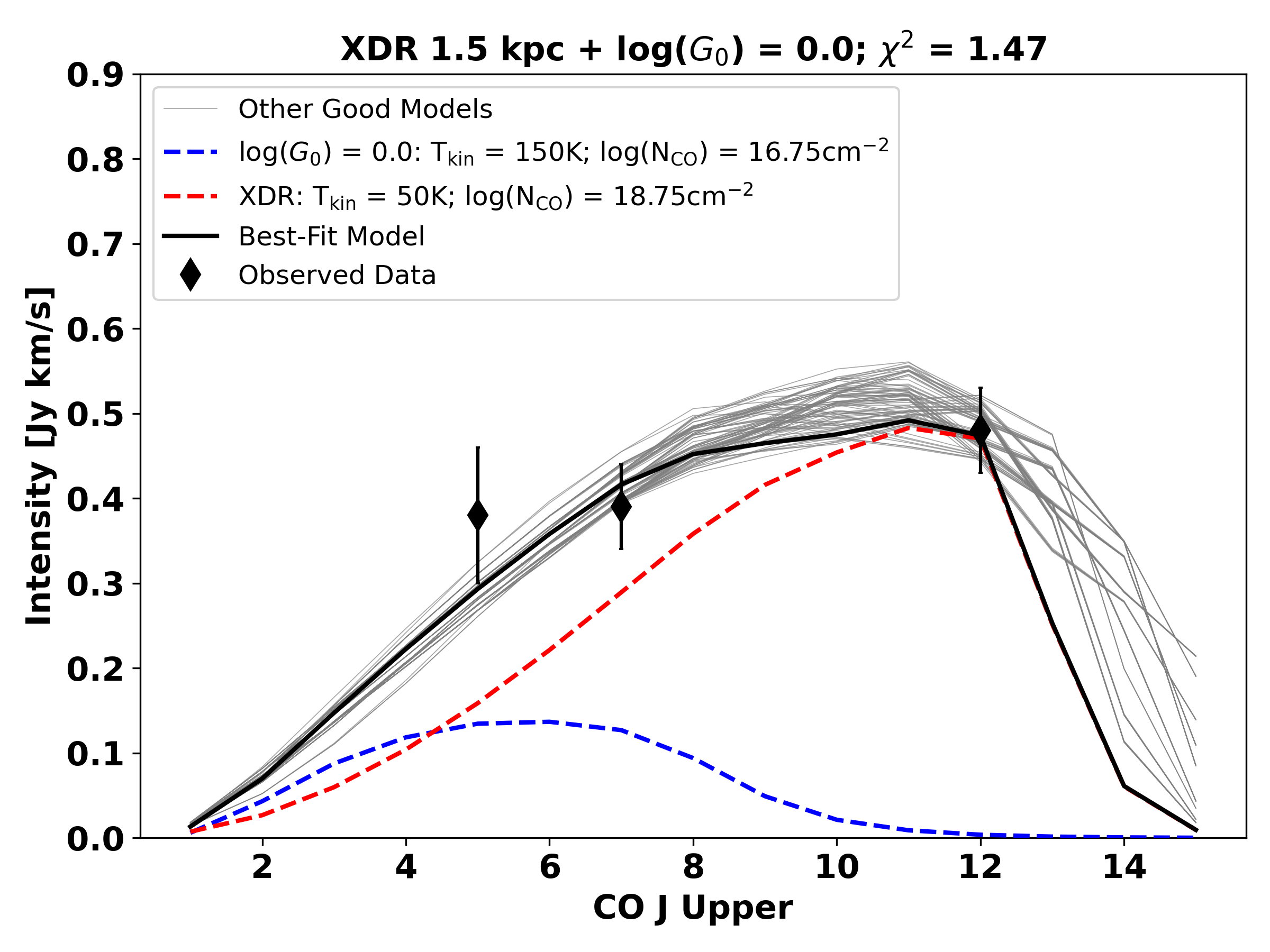}
    \includegraphics[width = 1.0\columnwidth]{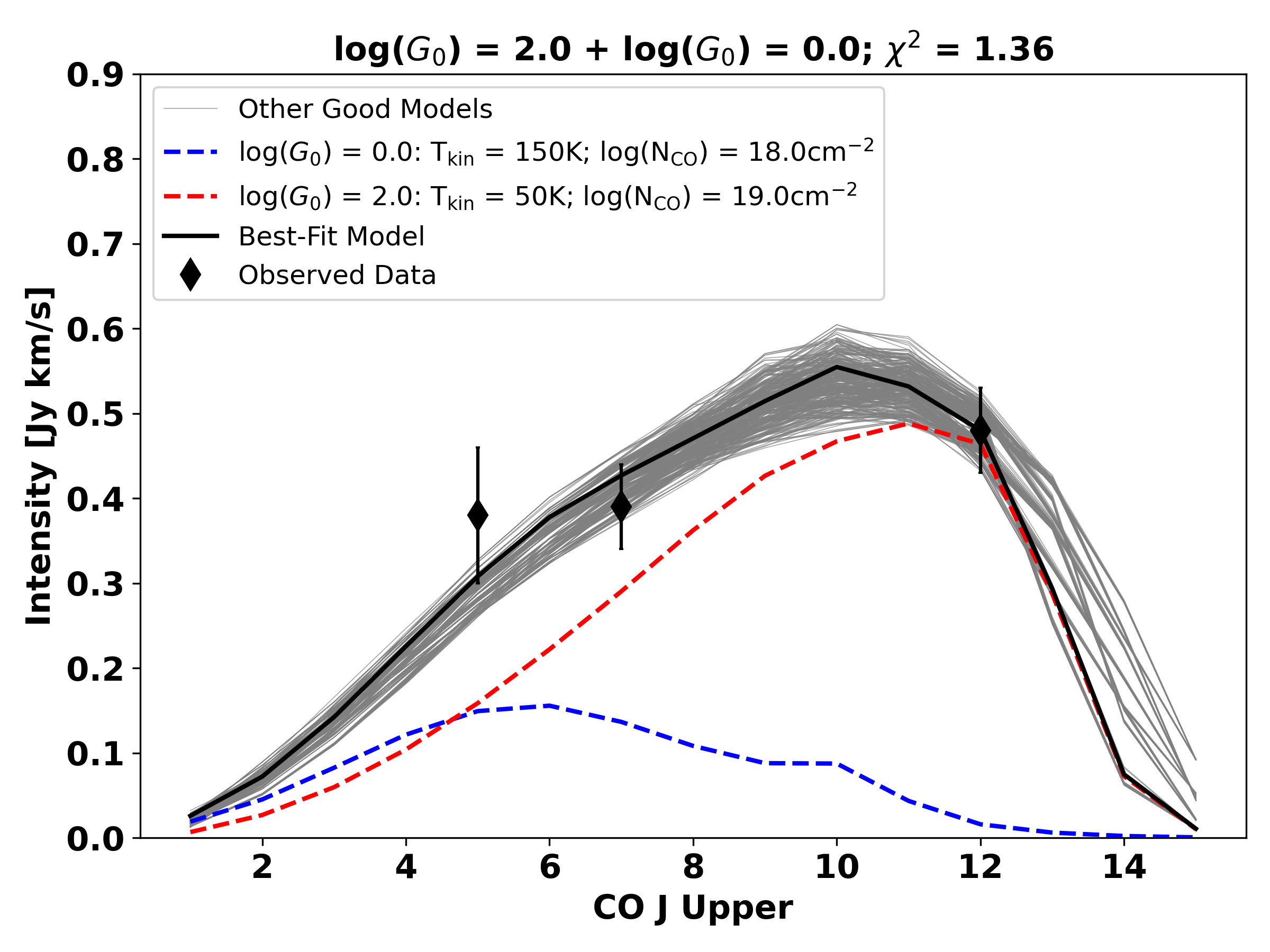}
    \includegraphics[width = 1.0\columnwidth]{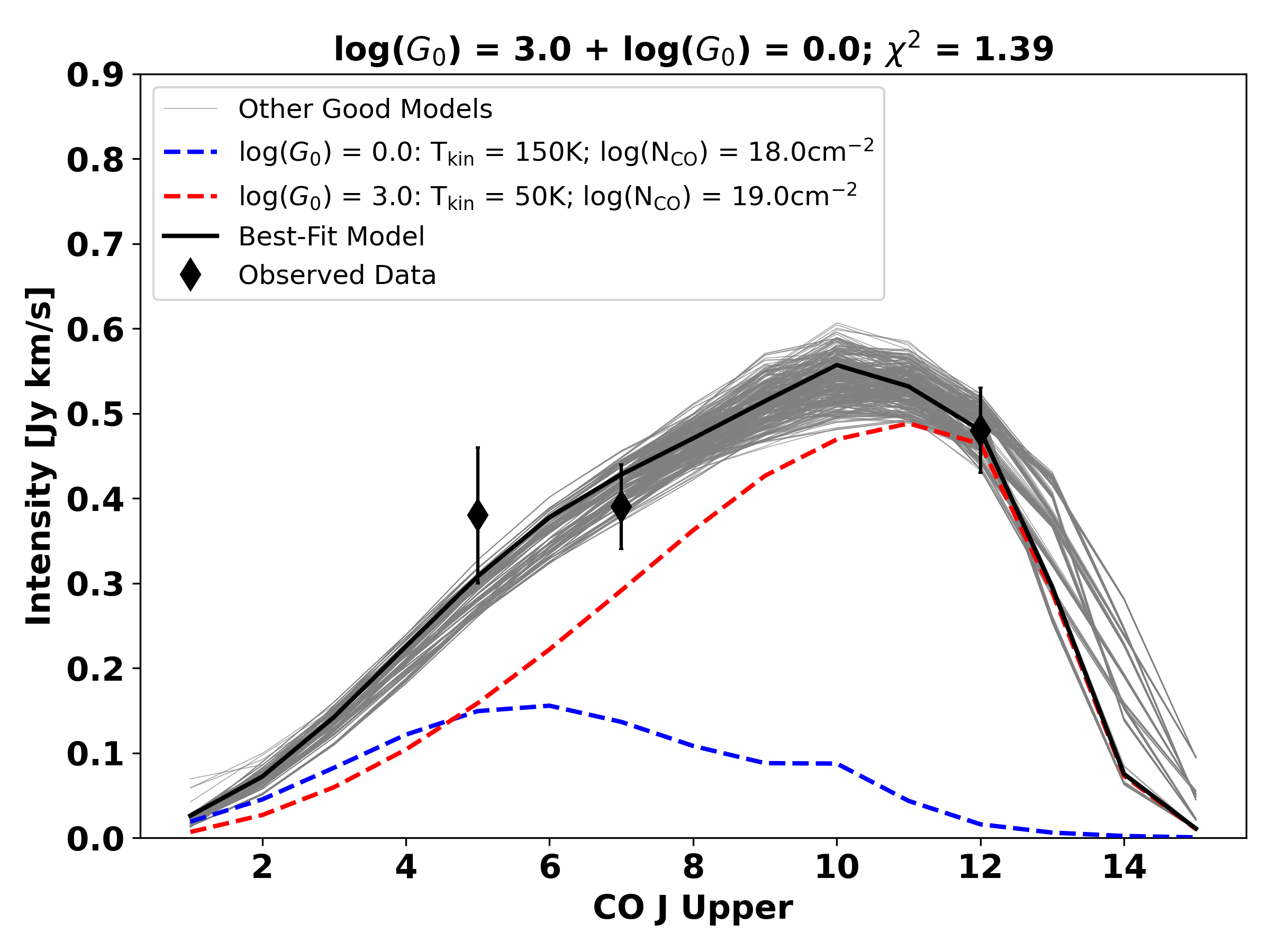}
    \includegraphics[width = 1.0\columnwidth]{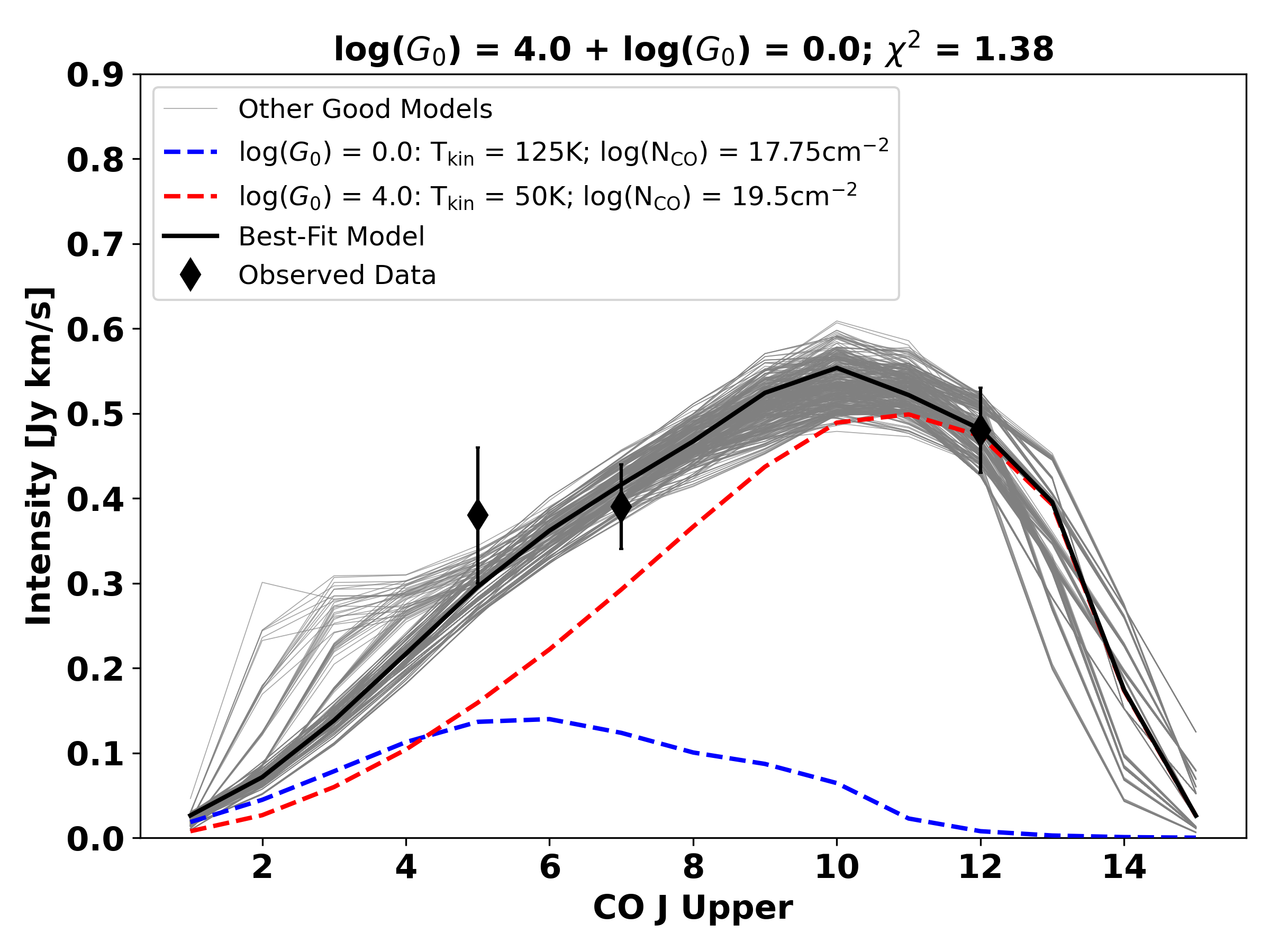}
    \caption{CO SLEDs using a two-component description of the ISM in which one component is represented by PDRs scaled to log($G_0$) = 0.0 (dashed blue line) and the other either PDRs scaled to log($G_0$) = 2.0, 3.0, and 4.0 or an XDR at distances from the line emitting region of 1.0\,kpc and 1.5\,kpc (dashed red line). The solid black line shows the best-fit composite model and the lighter grey lines show other models within $1\sigma$. We note that although the best-fit models appear to have been normalized to the CO(12--11) line, no such normalization was performed.}
    \label{fig:co_sleds_comb_1per}
\end{figure*}

%%%%%%%%%%%%%%%%%%%%%%%%%%%%%%%%%%%%%%%%%%%%%%%%%%%%%%%%%%%%%%%%%%%%%%%%%%
\subsection{Two-Component Models of the ISM} \label{sec:2comp}
%%%%%%%%%%%%%%%%%%%%%%%%%%%%%%%%%%%%%%%%%%%%%%%%%%%%%%%%%%%%%%%%%%%%%%%%%%
It is perhaps not surprising that models that employ a single heating mechanism are insufficient to reproduce the observations of the low-, mid-, and high-J CO transitions due to the different excitation requirements of these transitions. Studies of both local and high-redshift galaxies with high-$J$ detections similar to the CO(12--11) line in BRI\,0952 have been unable to successfully reproduce their observations without the use of a two component model \citep[e.g.,][]{Weiss07, vanderWerf10, Gallerani14, JYang19, Li20}. More specifically, in the case of the local quasar Mrk231 \citep{vanderWerf10}, the high-redshift quasars APM08279+5255 ($z = 3.9$) \citep{Weiss07} and J1148+5251 at ($z = 6.4$) \citep{Gallerani14} an AGN component contributing XDRs was a required to successfully model the CO SLED. However, as noted in \citet{Weiss07}, the hot component need not be directly correlated with the AGN and could instead be a result of for example intense star-formation. 

We explore two scenarios for the two-component models of the ISM; in these scenarios, the heating mechanisms are represented by either two PDRs with varying stellar radiation field intensities, or a combination of a PDR and XDR component. In the former case, one PDR component is modeled as log($G_0$) = 2.0, 3.0, and 4.0 to represent a stronger starburst region while the other component is modeled as log($G_0$) = 0.0. In the latter, the XDR is modeled as in Section \ref{sec:agn_heating} at distances of 1.0\,kpc and 1.5\,kpc from the emitting region, and the PDR is modeled as log($G_0$) = 0.0. In both cases the PDR, log($G_0$) = 0.0, component is expected to contribute more to the lower-$J$ transitions and the scaled PDR or XDR to the higher-$J$ transitions. The latter of these models is of particular interest as it is physically motivated by the high SFR found in \citet{Kade23} which, combined with the clear AGN signatures in the SED \citep{Kade23}, suggests that there is both AGN and star-formation activity influencing the ISM of the quasar. We require the PDR component (modeled as log($G_0$)= 0.0) to have $n < 4.5$\,cm$^{-3}$, physically motivated in our models to represent regions of the ISM where the gas physical properties are not as extreme \footnote{We base this assumption off of modeling from \citet{Vallini19} and \citet{Pensabene21} but also note that this is primarily an attempt to describe a physically motivated model of the ISM.}. We expect this component to contribute primarily to the lower-$J$ transitions. Additionally, we limit the scaled PDR and XDR component to have $T_{\mathrm{kin}} \leq 150$\,K. Although this temperature range goes above the typically expected range for giant molecular clouds (GMCs), we include the possibility of gas with an intrinsically higher temperatures which could be due to non-radiative processes such as shocks from intense star-formation, turbulence, or cosmic rays \citep[e.g.,][]{Ao13}. Given the companion sources, signatures of on-going star formation, and outflow detection, it is likely that these non-radiative mechanisms are also at work in the quasar. We show the best-fit models for these scenarios in Fig. \ref{fig:co_sleds_comb_1per}; given the high number of models when combining all scenarios for the two components we consider good models to be within $1\sigma$. The results of our analysis show that two-component model of the ISM provides a better fit to the observed CO emission than a single-component model.

The limited number of CO lines observed in BRI\,0952 does not allow for an interpretation of the exact heating mechanism in the warm component. Purely in terms of the best $\chi^2$ fit, the best model is composed of a combination of PDR and XDR heating where the XDR is located at a distance of 1.0\,kpc from the emitting region. In this case the physical conditions of the two components are $T_{\mathrm{kin, PDR}} = 125$\,K, log($N_{\rm CO, PDR}$/cm$^{-2}$) = 17, $T_{\mathrm{kin, XDR}} = 50$\,K, and log($N_{\rm CO, XDR}$/cm$^{-2}$) = 18.75. However, given the very small difference in $\chi^2$ between this model and the best-fit pure PDR two-component model of $\chi^2$ = 1.33 for the former and $\chi^2$ = 1.36 for the latter, either PDR or AGN heating can reproduce the CO SLED. We find that the best fit model for all five different two-component models have very similar physical conditions, suggesting that these conditions may be indicative of the intrinsic properties of the gas in BRI\,0952. However, the wide range for each parameter that still provides a fit within the $\chi^{2} \leq 1\sigma$ range (see Table \ref{tab:molpop_1sig_parameterrange}) suggests that the current data is not capable of providing tight constraints on the true intrinsic properties of the gas. Further, without additional constraints specifically regarding the turnover point of the CO SLED and the behavior of the CO SLED at $J > 12$ a preferred distinction cannot be made between these heating mechanisms. However, the partial wing detection of CO(16--15), not included in the radiative transfer modeling given the tentativeness of the detection, suggests that there is still significant emission out to $J_{\rm upper} = 16$ (see Section \ref{sec:co_sled}) and therefore the favored scenario would be that of a combination of XDR and PDR heating as those models show significant emission out to $J_{\rm upper} = 15$. The implications of this are further discussed in Section \ref{sec:ISM_BRI}. 

In order to better understand the similarity in best-fit physical parameters between different two-component models, we investigate the likelihood distribution for each parameter two-component regime. We follow the approach of \citet{Ward03}, \citet{Kamenetzky11}, and \citet{GonzalezAlfonso_21} for calculating the likelihood distributions for each model given a vector, \textit{a}, containing the different model parameters. The Bayesian likelihood of a specific model given a set of measurements, \textit{M}, and parameters \textit{a} is given by,

\begin{equation}
    P(a|M, \sigma) = \frac{P(a)P(M|a, \sigma)}{\int da P(a)P(M|a, \sigma)}
\end{equation}

where $\sigma$ is the measurement error and P(\textit{a}) is the prior probability density function. The probability density function, $P(M|a)$ , is given by the following,

\begin{equation}
    P(M|a, \sigma) = \prod_{i = 1}^{n} \frac{1}{\sqrt{2\pi}\sigma_{i}} \, \exp \left(  \frac{-1}{2} \left[ \frac{M^{obs}_i - M_{i}}{\sigma_i} \right] \right)
\end{equation}

Here we define the prior probability function P(\textit{a}) as P(\textit{a}) = 1 for $T_{\mathrm{kin}}$ and $\Delta L$ in the log(G$_0$) = 0.0 component and for the scaled PDR or XDR we define P(\textit{a}) = 1 for $\Delta L$ and $n(\rm H_2)$ and as the following for $T_{\mathrm{kin}}$,

\[
    P(a)_{n(\rm H_2, log(G_0) = 0.0}= 
\begin{dcases}
    1, & \text{if } n(\rm H_{2}) \geq 4.5\,cm^{-3} \\
    0,              & \text{otherwise}
\end{dcases}
\]

\[
    P(a)_{T_{kin, scaled PDR or XDR}}= 
\begin{dcases}
    1, & \text{if } T_{kin} \leq 150\,K \\
    0,              & \text{otherwise}
\end{dcases}
\]

\begin{figure*}
    \centering
    \includegraphics[width = 1.0\linewidth]{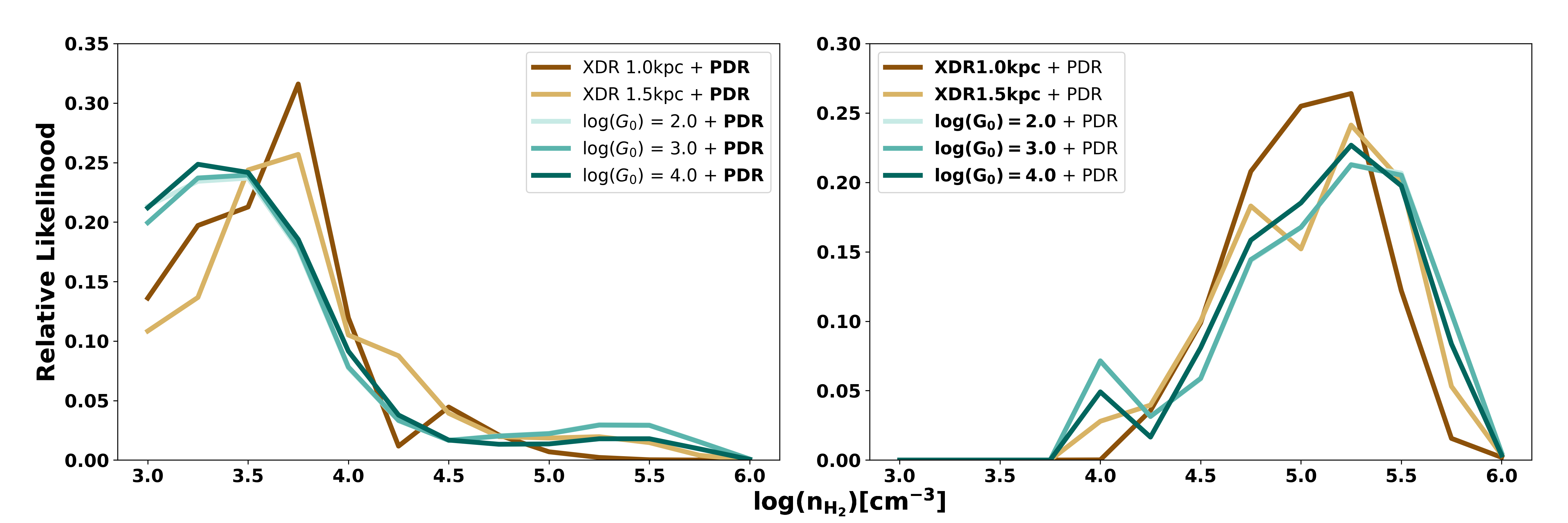}
    \includegraphics[width = 1.0\linewidth]{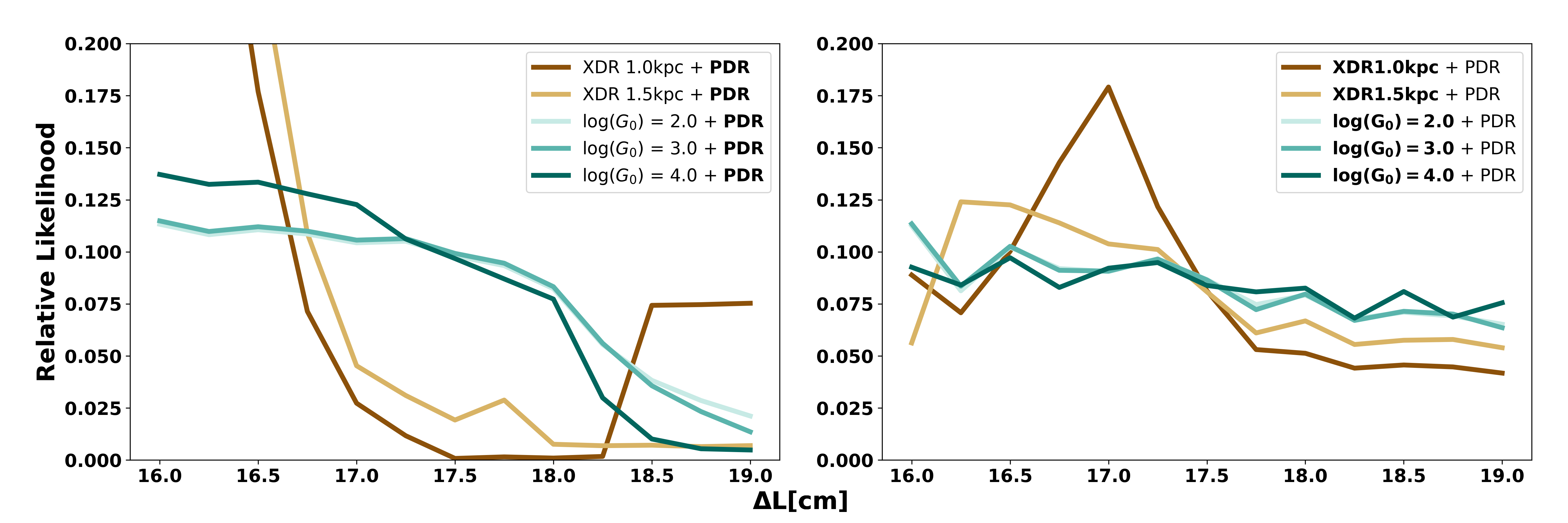}
    \includegraphics[width = 1.0\linewidth]{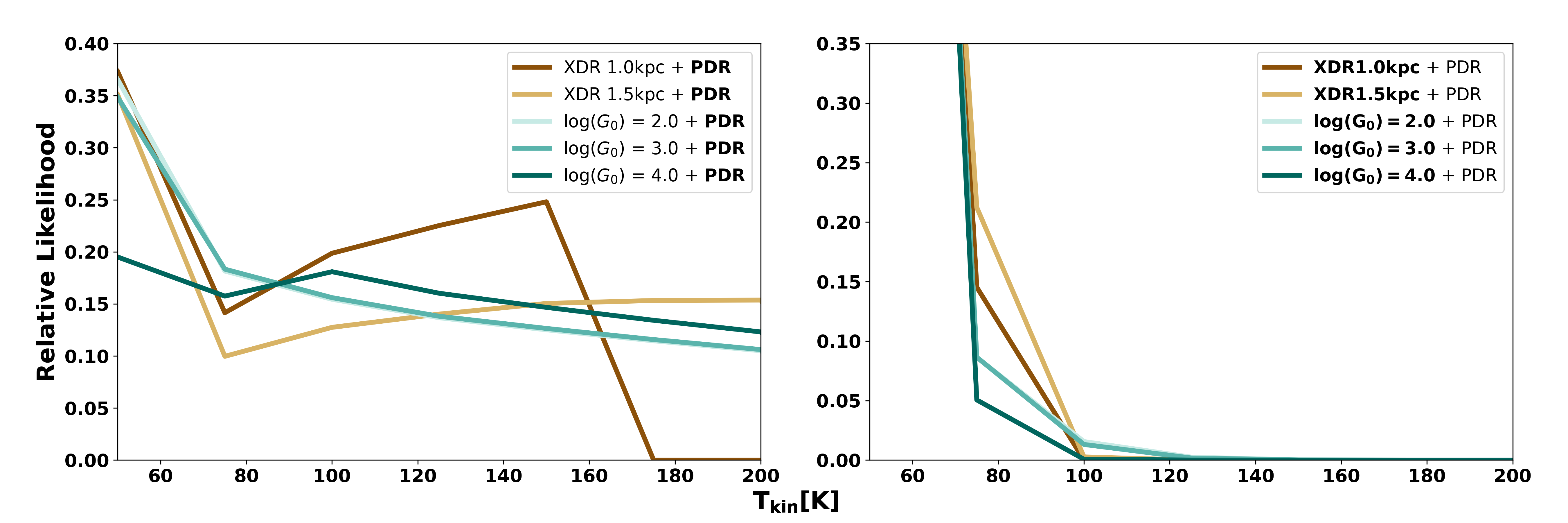}
    \caption{Relative likelihood of each input parameter to MOLPOP-CEP in different heating scenarios. The left column shows the relative likelihood of each parameter in the PDR component modeled by log($G_0$) = 0.0 (in bold in the legend) and the right shows the same for PDRs scaled to log($G_0$) = 2.0, 3.0, and 4.0 or an XDR at distances from the line emitting region of 1.0\,kpc and 1.5\,kpc (in bold in the legend). These models correspond to those shown in Fig. \ref{fig:co_sleds_comb_1per}.}
    \label{fig:likelihood}
\end{figure*}

We show the relative likelihood for each parameter ($n(\rm H_2)$, $\Delta L$, and $T_{\mathrm{kin}}$) in both the scaled PDR or XDR component and the PDR, log($G_0$) = 0.0, component in Fig. \ref{fig:likelihood}. It is clear that the preferred kinetic temperature and zone gas volume density for both components remains relatively constant across the different heating scenarios. The only parameter that shows variation across the different heating scenarios is the zone width, wherein the log($G_0$) = 0.0 PDR component favors lower zone widths in the XDR regime but is relatively flat for the PDR+PDR case. The small discrepancies between different heating scenarios provide further evidence that there is little statistical difference between scenarios; however, we maintain that the most physically motivated description is that of a combination of XDR and PDR heating mechanisms.

\subsection{Modeling of \cii and \ci}

\begin{figure*}
    \centering
    \includegraphics[width = 0.99\columnwidth]{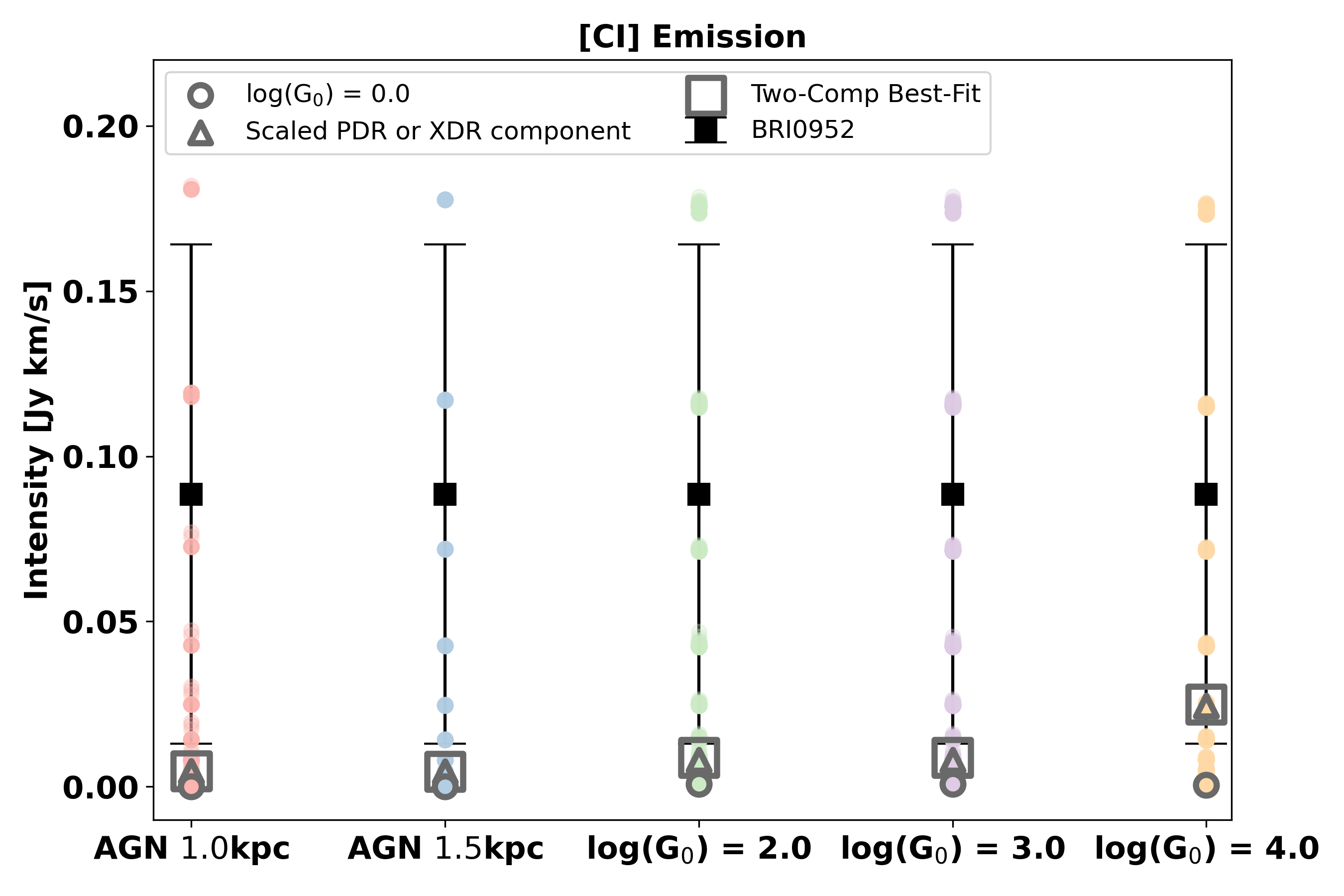}
    \includegraphics[width = 0.99\columnwidth]{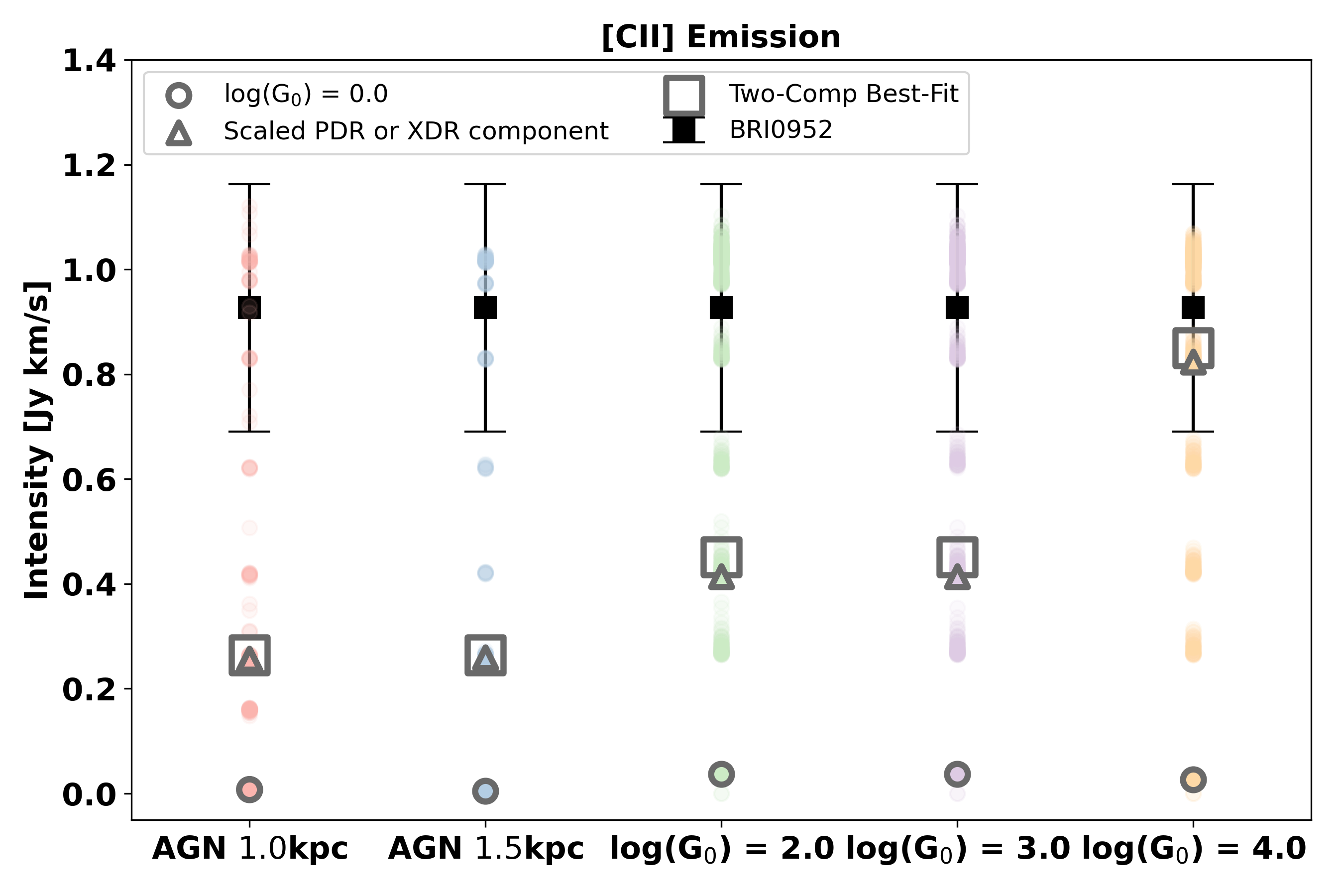}
    \caption{Best-fit MOLPOP-CEP CO model applied to the \ci and \cii emission. The different two-component model is given in the x-axis of the plot, the squares show the absolute lowest $\chi^{2}$ model from the CO emission (i.e., the black lines in Fig. \ref{fig:co_sleds_comb_1per}), the triangle shows the contribution from the scaled PDR or XDR model contributing to the high-$J$ transitions, likewise the circle shows the contribution from the PDR log($G_0$) = 0.0 component, and the lightly colored circles show models within $1\sigma$ of the lowest $\chi^2$. Note that the error on the magnification factor is not propagated through to the error on the intensity.}
    \label{fig:cii_ci_molpop}
\end{figure*}

We extend out modeling of the species detected to the \ci and \cii line detections using the same \molpop models as used for the CO emission. We evaluate the optimal models derived from various two-component scenarios explored for the CO emission, taking into account models with $\chi^2 < 1\sigma$, shown in Fig. \ref{fig:cii_ci_molpop}. Generally, the best-fit models from the CO underestimate both the \ci and \cii emission. We find that only the PDR+PDR model, composed of one component with log(G$_0$) = 0.0 and one with log(G$_0$) = 4.0, reproduces the \ci and \cii emission within the margin of error. We note that in all two-component scenarios, the scaled PDR or XDR component contributes more to the emission than the log(G$_0$) = 0.0 PDR component, indicating that the majority of the \cii and \ci emission comes from the more intense heating source. In the case of the \cii emission, this suggests, given that \cii emission traces colder and more diffuse gas including the neutral ISM, the best-fit CO models likely do not describe the intrinsic properties of the gas responsible for the \cii emission. Indeed, the low level of emission from the PDR log(G$_0$) = 0.0 component also suggests that applying the best-fit CO model directly to the \cii does not provide a reasonable model of the intrinsic conditions responsible for the \cii emission in BRI\,0952. Regarding the \ci emission, the larger contribution from the scaled PDR or XDR component is more expected as \ci is generally thought to trace the warmer molecular gas (see Section \ref{sec:gas_mass}). However, the best-fit CO models also do not reproduce the \ci emission with the exception of the single aforementioned model. However, as we show in Fig. \ref{fig:cii_ci_molpop}, the additional models within $\chi^2 < 1\sigma$ reproduce the \ci and \cii emission. This is not surprising given the large range in physical conditions that fall within $1\sigma$ for the CO emission, as given in Table \ref{tab:molpop_1sig_parameterrange}, leaving the intrinsic physical conditions of the \ci and \cii emitting regions unconstrained.

Previous works by \citet{Venemans17}, \citet{Novak19}, and \citet{Pensabene21} have utilized the \cii/\ci line luminosity ratio as a diagnostic tool to distinguish between PDR and XDR heating regimes. We find \cii/\ci$\sim100$, well above the expected ratio in the case of XDR heating found by \citet{Venemans17} and \citet{Pensabene21} (see Fig. 6 and Fig. 8, respectively). In the PDR case, modeling from \citet{Pensabene21} suggests that very high densities and radiation fields would be necessary to reproduce the ratios found in this work. This supports our findings from \molpop suggesting that intense PDR radiation is the only model able to reproduce the \cii and \ci emission detected in BRI\,0952; however, we note that the radiation field strengths from \citet{Pensabene21} necessary to reproduce the ratio found here are significantly higher than those we utilized in our \molpop modeling. Overall, these results suggest that the \cii and \ci emission is associated with on-going star formation in the quasar host galaxy rather than being associated with AGN activity. 

%%%%%%%%%%%%%%%%%%%%%%%%%%%%%%%%%%%%%%%%%%%%%%%%%%%%%%%%%%%%%%%%%%%%%%%%%%
\section{Discussion} \label{sec:discussion}
%%%%%%%%%%%%%%%%%%%%%%%%%%%%%%%%%%%%%%%%%%%%%%%%%%%%%%%%%%%%%%%%%%%%%%%%%%

%%%%%%%%%%%%%%%%%%%%%%%%%%%%%%%%%%%%%%%%%%%%%%%%%%%%%%%%%%%%%%%%%%%%%%%%%%
\subsection{The \water-IR luminosity relation} \label{sec:h2o_lir}
%%%%%%%%%%%%%%%%%%%%%%%%%%%%%%%%%%%%%%%%%%%%%%%%%%%%%%%%%%%%%%%%%%%%%%%%%%
Sub-mm and far-infrared observations of water with telescopes such as \textit{Herschel}, ALMA, and NOEMA have shown that water can be a powerful tool for probing the ISM properties of galaxies \citep[e.g.,][]{GA10, vanderWerf11, Falstad15, Yang16, Konig17, Liu17}. Indeed, studies have shown that water lines can be, in some cases, of equal strength as mid-J CO lines \citep{Yang13, Omont13}. In addition, there has been mounting evidence that there is a correlation between \water and infrared luminosity \citep[e.g.,][]{Yang13, Yang16, Omont13, Jarugula21, Pensabene22}. This correlation is due, in part, to the connection between the excitation of certain water transitions and infrared pumping \citep{GA10, GA12, GA22}. In the case of the \water $2_{11} - 2_{02}$ line, $101 \rm \mu m$ photons excite \water $1_{11}$ to the \water $2_{20}$ level, which cascades down to the \water $2_{11}$ level and subsequently the \water $2_{02}$ transition observed in BRI\,0952. In addition to being pumped by far-infrared photons, the \water $2_{11}-2_{02}$ line likely originates from the warm component of the ISM, that is the component with $n(\rm H) \sim 10^{5} - 10^{6} \rm \, cm^{-3}$ and $T_{\rm dust} \sim 40-70\,K$ based on modeling by \citet{Liu17}. As these conditions are the same as those required to excite CO(7--6), it is not surprising that the line profile of the water line is quite similar to that of the CO(7--6) line as both likely originate from similar regions within the galaxy (Fig. \ref{fig:h2o_co_spec}). 

\begin{figure}
    \centering
    \includegraphics[width = 1.0\linewidth]{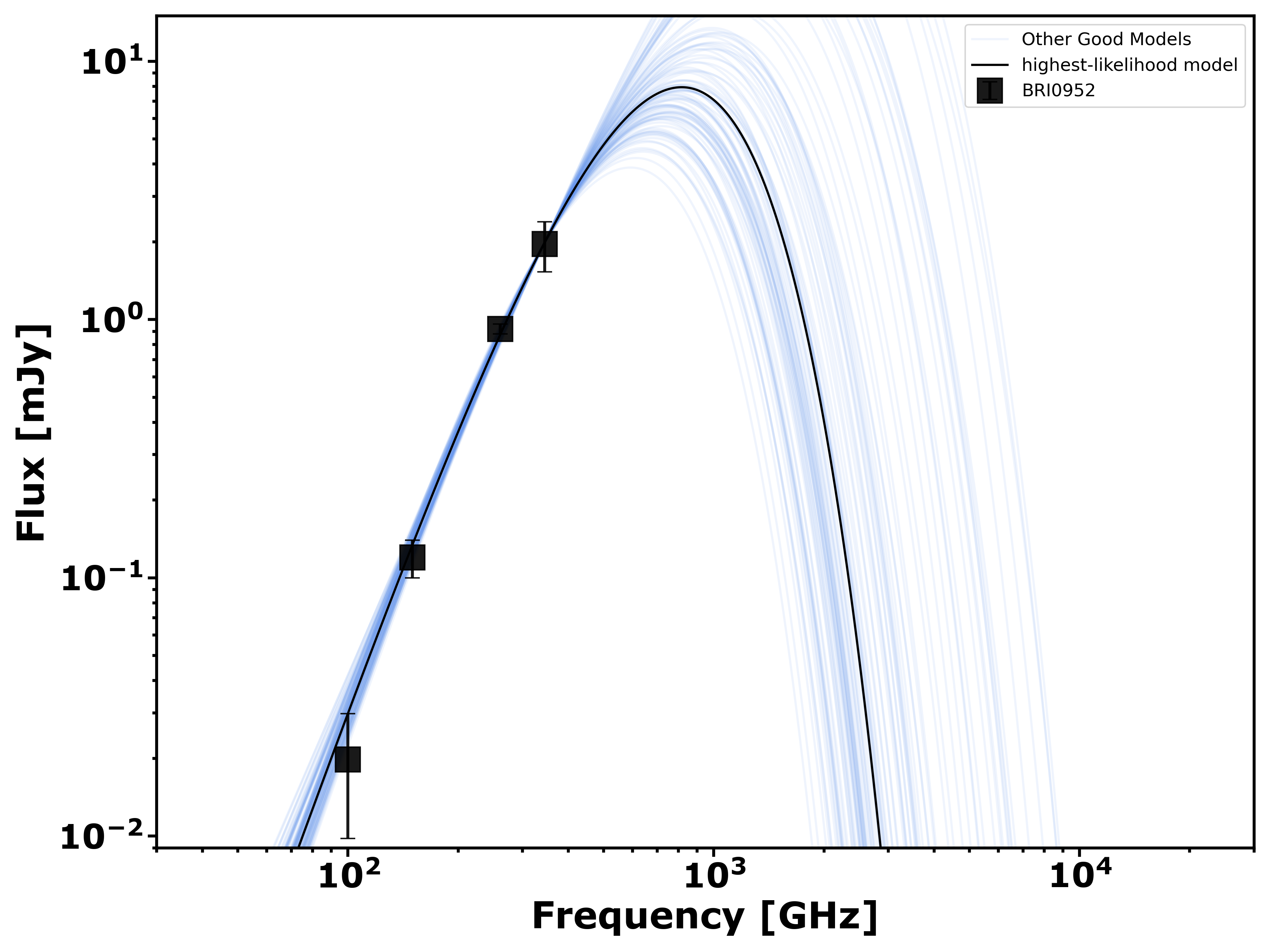}
    \includegraphics[width = 1.0\linewidth]{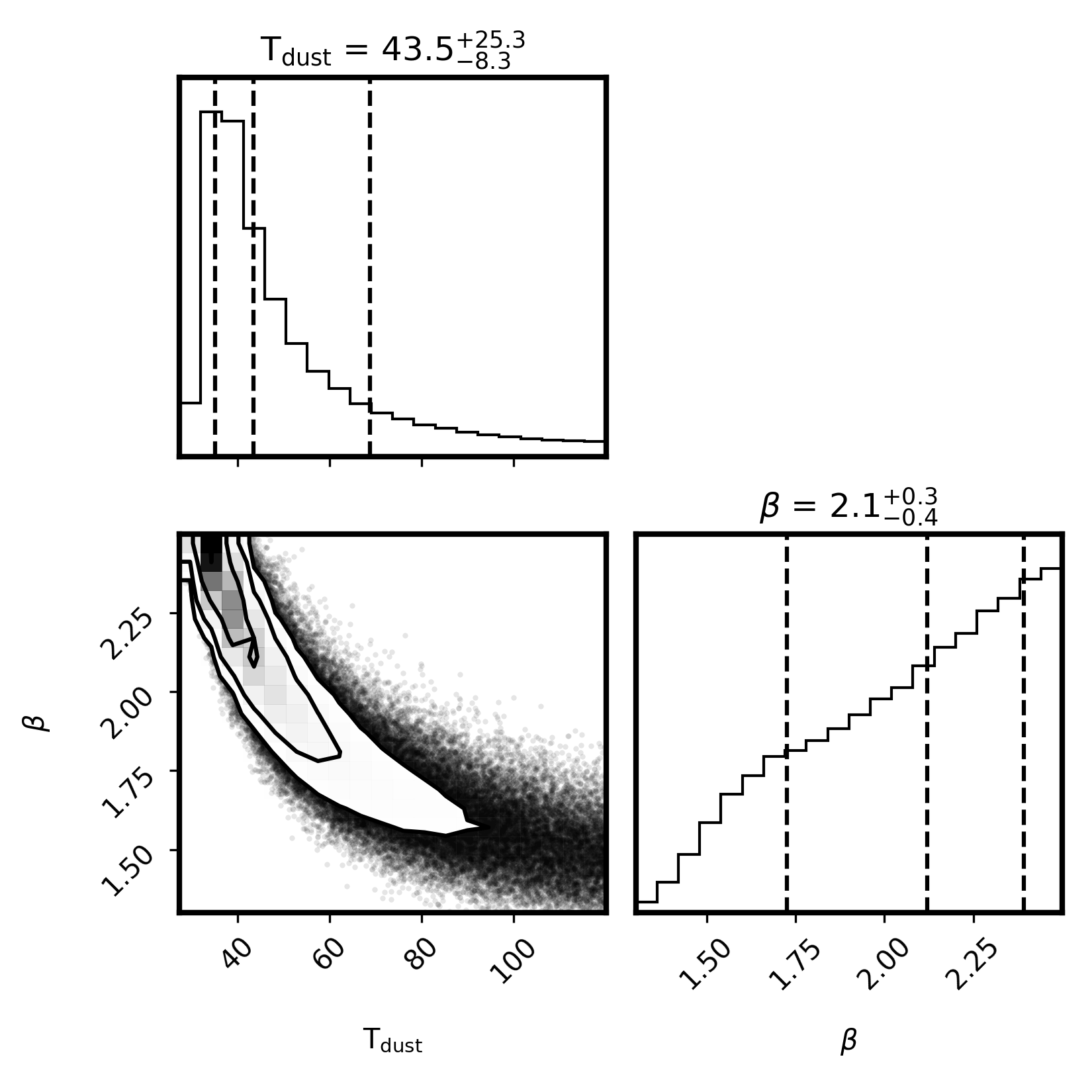}
    \caption{Top: Modified black-body SED of BRI\,0952 used to obtain the dust temperature. The best-fit model returns a dust temperature of $T_{\rm dust} \sim 45$\,K and $\beta \sim 2.0$. Bottom: Corner plot from the MCMC fitting showing the 16th, 50th, and 84th quartiles. }
    \label{fig:Tdust}
\end{figure}

We make a very brief analysis of the conditions under which this water line forms. We evaluate the temperature of the dust using two different methods under the assumption that the dust is the main heating mechanism responsible for exciting this molecule. First, we consider the dust temperature from the SED as the continuum measurements in this paper mainly probe the colder dust. Secondly, we compare our SED-estimated dust temperature to MOLPOP-CEP modeling of the \water line where the \water emission is modeled using different dust temperatures. We note that \citet{Kade23} fit an SED; however, the dust temperature was not a parameter fit in the previous iteration of the SED. As an alternative to estimate the temperature of the colder dust, we use the python package ({\sc dust emissivity}, \cite{dust_emm}) to fit a modified black body function where the modified blackbody function is of the form,

\begin{equation}
    \centering
    S_{\nu} = \frac{2h\nu^3}{c^2} \frac{1-e^{-\tau_{\nu}}}{e^{\textit{h}\nu/\kappa T} -1}, 
\end{equation}

\noindent where $S_\nu$ is the flux density, $h$ is the Planck constant, $\tau_\nu$ is the opacity at frequency $\nu$, T is the temperature, and $\kappa$ is the Boltzmann constant. The functional form of the equation does not including heating from the CMB, however including the background contribution from the CMB in the form $B_{\nu}[T = T_{CMB(z)}]$ does not alter the output dust temperature; at z = 4.432 it is unlikely that the CMB would have a large affect on the shape of the SED \citep[e.g.,][]{daCunha13} and therefore we chose not to include it in our modified blackbody fit. 
This functional form allows for variation in the dust temperature and the $\beta$ value. We explored the parameter space of both of these values using a Markov chain Monte Carlo (MCMC) approach in which the dust temperature was allowed to vary between 14.742\,K and 120\,K where the lower limit is set by the CMB temperature at $z = 4.432$ and $\beta$ was allowed to vary between 1.3 and 2.5. Both are typical parameter ranges for high-redshift galaxies \citep[e.g.,][]{Beelen06, Leipski13, Pensabene22}. 
%We sample the parameter space using 1000 walkers and 1000 iterations. 
The best-fit parameters returned from this method were $T_{\rm dust} \sim 45$\,K and $\beta \sim 2.0$, with a large margin of error. We note that the dust temperature from the SED fitting cannot be well constrained with the available data, as can be shown in Fig. \ref{fig:Tdust} by the wide range of temperature and beta values that produce a reasonable fit.

\begin{figure}[h!]
    \centering
    \includegraphics[width =1.0\linewidth]{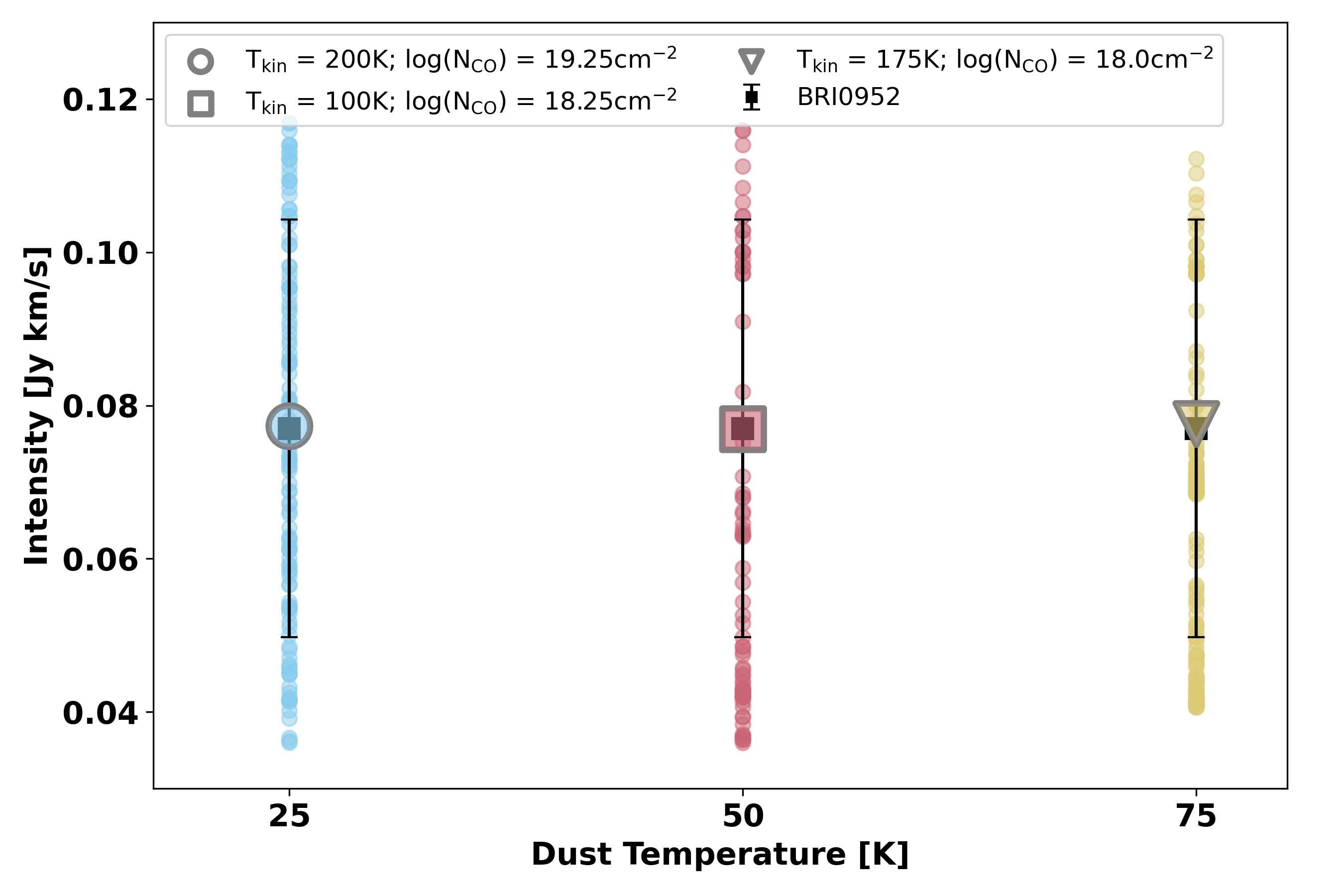}
    \caption{MOLPOP-CEP modeling of the \water($2_{11} - 2_{02}$) line as a function of different modeled dust temperatures. The grey circle, square, and triangle show the absolute best fit model, respectively. The smaller colored circles show the $1\sigma$ distribution. Note that the error on the magnification factor is not propagated through to the error on the intensity.}
    \label{fig:h2o_tdust}
\end{figure}

In the case of the MOLPOP-CEP models, we consider the only heating mechanism to be the dust temperature and assume an opacity of $\tau = 0.3$ \footnote{The $\tau$ assumption is based on a combination of the extinction opacity, $\kappa$, and emissivity index $\beta$ from \citet{draine2007} and \citet{galliano18} combined with an assumption that the molecular gas column density is $\log(N_{\rm H_2}/[{\rm cm^{-2}]}) \sim 22.2-22.7$ (consistent with the column densities from the analysis in subsect.~\ref{sec:2comp}), from which we estimate $\tau$ in the range $\sim$0.1-0.7.}. We model the \water $2_{11} - 2_{02}$ line using three different dust temperatures, $T_{\rm dust} = 25, 50, 75$\,K and using the same grid of parameters for $\Delta L$, $T_{\rm kin}$, and $n(\rm H_2)$ as for the CO (Section \ref{sec:Molpop}). We evaluated the models using the same methodology as for the CO. We show the best-fit model and other good models within 1$\sigma$ in Fig. \ref{fig:h2o_tdust}. We find that all of the explored dust temperatures are able to reproduce the observed \water flux, however, correlating the kinetic temperature, volume density, and zone width with modeling from \citet{Liu17}, which suggests that \water($2_{11} - 2_{02}$) comes from regions with $n(\rm H) \sim 10^5 - 10^6$\,cm$^{-3}$ and $T_{\rm dust} \sim 40-70$\,K, giving an indication that $T_{\rm dust} \sim 50$\,K may be the most physically motivated model. This is in close agreement with the dust temperature from the SED of $T_{\rm dust} \sim 45$\,K. However, this analysis and limited by a lack of continuum measurements and a single \water detection. Thus the dust temperature should therefore only be seen as a first estimate.

\begin{figure}[h]
    \centering
    \includegraphics[width = 1.0\linewidth]{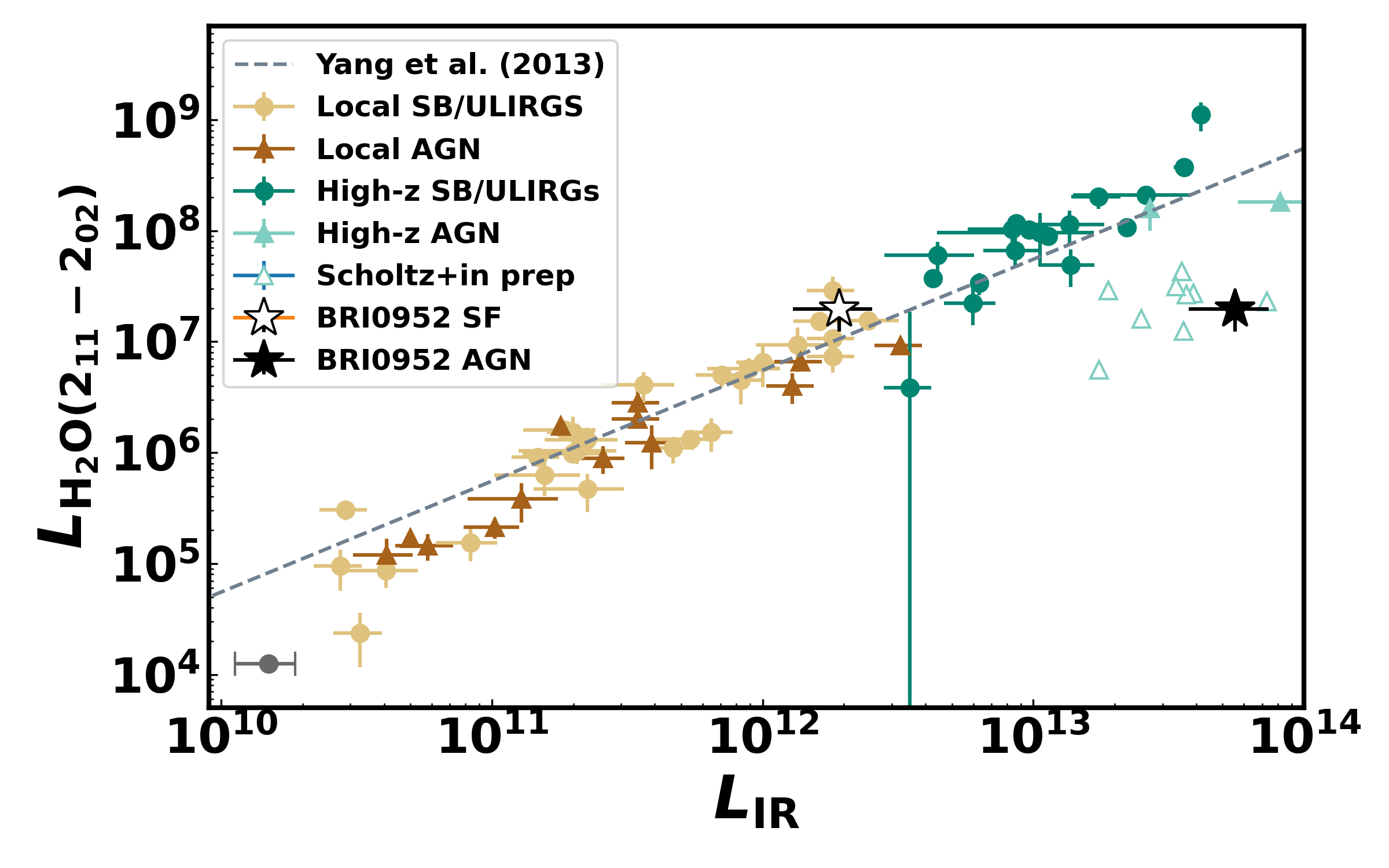}
    \caption{The \water-$L _{\rm IR}$ relationship with BRI\,0952. The dark grey point in the bottom left of the plot shows the indicative error of $30\%$ on the $\rm L_{IR}$ due to discrepancies in reporting  $\rm L_{IR}$ vs. $\rm L_{FIR}$ in the literature. The samples shown are from \citet{Riechers13}, \citet{Yang13, Yang16}, \citet{Falstad15}, \citet{Liu17}, \citet{Casey19}, \citet{Apostolovski19}, \citet{Pensabene22}, \citet{Scholtz23}, Scholtz et al. (in prep), and Liu et al. (in prep).}
    \label{fig:H2O_LIR}
\end{figure}

We compare the ratio of the $\rm H_{2}O$/$L_{\rm IR}$ to the $L_{\rm IR}$ for BRI\,0952 to other low- and high-redshift galaxies in Fig. \ref{fig:H2O_LIR}. We distinguish between sources with a known AGN contribution and those without. The majority of the sample of high-$z$ AGN are extremely red quasars from Scholtz et al. (2023; in prep), shown in unfilled blue triangles, and the $L_{\rm IR}$ should be considered as an upper limit. We show BRI\,0952's position relative to other high- and low-redshift galaxies using both the star-formation dominated infrared luminosity and the infrared luminosity which includes the contribution from the AGN. For sources with reported far-infrared luminosities we correct these to $L_{\rm IR}$ as $L_{\rm FIR} \sim 0.75 \times L_{\rm IR}$ \citep[e.g.,][]{Decarli17}. Due to the discrepancies between sources and samples in reporting IR vs. FIR luminosities, we include a conservative indicative error of 30\% on the $\mathrm{L_{IR}}$. 

\citet{Kade23} decomposed the $L_{\rm IR}$ of BRI\,0952 into AGN versus star-formation components, respectively and we consider both components when comparing to other low- and high-redshift sources. We find that BRI\,0952 follows the $L_{\rm \water}$ vs. $L_{\rm IR}$ trend from \citet{Yang13} only when using the decomposed $L_{\rm IR_{SF}}$. When using the $L_{\rm IR_{AGN}}$, the quasar lies distinctly below the trend line. However, it should be noted that there is a relative lack of detections of the \water ($2_{11} - 2_{02}$) line in high-redshift AGN. BRI\,0952 lies with other AGN-dominated sources from Scholtz et al. (2023; in prep). Considering IR-pumping with the current SED decomposition, the 101\,$\mu$m photons needed to pump the \water $2_{11}-2_{02}$ line could originate from either AGN or SF activity. However, given the placement of BRI\,0952 in relation to other high-redshift quasars and star-forming galaxies and models of the \water $2_{11}-2_{02}$ line, there are strong indications that the line emission is correlated with star formation rather than AGN activity. 

%%%%%%%%%%%%%%%%%%%%%%%%%%%%%%%%%%%%%%%%%%%%%%%%%%%%%%%%%%%%%%%%%%%%%%%%%%
\subsection{The CO SLED of BRI\,0952 compared to low- and high-redshift galaxies} \label{sec:co_sled}
%%%%%%%%%%%%%%%%%%%%%%%%%%%%%%%%%%%%%%%%%%%%%%%%%%%%%%%%%%%%%%%%%%%%%%%%%%

\begin{figure}[h]
    \centering
    \includegraphics[width = 1.0\linewidth]{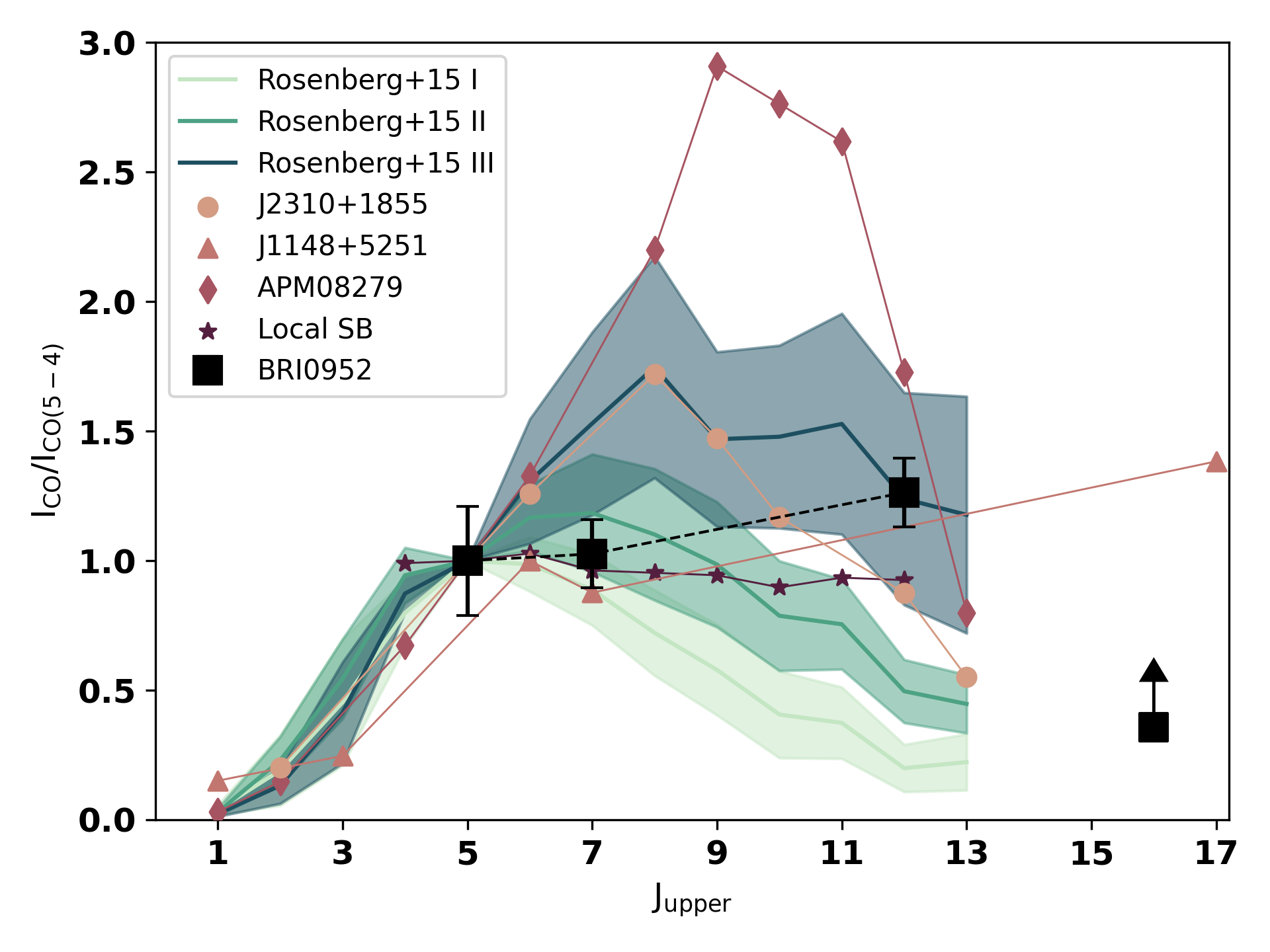}
    \caption{CO SLED of BRI\,0952, normalized to the CO(5--4) line. Local universe representatives are a sample of ULIRGs from \citet{Rosenberg15} where $1\sigma$ errors are shown by the shaded regions for class I, II, and III (purple diamonds), a sample of normal and starburst galaxies from \citet{Liu17} (pink pentagons). High-redshift sources are J2310+1855 \citep{Li20}, J1148++5251 \citep{Gallerani14}, and APM 08279+5255 \citep{Weiss07, Bradford11}. J1148+5251 has been normalized to the CO(6--5) transition. APM 08279+5255 has been normalized to an inferred value for the CO(5--4) transition by averaging the adjacent CO transitions (namely the CO(4--3) and CO(6--5) transitions). We show BRI\,0952 including a lower limit on the CO(16--15) from the partial line wing detected in the band 7 data.}
    \label{fig:CO_SLED}
\end{figure}

The lensing-corrected CO SLED of BRI0592 in comparison to other low- and high-redshift sources and samples is shown in Fig. \ref{fig:CO_SLED}. We place a lower limit on the CO(16--15) emission detected at the edge of the spectral window containing the OH emission, shown in Fig. \ref{fig:BRI_spec} by assuming that the peak line flux is that which is measured in the spectra ($\sim 2.0$\,mJy) and a line width of $250$\,km\,s$^{-1}$, consistent with the observed lower-$J$ CO lines. We note that the partial detection is not robust enough for further analysis; however, we include the line as a lower limit when comparing the CO SLEDs as it is indicative of highly excited emission in BRI\,0952. 

We limit our comparison sample to those galaxies that have detections of CO(5--4) in order to remove as much bias from the normalization as possible; however, J1148+0875 is normalized to CO(6--5) and APM 08279+5255 is normalized to an inferred value for the CO(5--4) transition by averaging the adjacent CO transitions (namely the CO(4--3) and CO(6--5) transitions). The flux of the CO(12--11) line in BRI\,0952 is elevated compared to the average of the class III galaxies studies in \citet{Rosenberg15} in which extreme heating mechanisms (AGN or mechanical processes) were present, and significantly higher than class I and II sources. The CO SLED of BRI\,0952 follows that of the median CO SLED of galaxies with mid-IR AGN contributions of $\geq 50\%$ from \citet{Kirkpatrick19}; however, it should be noted that \citet{Kirkpatrick19} point out that the statistical difference between the CO SLEDs of galaxies with mid-IR AGN contributions $<50\%$ and those with $\geq 50\%$ is minimal. Comparisons between the CO transitions detected in BRI\,0952 and those reported in \citet{Kirkpatrick19} are also limited as the median CO SLED from \citet{Kirkpatrick19} does not extend beyond $J_{\mathrm{upper}} = 7$. 

The flux of the CO(12--11) transition observed in BRI\,0952 is significantly higher than other samples which extend out to the $J_{\mathrm{upper}} = 12$ transition. One exception is the quasar APM\,08279+5255 at $z = 3.9$ \citep{Weiss07}. The CO SLED of BRI\,0952 broadly follows the shape of the SLED of APM\,08279+5255; however with only three robustly observed CO transitions in BRI\,0952 it is not possible to determine if the CO SLED turns over between $J = 7$ and $J = 12$ as is the case for APM\,08279+5255, where \citet{Weiss07} show that a two component model including an XDR of the ISM is a best-fit to the CO SLED. On the other hand, while the CO(5--4) emission detected in BRI\,0952 roughly match those of the quasar J2310+1855 at $z = 6.003$ \citep{Li20}, the SLED of J2310+1855 does not exhibit highly excited high-$J$ transitions and the flux of the CO(12--11) emission is significantly lower than that observed in BRI\,0952. Regarding the turnover point of BRI\,0952, without the partial detection of CO(16--15) the CO SLED could match the shape of either APM\,08279+5255 or J2310+1855 but the partial detection of such a high-$J$ CO line raises uncertainty about the turnover point and suggests that the shape may follow that of the quasar J1148+5251 at $z = 6.4$ where the model for that CO SLED also required an XDR region \citep{Gallerani14}. Without additional CO detections between $J_{\rm upper} = 7$ and $J_{\rm upper} = 12$ as well as at $J_{\rm upper} > 12$, precise conclusions about the shape of the CO SLED of BRI\,0952 cannot be drawn. 

We highlight again the uncertainties of correcting the CO line emission for magnification using the \cii emission, as discussed in Section \ref{sec:lensing_reconstruction}. It is possible that the CO(12--11) emission has a higher magnification factor, depending on the location of the emission, than the \cii emission, causing the CO(12-11) flux to be lower and the turnover point of the CO SLED to shift to lower-$J$ transitions. However, This change would have to be significantly different to affect the conclusions of this paper. 

%%%%%%%%%%%%%%%%%%%%%%%%%%%%%%%%%%%%%%%%%%%%%%%%%%%%%%%%%%%%%%%%%%%%%%%%%%
\subsection{The ISM of BRI\,0952} \label{sec:ISM_BRI}
%%%%%%%%%%%%%%%%%%%%%%%%%%%%%%%%%%%%%%%%%%%%%%%%%%%%%%%%%%%%%%%%%%%%%%%%%%
In a commonly accepted paradigm of massive galaxy evolution, massive galaxies merge together causing burst in star formation and a subsequent transition to a quasar, whereby the galaxy then turns into a elliptical galaxy \citep[e.g.,][]{Hopkins08}. Throughout this process the star formation in the galaxy is `quenched' by either feedback processes, be it from mechanical processes such as intense star formation or AGN feedback \citep[e.g.,][]{Fabian12, Ishibashi16}. In high-redshift galaxies, it is difficult to quantify the exact contribution of the underlying heating mechanisms present in the ISM. Indeed, it is also challenging to place constraints on the precise evolutionary stage of individual galaxies. This analysis of both archival and high-resolution data allows for an in-depth analysis of a variety of probes of the ISM in the quasar BRI\,0952. However, as we have shown throughout the paper and discuss further here, even with these probes many facets of the ISM remain unclear. 

BRI\,0952 is a radio-quiet quasar \citep{Omont96} which has been shown through SED decomposition to host a dominant AGN contribution \citep{Kade23}. Star formation tracers including dust (IR luminosity) and \cii emission suggest that the quasar also hosts intense star formation of a few $\times 100 - 1000$\,M$_{\odot}$\,yr$^{-1}$. The three CO line detections reported here give the CO SLED a distinctive shape. The CO(5--4) and CO(7--6) emission are comparable, but the CO(12--11) emission is brighter than either of the lower-$J$ transitions. This, coupled with the partial detection of CO(16--15), is indicative of an extreme underlying heating mechanism within the quasar. \molpop modeling required a two-component model of the ISM in order to reproduce the observed CO emission where the preferred scenario included a PDR and XDR contribution, suggesting that both star formation and the AGN play an active role in heating the gas of the quasar, but without additional CO line detections we cannot come to strong conclusions about individual effect of these mechanisms. 

As an indication of AGN activity we consider the tentative \ohp emission. In the quasar Mrk231, strong \ohp emission was an indication of AGN activity \citep{vanderWerf10} and has been detected in the high-redshift quasar J1148+5251 \citep{Gallerani14}. Radiative transfer modelling from \citet{GA13} of NGC4418 and Arp220 suggested that the primary production mechanism for H$^+$, a precursor to OH$^{+}$, was through XDRs or cosmic rays rather than FUV ionizing photons. This result lends strength to the AGN in BRI\,0952 being responsible for the \ohp emission. 

The total molecular gas mass, inferred from the \ci(2-1) luminosity, suggests that the quasar has a lower total molecular gas mass (of a factor $\sim 100$ in all cases) compared to higher-redshift star-forming or star-bursting galaxies such as HFLS3 \citep[$z = 6.34$][]{Riechers13}, SPT0311-58 \citep[$z = 6.9$][]{Jarugula21}, AzTEC-3 \citep[$z = 5.3$][]{Riechers20}, and Mambo-9 \citep[$z = 5.85$][]{Casey19}. The total molecular gas mass is also lower than the value found for quasars and star-forming galaxies at $z \sim 3-4$ \citep[e.g.,][]{Lu18, Gururajan22}. The relatively lower molecular gas mass, coupled with the high star-formation rate, suggests that the quasar could be using up its star-forming fuel supply at a tremendous rate. 

We investigate the possibility that BRI\,0952 has begun to quench its star formation by calculating the depletion time of BRI\,0952. The depletion time of a galaxy, defined as $t_{\rm dep} = M_{\rm gas}$/SFR, is the length of time for which a galaxy can maintain its observed rate of star formation. The depletion time of the local main-sequence galaxies has been extensively studied \citep[e.g.,][]{Saintonge11, Tacconi18, Tacconi20}, and there exists a main-sequence trend for local galaxies, although there is ongoing debate about the evolution of depletion time with redshift. We calculate the depletion time of BRI\,0952 using both the $L_{\rm IR}$ and \cii SFRs and find $t_{\rm dep_{L_{\rm IR}}} \sim 8$\,Myr and $t_{\rm dep_{[CII]}} \sim 25$\,Myr. This is significantly lower than DSFGs at $z\sim3$, those found for the $z = 6.9$ DSFGs studied in \citet{Jarugula21}, and the depletion timescales found for the sample in \citet{Stanley21}. This relatively low depletion time provides an indication that a quenching process may be operating within the quasar. 

BRI\,0952 is also known to have two companions detected in \cii emission within 2.5 arcseconds of the northern image of the quasar where faint velocity gradients between both companions and the quasar have been noted \citep{Kade23}. In addition, the high ratio of \ci/CO(7--6) of $\sim 4.8$ suggests that the source may be merger driven as lower ratios are associated with disc-like rotating galaxies. The lensing-corrected spectral line profiles of the CO(7--6) and CO(12--11) emission appear complex and the velocity gradient across the \cii line profile from lensing (Fig. \ref{fig:Lensing_pv}) provide indications of the kinematics operating within BRI\,0952, be it regular rotation or merger activity. The \cii emission line profile exhibits a broad component indicative of an outflow rate of $\dot{M}_{\rm out} = 98 \pm 19$\,M$_{\odot}$\,yr$^{-1}$; although it should be noted that the outflow cannot be confirmed without observations of a robust outflow tracer such as P-Cygni absorption line profiles in for example the OH $119\,\mu$m or \ohp($1_1-0_1)$ lines \citep[e.g.,][]{Spilker20, Riechers21}. Considering the \cii outflow as a source of mass loss, the depletion time would be shortened with respect to the value reported above.

Given the high SFR, low gas mass and thereby low depletion time, highly excited CO SLED, outflow evidence, and companions we postulate that BRI\,0952 is in a transitionary stage between a galaxy dominated by star-formation activity to one dominated by AGN activity. However, we note the limitations of the data used in this analysis. Although we have a number of observations of atomic and molecular transitions observed towards BRI\,0952, these observations are not standardized and suffer from angular resolution limitations, particularly when extrapolating between the data sets. Additionally, our analysis of the CO SLED has shown that without further detections of CO between $J_{\rm upper} = 7$ and $J_{\rm upper} = 12$ as well as confirmation of the CO(16--15) emission, it is not possible to robustly determine the heating mechanisms at work in the quasar. Hence, a standardized analysis of CO in BRI\,0952 would be necessary to leverage the full power of the CO SLED for the quasar and coming to strong conclusions about the contribution of star formation versus AGN activity to the excitation would likely require an extension beyond CO. 

\section{Conclusions} \label{sec:conclusions}

We report detections of CO(5--4), CO(7--6), CO(12--11), \ci(2--1), OH $^{2}\Pi_{1/2} (3/2-1/2)$, \water($2_{11} - 2_{02}$), a tentative detection of \ohp, and a partial wing detection of CO(16--15) in the high-redshift quasar BRI\,0952 at $z = 4.432$. We update the lensing model from \citet{Kade23} using the \cii line emission to extend the lensing model to the other line detections reported in this paper. We have performed an analysis of the ISM of the quasar using the radiative transfer tool MOLPOP-CEP to investigate the heating mechanisms at work. Our conclusions are provided below. 

\begin{enumerate}
    \item We further improve the lens model from \citet{Kade23} using \cii observations. We perform lensing modeling using the code {\sc Visilens} \citet{Spilker16} and obtain a magnification factor per channel across the \cii line which we use to correct the spectral line profiles of the \cii, CO(5--4), CO(7--6), and CO(12--11) emission. The broad component in the lensing corrected \cii line profile remains clear, confirming the likely presence of an outflow and indicating that lensing effects are not the cause of this component. The lens modeling suggests a velocity gradient across the line profile of the \cii emission that could be indicative of ordered rotation or originate from either interactions with the companions found in \citet{Kade23}. 
    
    \item We detect \water($2_{11} - 2_{02}$) in the quasar and confirm that the quasar exhibits a similar trend to those previously found in the literature regarding a correlation between $L_{\rm IR}$ and \water line luminosity. We use the $L_{\rm IR}$ from star formation and that from AGN and show that in the case of $L_{\rm IR_{SF}}$ the source lies in a similar region to local AGN and star-bursting galaxies (such as ULIRGs) whereas in the case of $L_{\rm IR_{AGN}}$ the source falls within the region of high-$z$ AGN-dominated sources, but below the trend found for previous star-formation dominated sources. This suggests that the \water emission likely originates from star-forming regions of the ISM.
    % Thus, using the relationship between both water and IR luminosity and IR luminosity and SFR, we calculate a SFR $\sim \sim 1540 \pm 820 M_{\odot} yr^{-1}$, falling narrowly within the error range of the previously reported SFR for BRI\,0952. 

    \item We detect \ci(2-1) emission in the quasar, allowing for a calculation of the total gas mass within the quasar. Through this approach we find an $\rm H_{2}$ gas mass of $(6.2 \pm 4.9) \times 10^{11} M_{\odot}$, similar to previously reported values for the quasar. We calculate the depletion time using the SFR from both $L_{\rm IR}$ and \cii and find $t_{\rm dep_{L_{\rm IR}}} \sim 8$\,Myr and $t_{\rm dep_{[CII]}} \sim 25$\,Myr, respectively. 

    \item We use the radiative transfer modeling code MOLPOP-CEP to model the excitation conditions of BRI\,0952, specifically with regard to the CO SLED. We find that a two-component model of the ISM gives the best fit to the observed data, specifically a combination of PDR and XDR heating where the XDR is located at a distance of 1.0\,kpc from the emitting region with physical conditions $T_{\mathrm{kin, PDR}} = 125$\,K, log($N_{\rm CO, PDR}$\/cm$^{-2}$) = 17, $T_{\mathrm{kin, XDR}} = 50$\,K, and log($N_{\rm CO, XDR}$/cm$^{-2}$) = 18.75. However, the precise heating mechanism cannot be determined with the current data as both stellar heating and XDRs can reproduce the current observations. 
    
    \item We compare the CO SLED of BRI\,0952 to other local and high-redshift sources and find that the CO SLED appears to increase up to CO(12--11), but the turnover point is unclear with only the three observed transitions. We note that the CO(12--11) transition is bright compared to other detections, but could follow the trend of other high-redshift quasars where CO(12--11) represents the peak of the SLED.    
\end{enumerate}

Combined, we suggest that BRI\,0952 is in a transitionary evolution phase, moving from a starburst-dominated phase to an AGN-dominated one. The improved resolution and technical capabilities of ALMA will allow for increased studies focusing on a variety of atomic and molecular species, as well as the kinematics and environment of high-redshift galaxies and eventually allow for a full decomposition of the true influence of star formation versus that of AGN activity.

\begin{acknowledgements}
The authors thank the referee for the useful comments and constructive feedback that improved the paper. Kiana Kade acknowledges support from the Nordic ALMA Regional Centre (ARC) node based at Onsala Space Observatory. The Nordic ARC node is funded through Swedish Research Council grant No 2017-00648. Kirsten Knudsen acknowledges support from the Swedish Research Council and the Knut and Alice Wallenberg Foundation. Sabine K\"onig gratefully acknowledges funding from the European Research Council (ERC) under the European Union's Horizon 2020 research and innovation programme (grant agreement No. 789410).
This paper makes use of the following ALMA data: 2017.1.01081.S, 2015.1.00388.S, and 2018.1.01536.S. ALMA is a partnership of ESO (representing its member states), NSF (USA), and NINS (Japan), together with NRC (Canada), MOST and ASIAA (Taiwan), and KASI (Republic of Korea), in cooperation with the Republic of Chile. 
This research made use of APLpy, an open-source plotting package for Python \citep{Robitaille12}.
\end{acknowledgements}

\bibliographystyle{aa}
\bibliography{ref}

\newpage

\begin{appendix}

\section{Un-tapered Band 3 Continuum Image}
This appendix includes the non-tapered band 3 continuum image.

\begin{figure}
    \centering
    \includegraphics[width = 1.0\columnwidth]{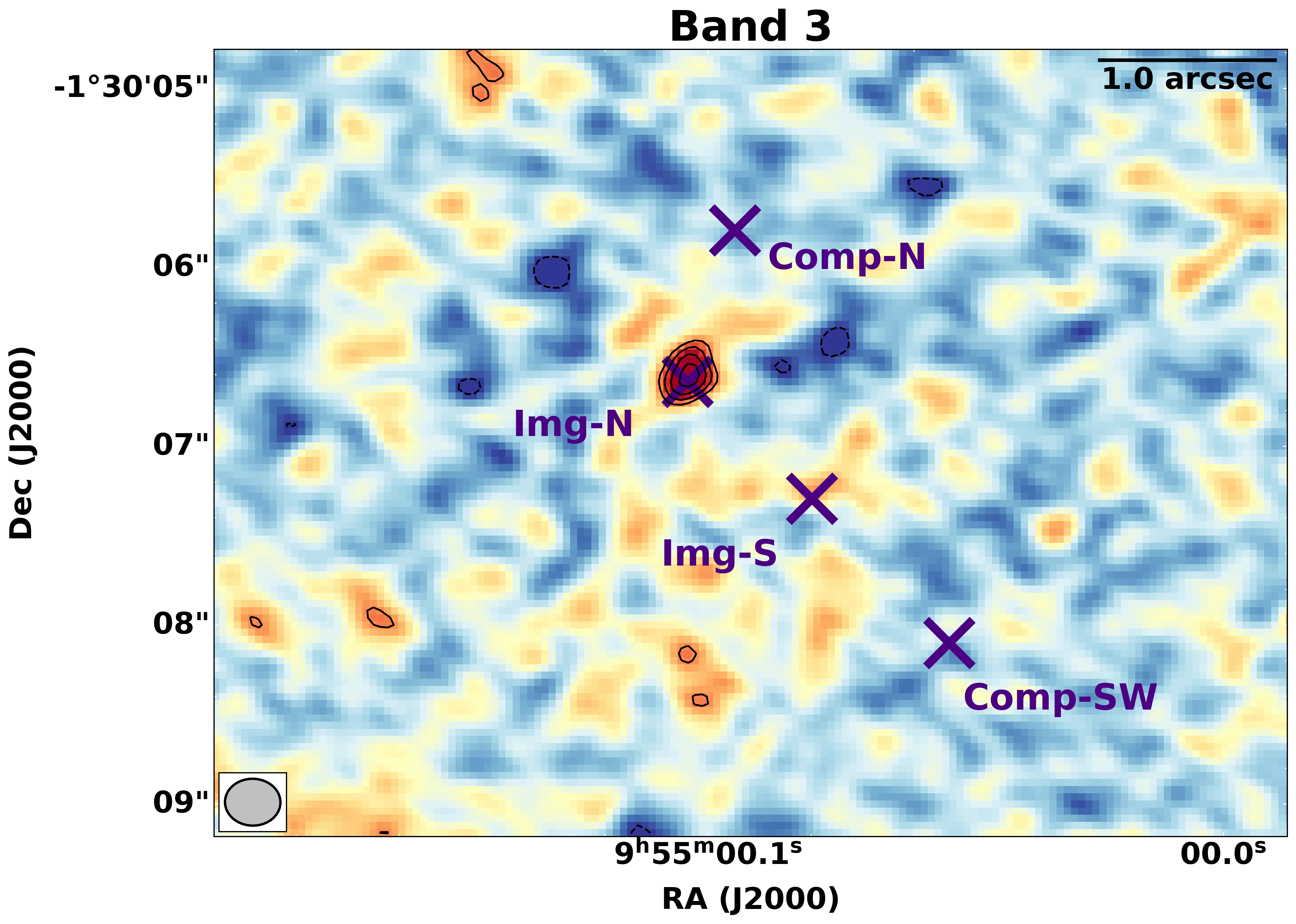}
    \caption{Non-tapered continuum image of the band 3 continuum data. The contours are shown at $-3, 3, 4, 5, 6, 7, 8, 9, and 10\sigma$ levels. The purple `X's show the position of the companion sources, Comp-N and Comp-SW, detected in \citet{Kade23} as well as the position of Img-N and Img-S, the two images of the quasar BRI\,0952. The synthesized beam is shown in the bottom left of each image.}
    \label{fig:b3_cont_untapered}
\end{figure}

\section{Spectra of detected species.}
This appendix includes the lensing-corrected spectra for the line profiles with sufficient signal-to-noise ratios to perform a channel-by-channel magnification correction.

\begin{figure*}[h!]
\centering
\includegraphics[width=0.8\columnwidth]{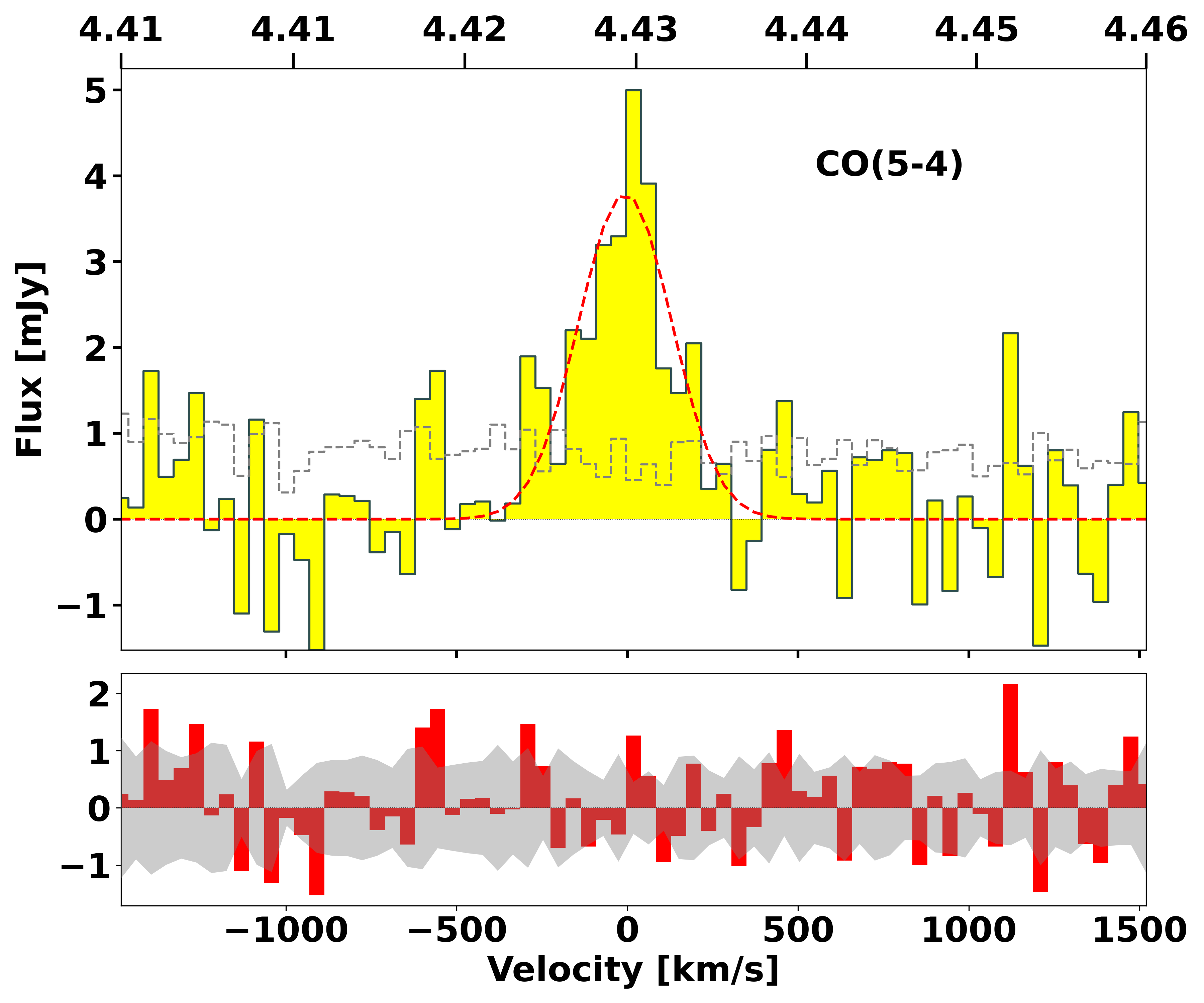}
\includegraphics[width=0.8\columnwidth]{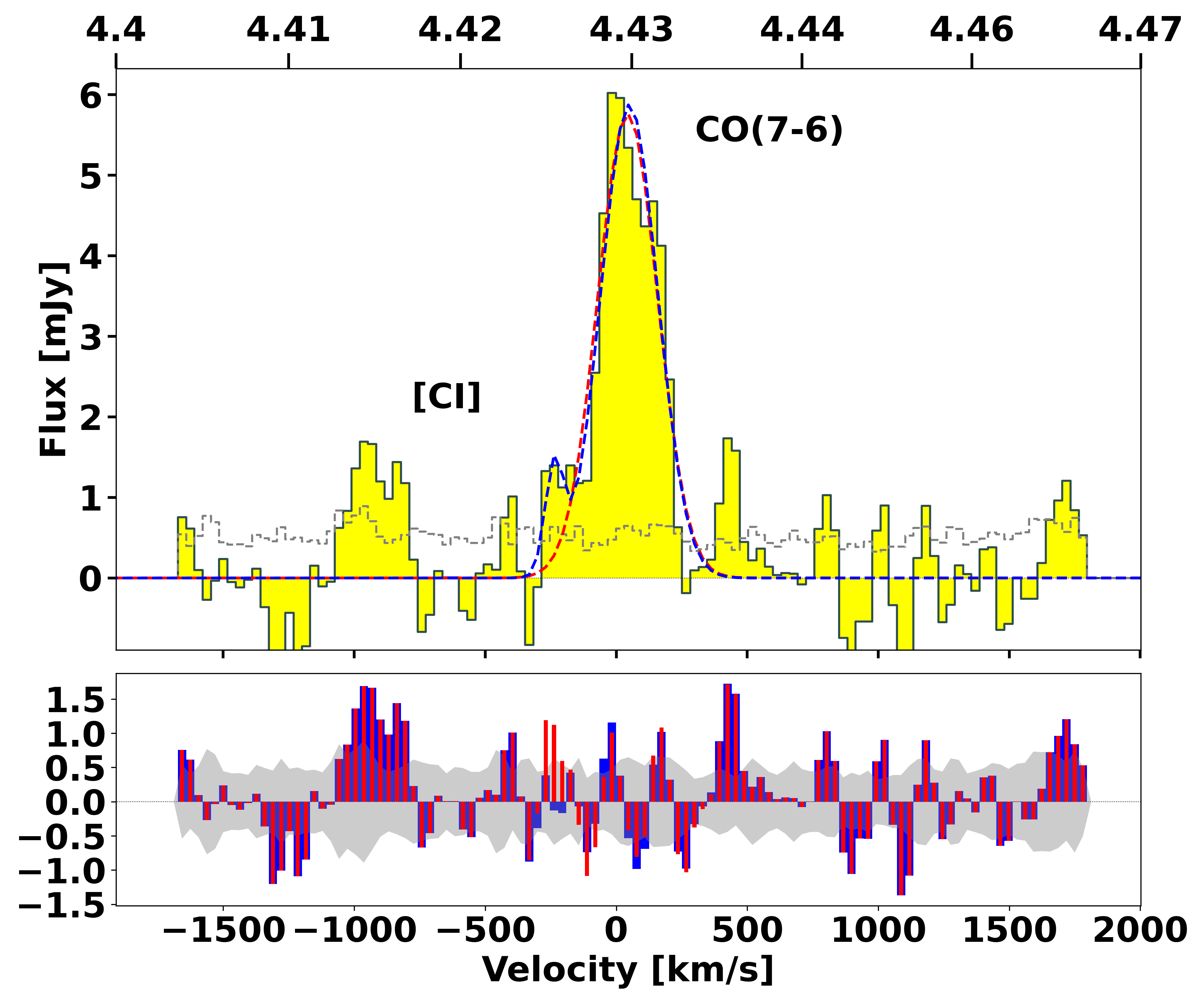}
\includegraphics[width=0.8\columnwidth]{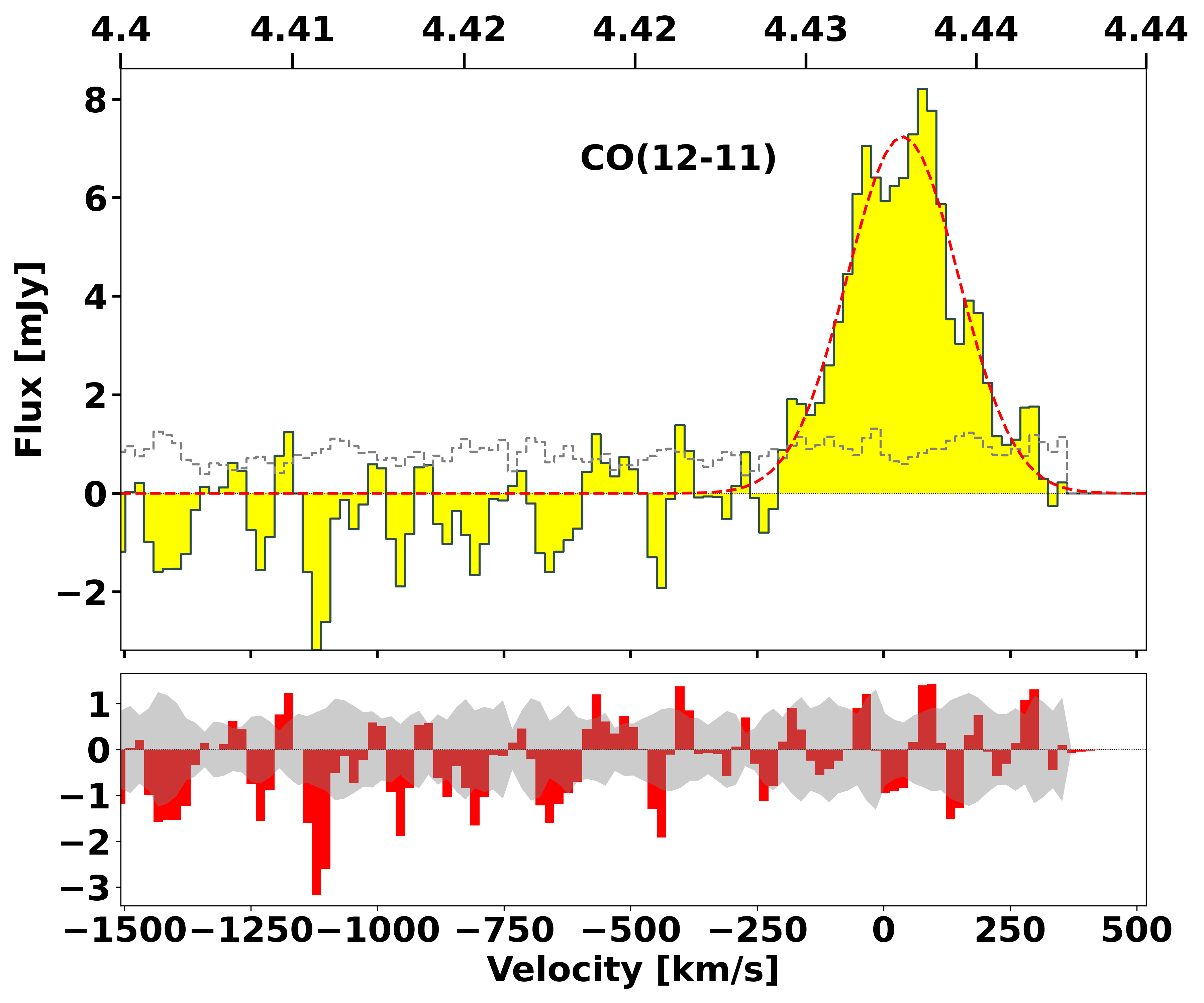}
\includegraphics[width=0.8\columnwidth]{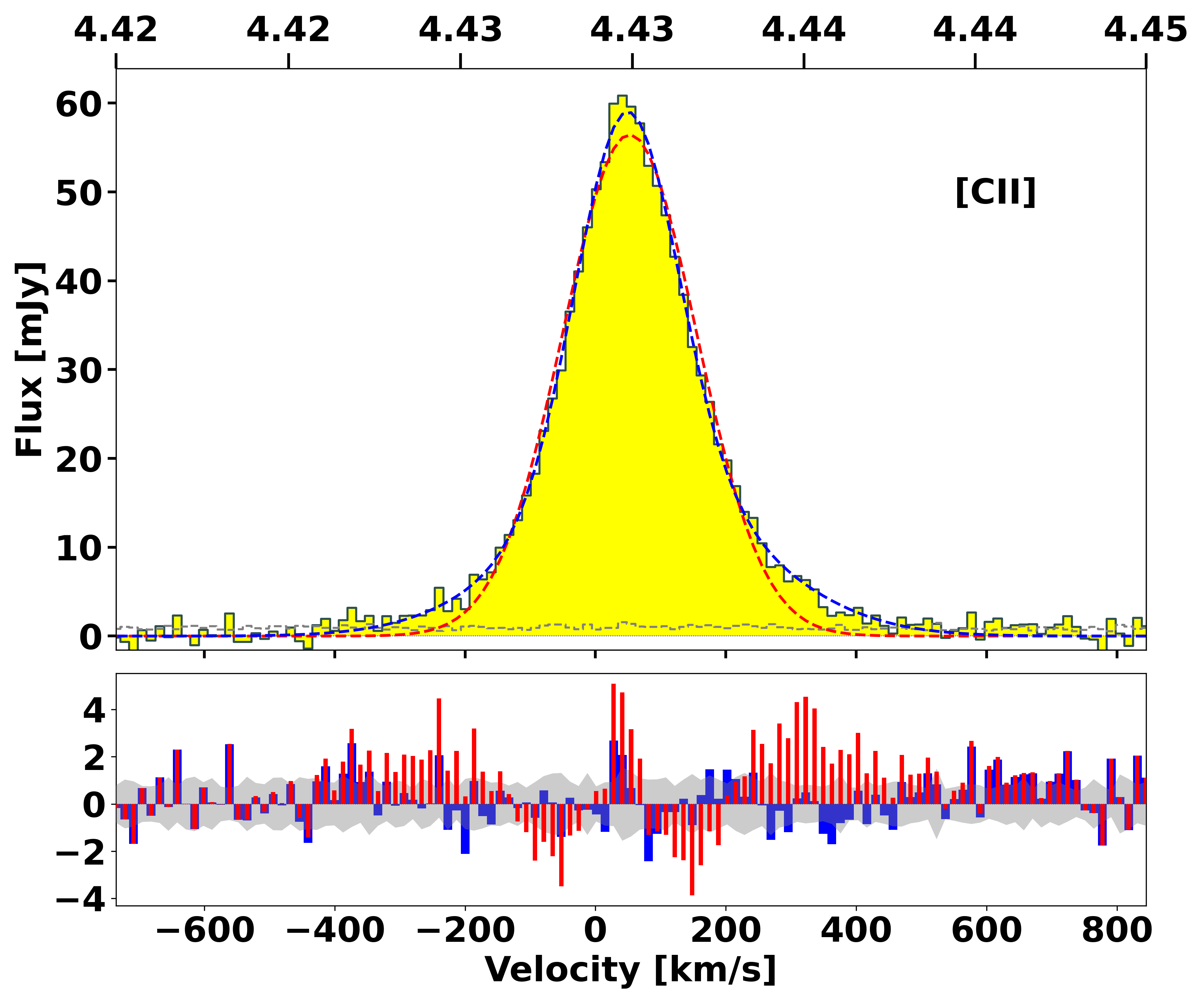}
\includegraphics[width=0.8\columnwidth]{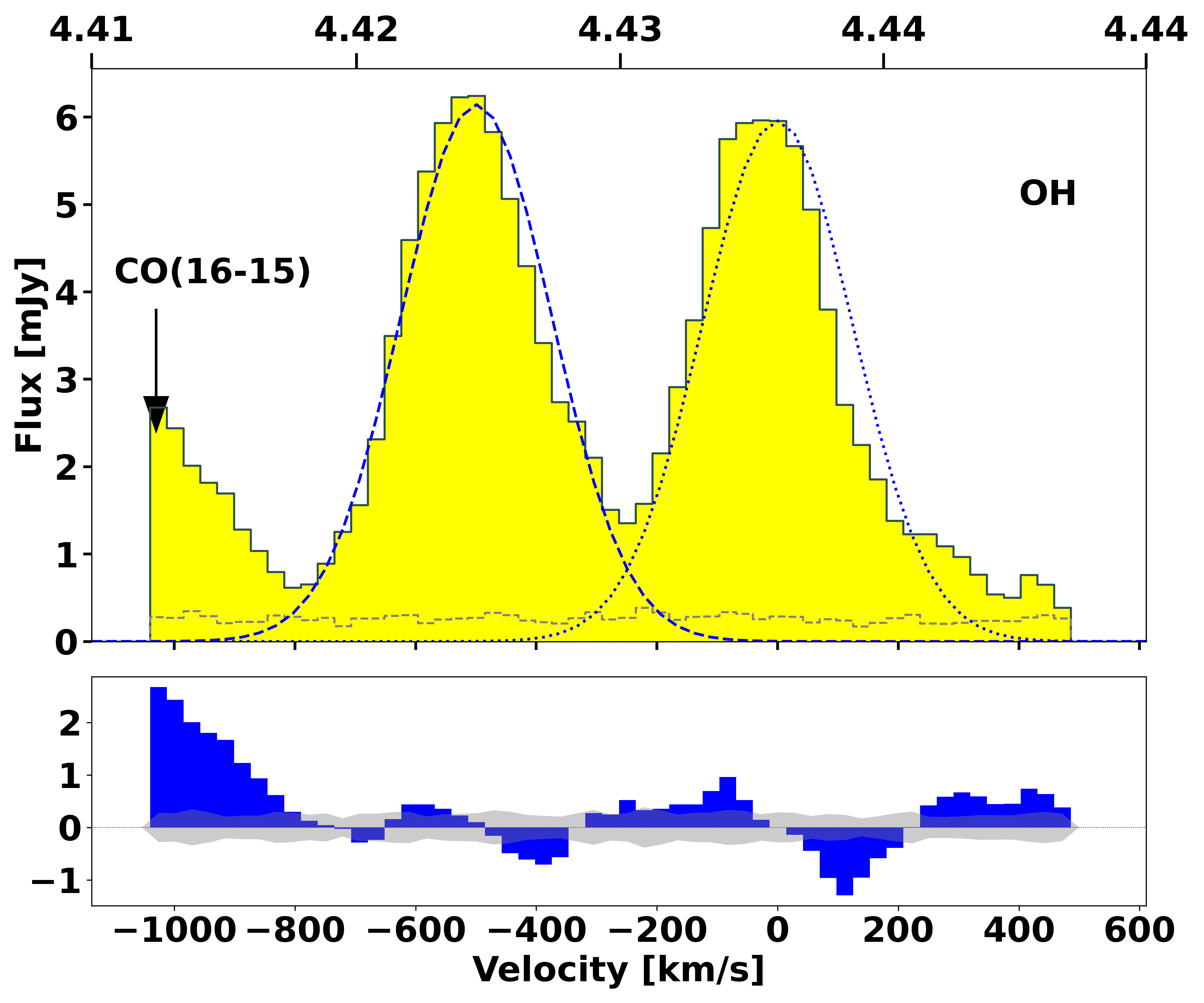}
\caption{Spectra of the molecular and atomic species detected in BRI\,0952, prior to correction for gravitational lensing. The top panel shows the observed spectra. Single Gaussian fits are shown in red and double Gaussian fits are shown in blue. The dashed gray line represents the rms. The bottom panel shows the residuals from the Gaussian fits where the red corresponds to single fits and the blue to fits using two Gaussian profiles and the gray shaded region represents the rms.  We fit the OH spectra with two single Gaussian components where the systemic velocities are fixed (see Section \ref{sec:OH}).}
\label{fig:BRI_spec}
\end{figure*} 

\section{Visilens Lens Modelling}
This appendix includes figures detailing the change in the different lensing (position, axis ratio, major axis, flux, and S\'ersic index) parameters across the \cii line for each of the four models described in Section \ref{sec:visilens_modeling}. 

\begin{figure*}[h!]
\centering
\includegraphics[width=0.8\columnwidth]{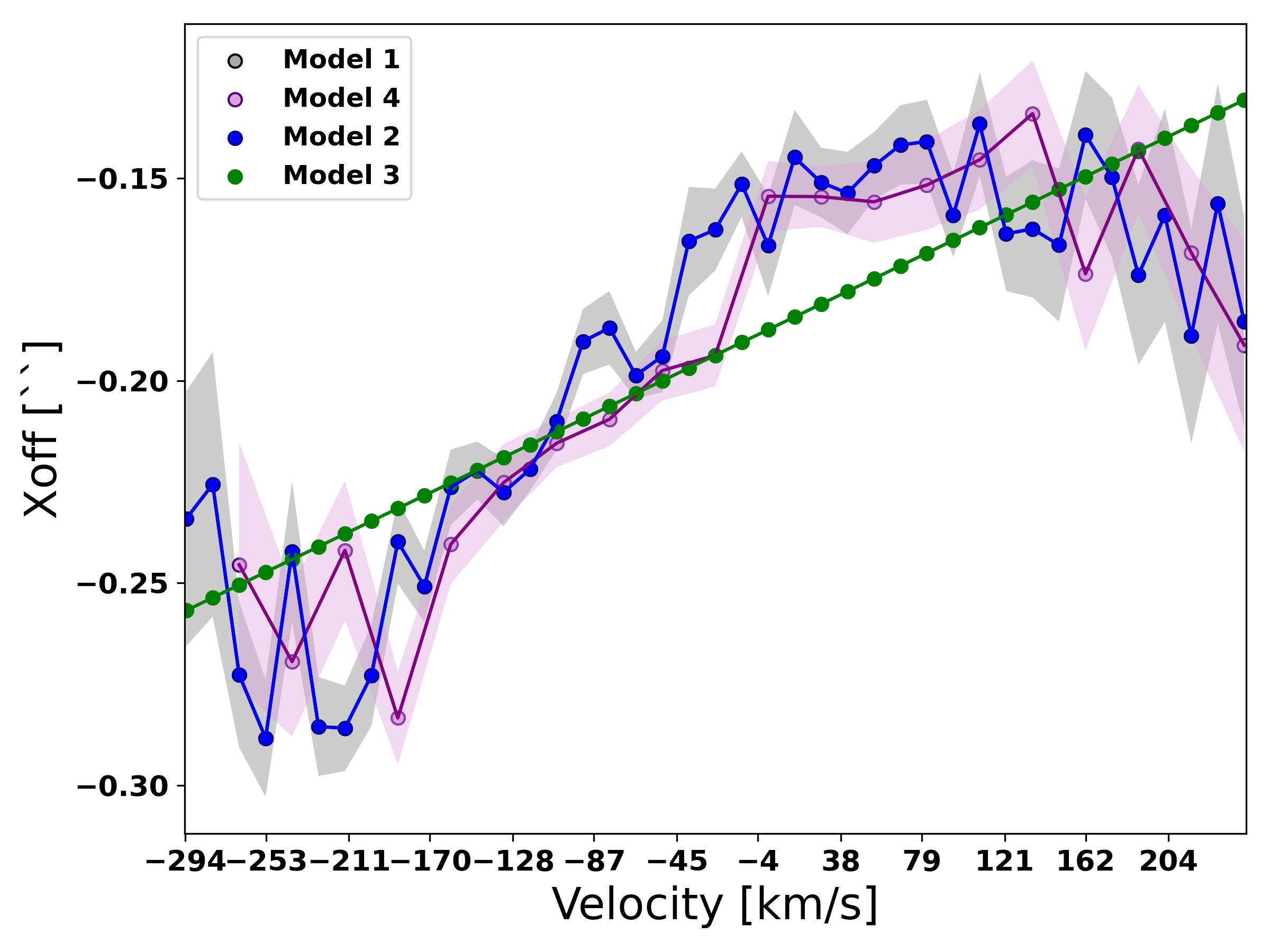}
\includegraphics[width=0.8\columnwidth]{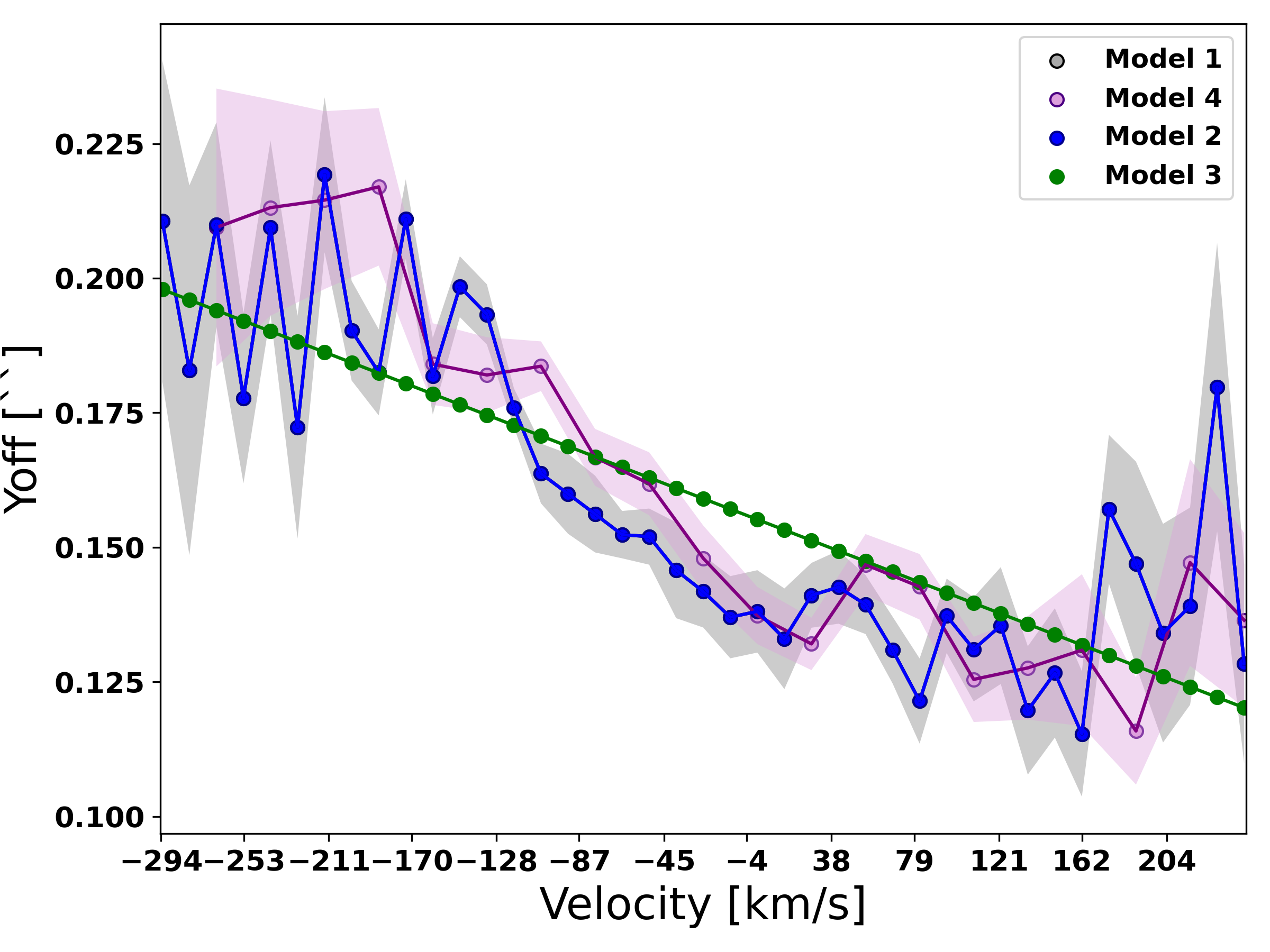}
\includegraphics[width=0.8\columnwidth]{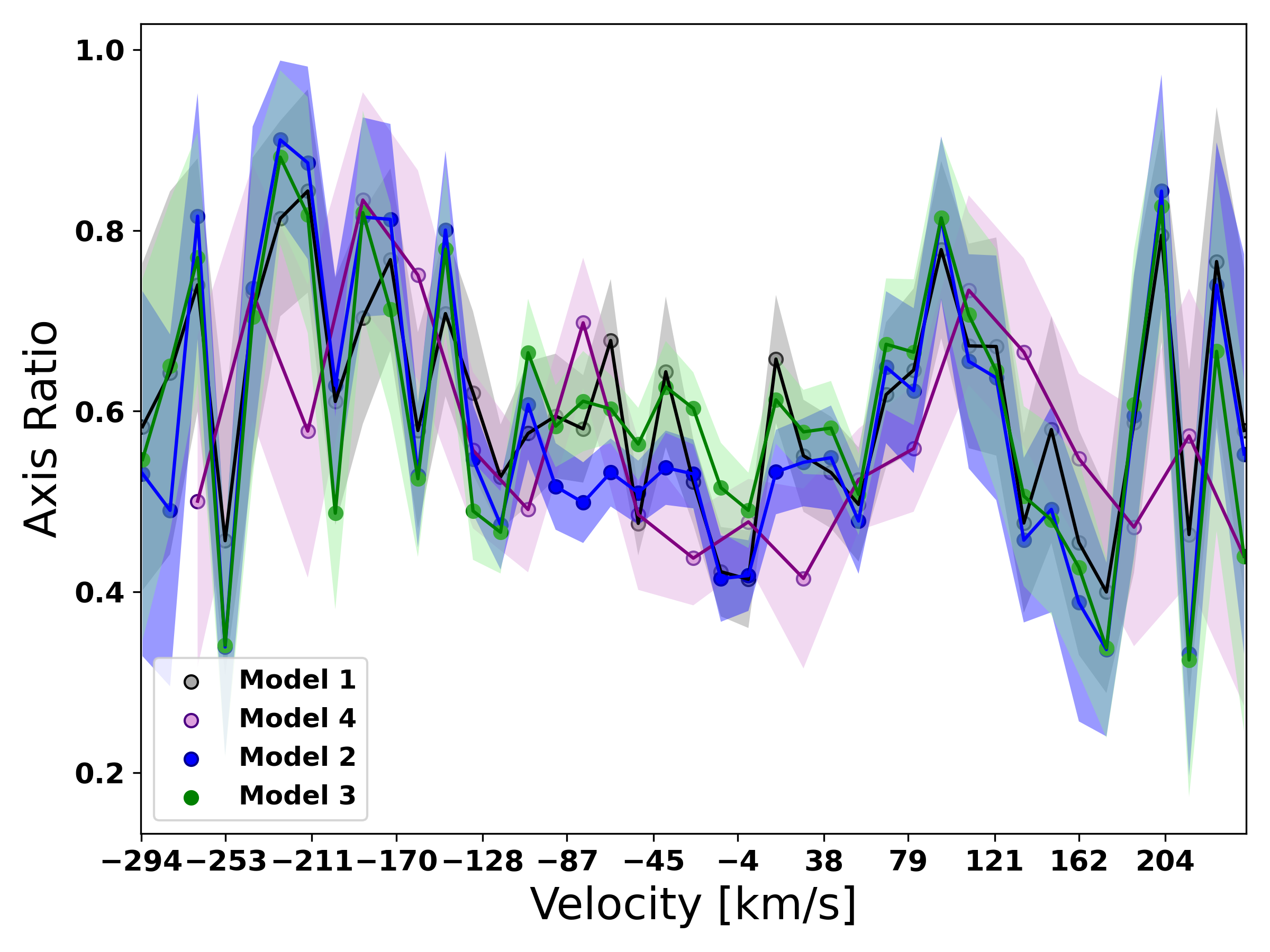}
\includegraphics[width=0.8\columnwidth]{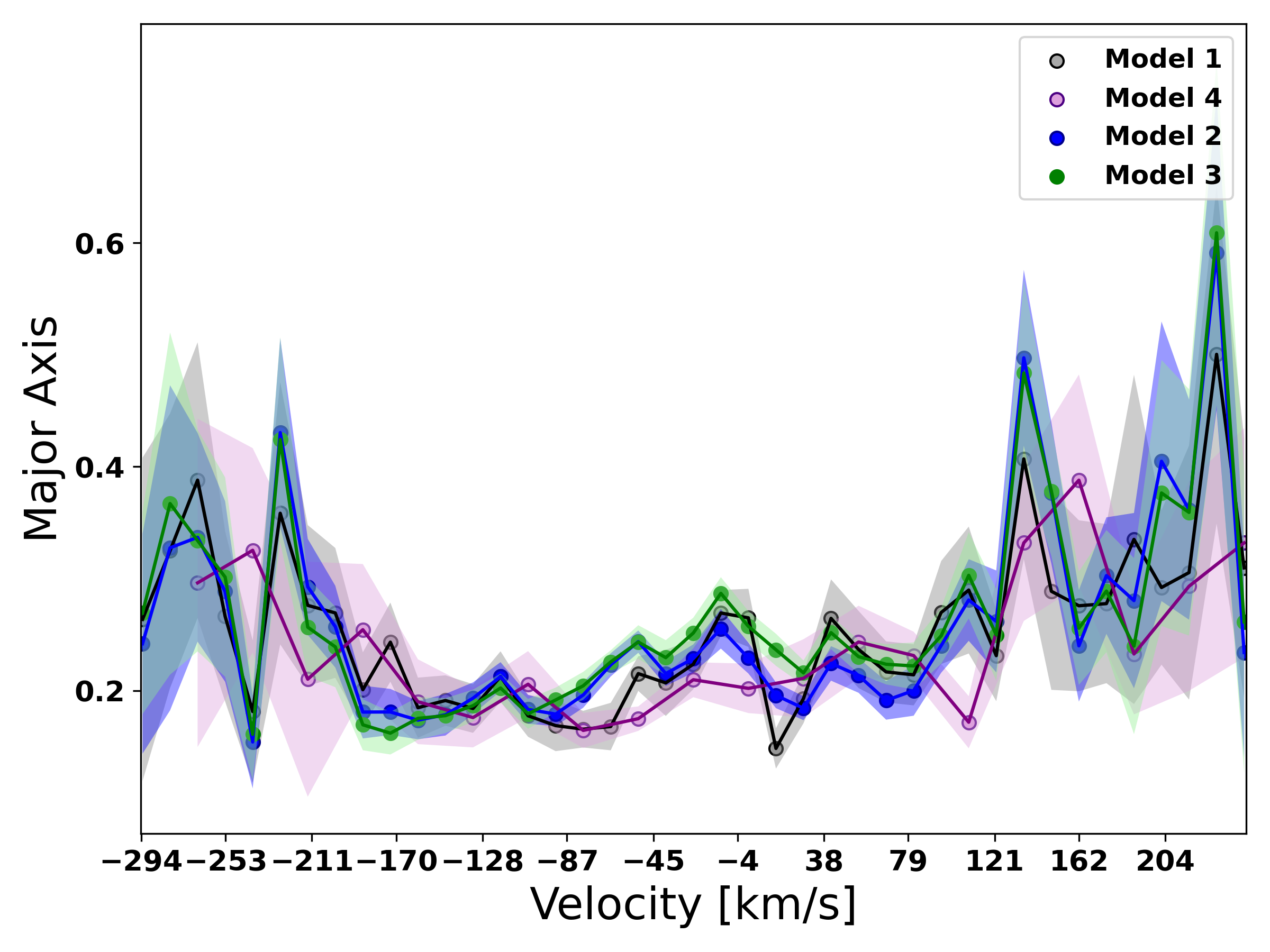}
\includegraphics[width=0.8\columnwidth]{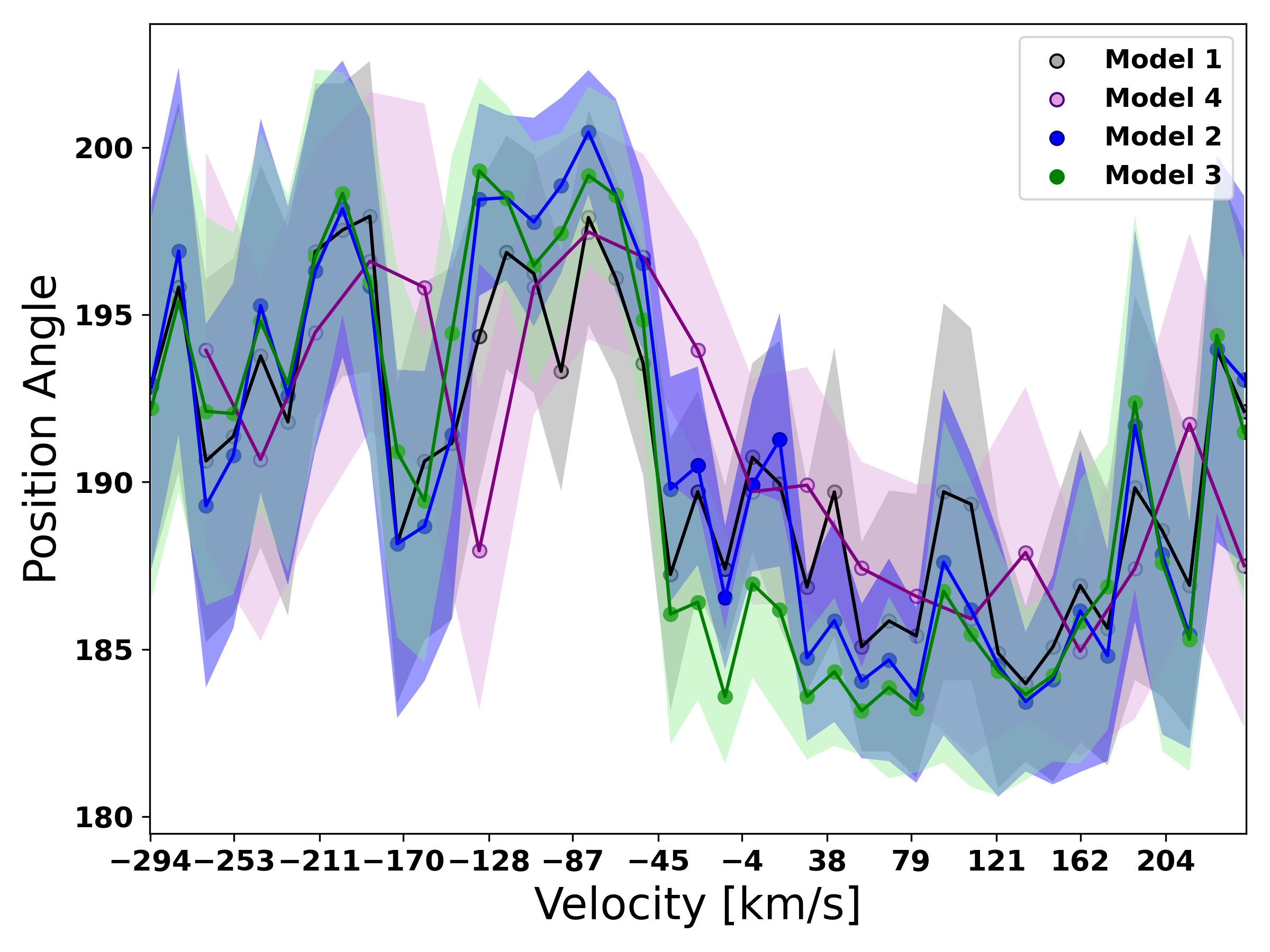}
\includegraphics[width=0.8\columnwidth]{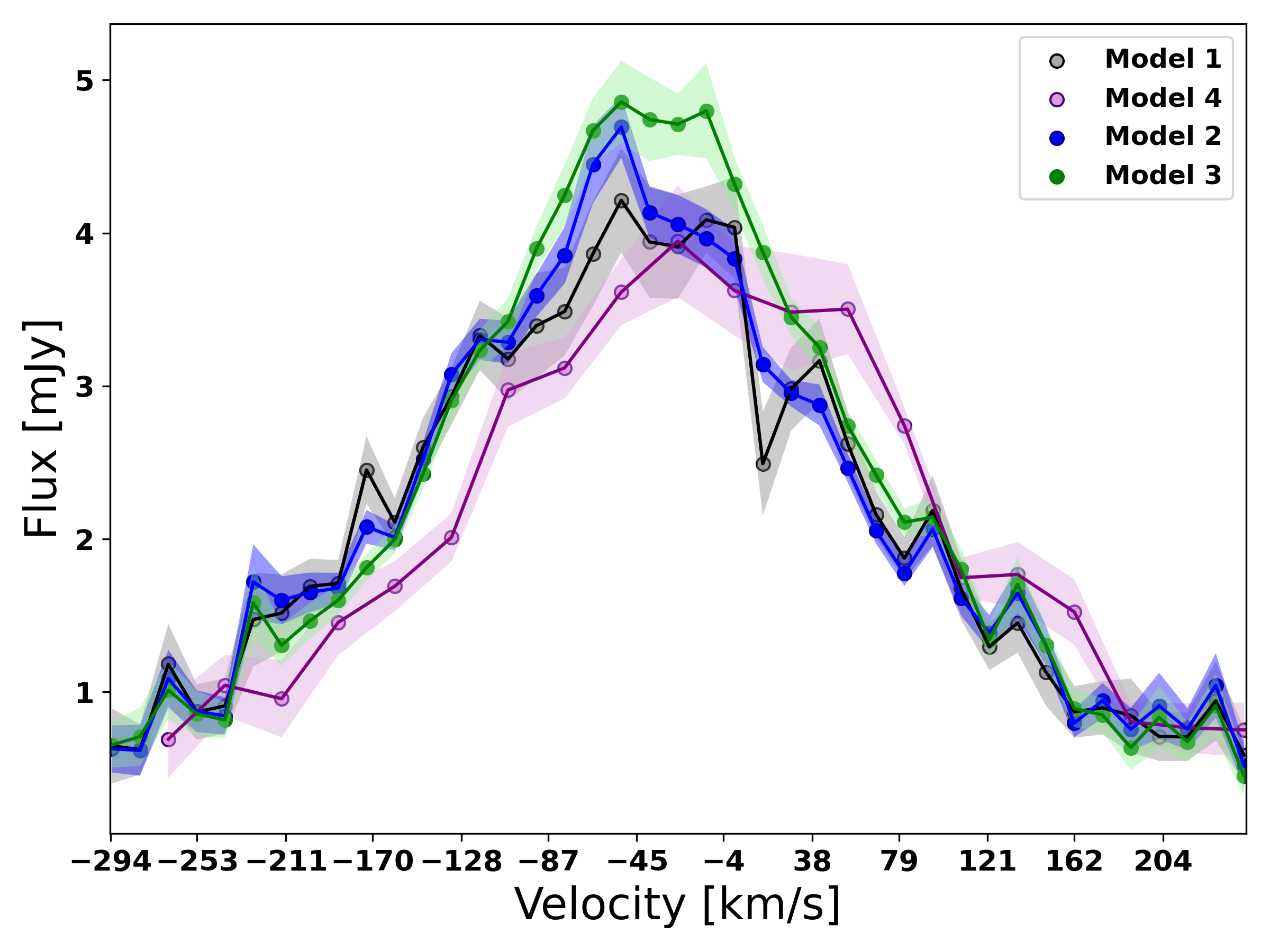}
\includegraphics[width=0.8\columnwidth]{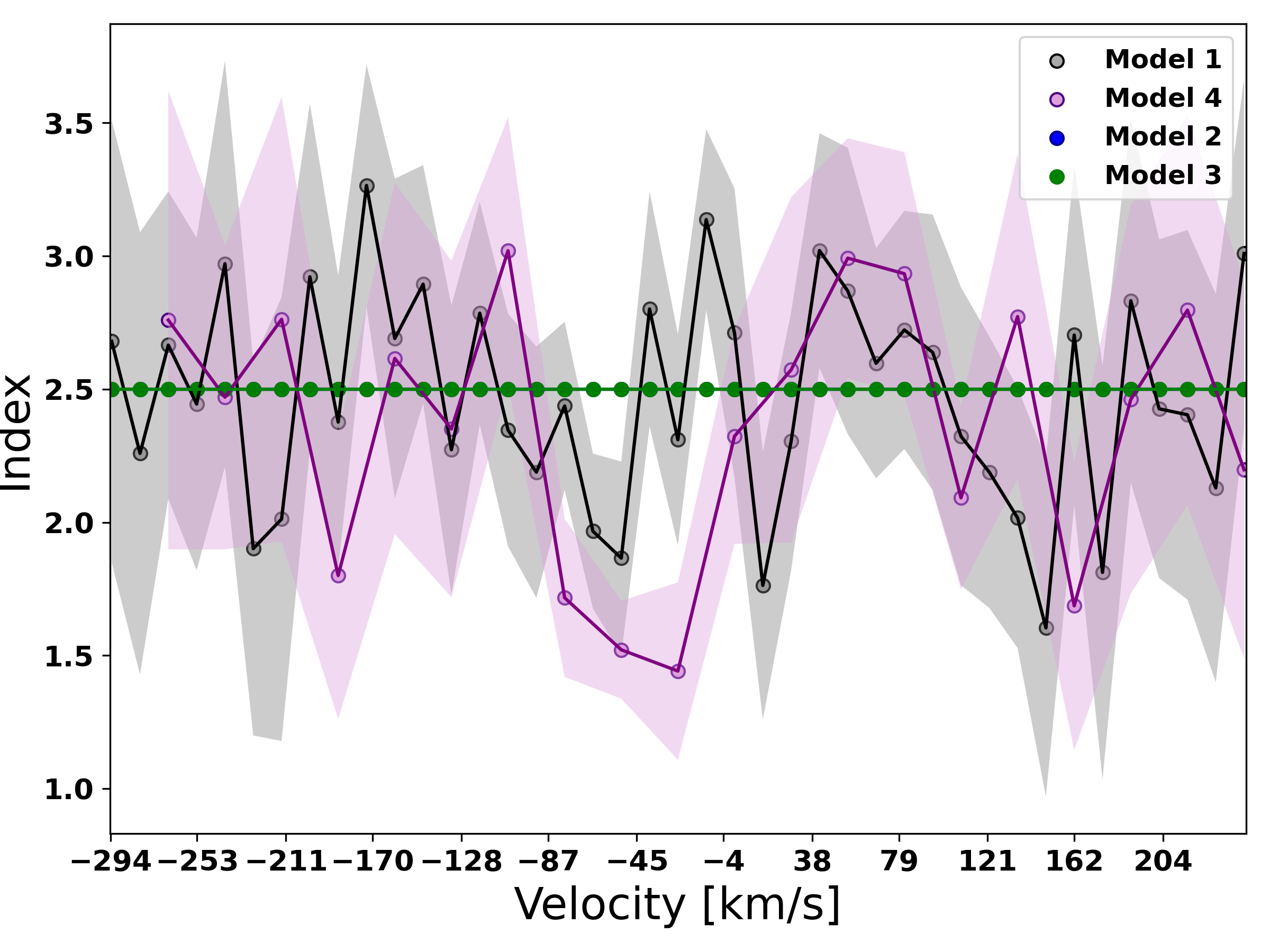}
\caption{Visilens parameters as a function of velocity for the four different lensing models described in \ref{sec:visilens_modeling}.}
\label{fig:BRI_lensing}
\end{figure*} 

\section{Single Component MOLPOP-CEP Models}
This appendix includes figures detailing the output of single component models of the CO SLED of BRI\,0952 from MOLPOP-CEP as described in \ref{basic}. 

\begin{figure*}[t!]
    \centering
    \includegraphics[width = 0.9\columnwidth]{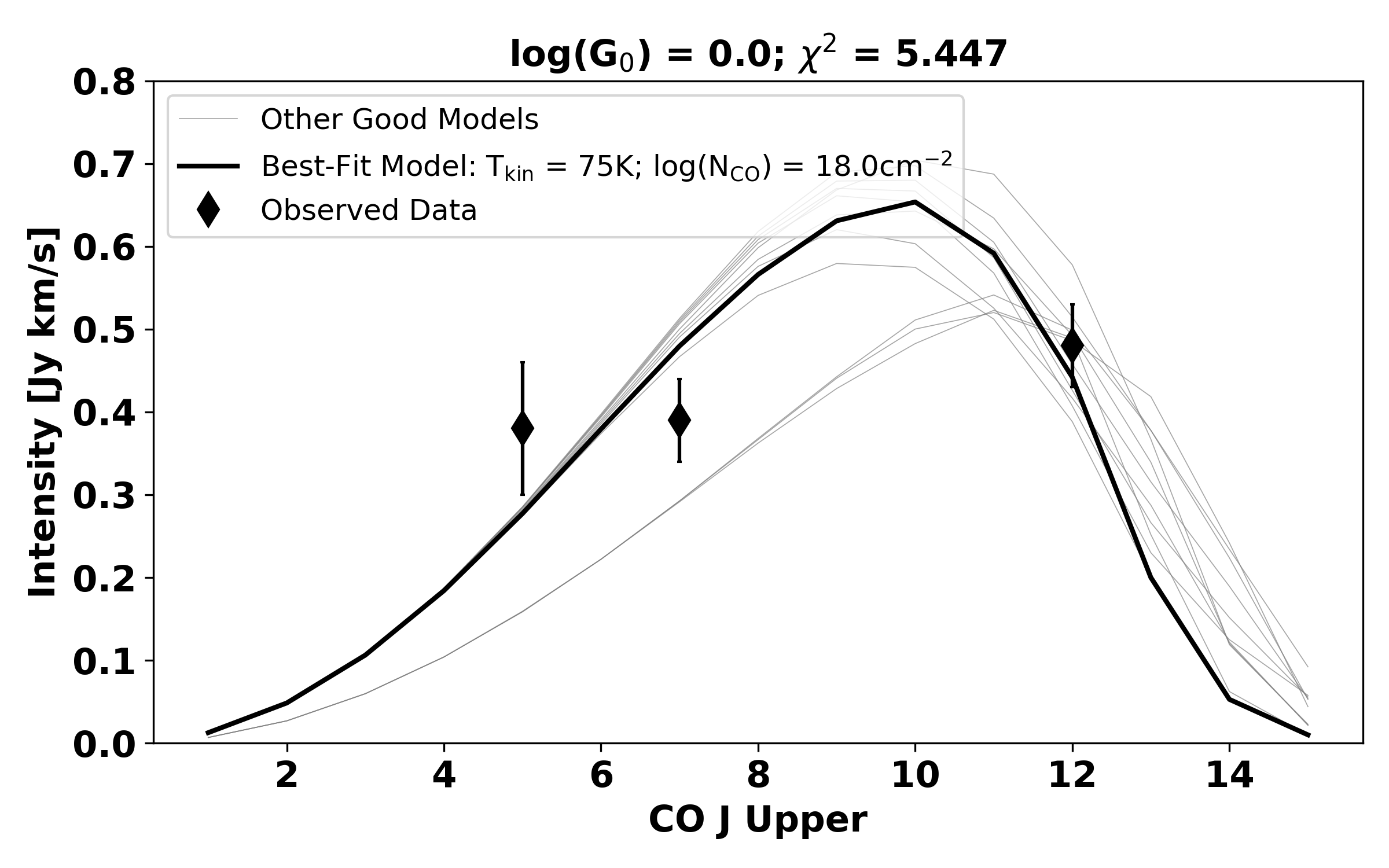}
    \includegraphics[width = 0.9\columnwidth]{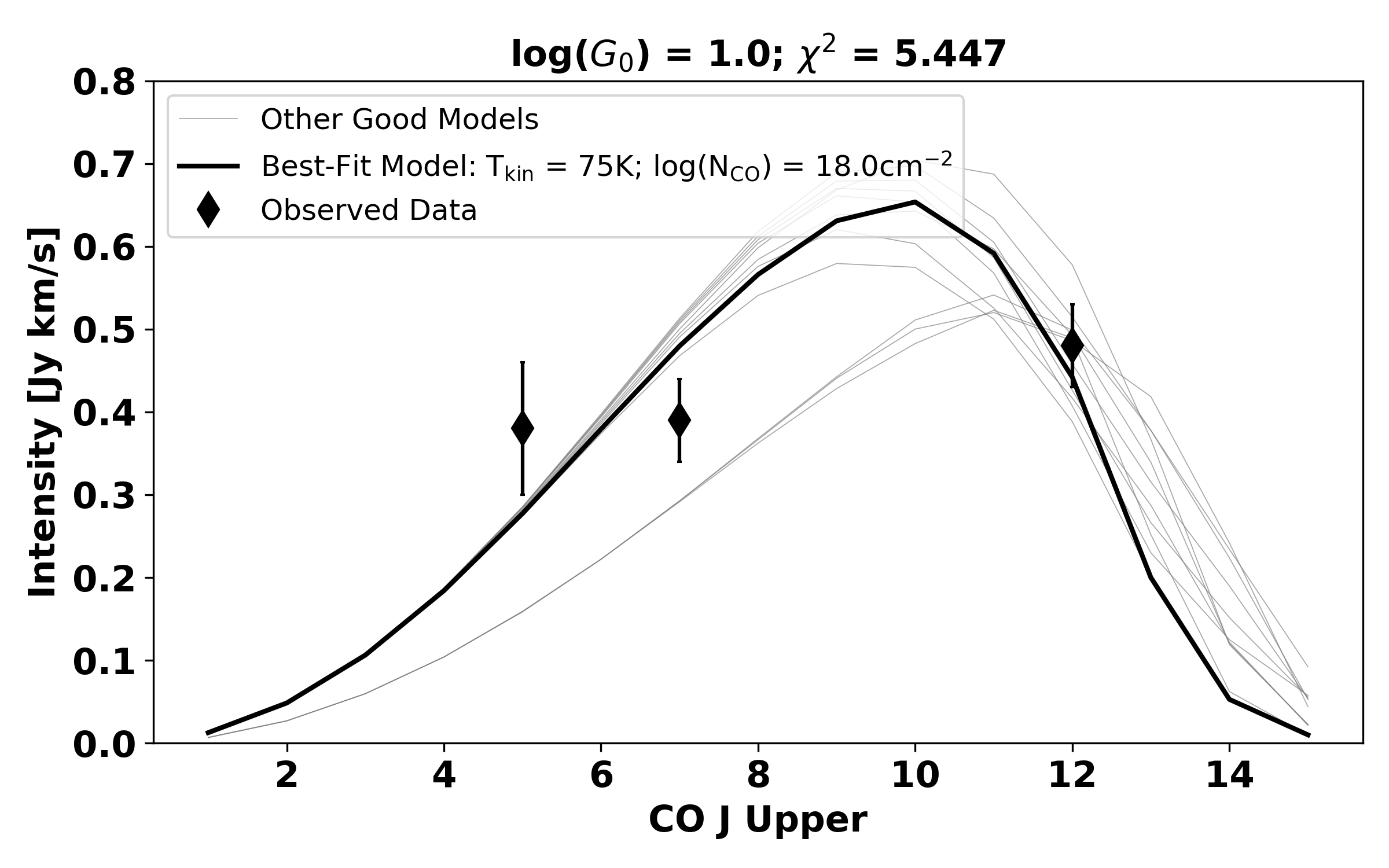}
    \includegraphics[width = 0.9\columnwidth]{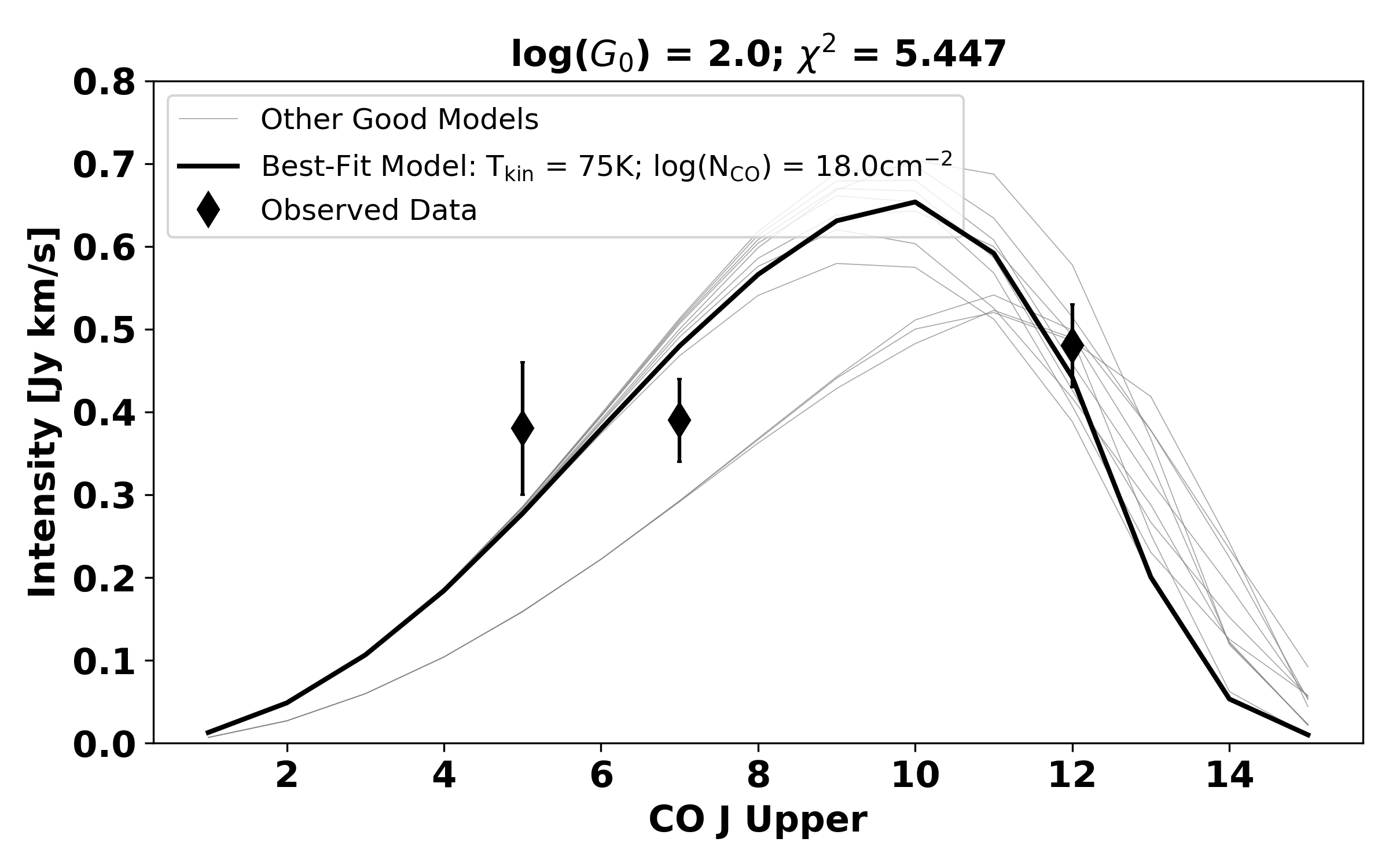}
    \includegraphics[width = 0.9\columnwidth]{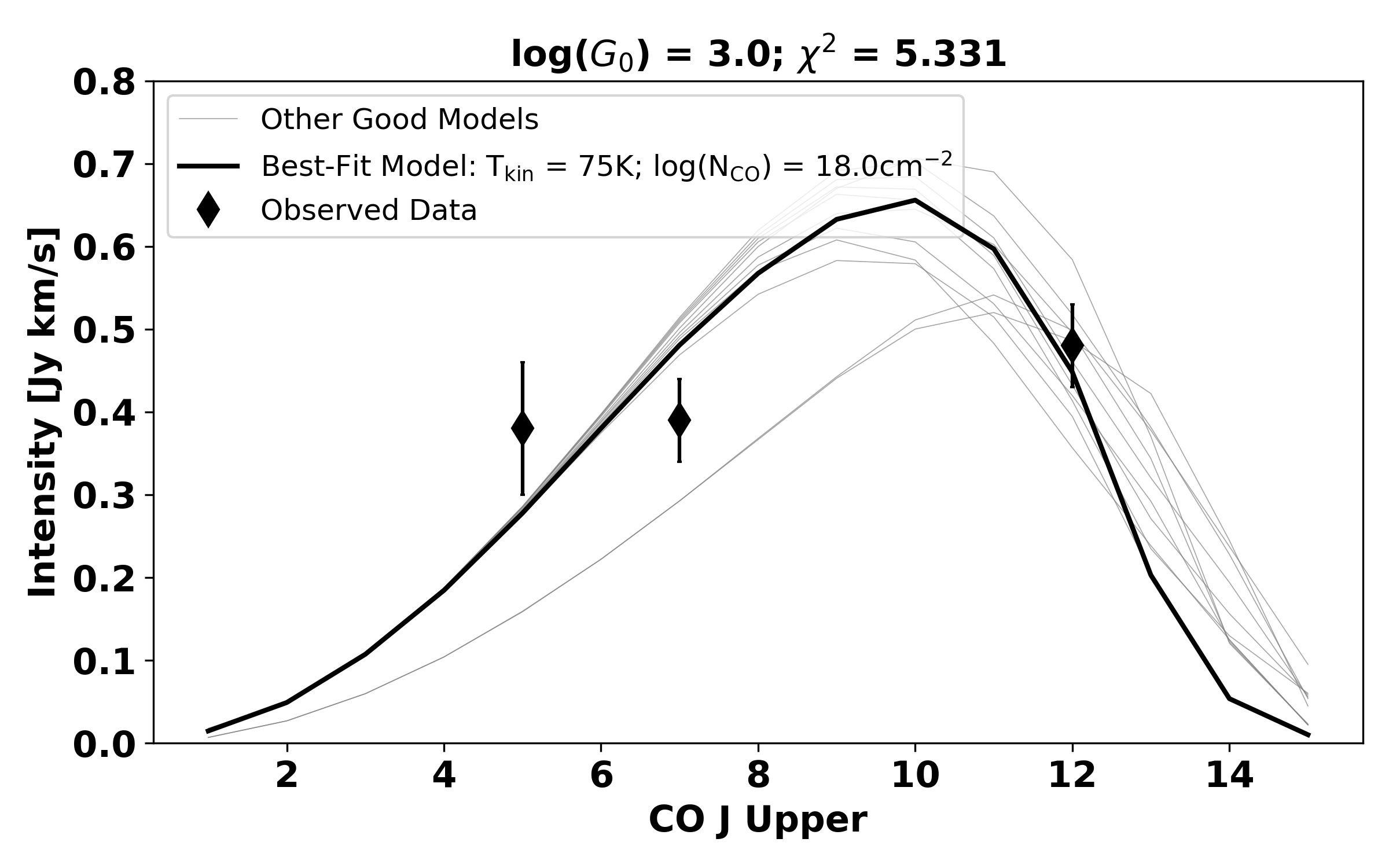}
    \includegraphics[width = 0.9\columnwidth]{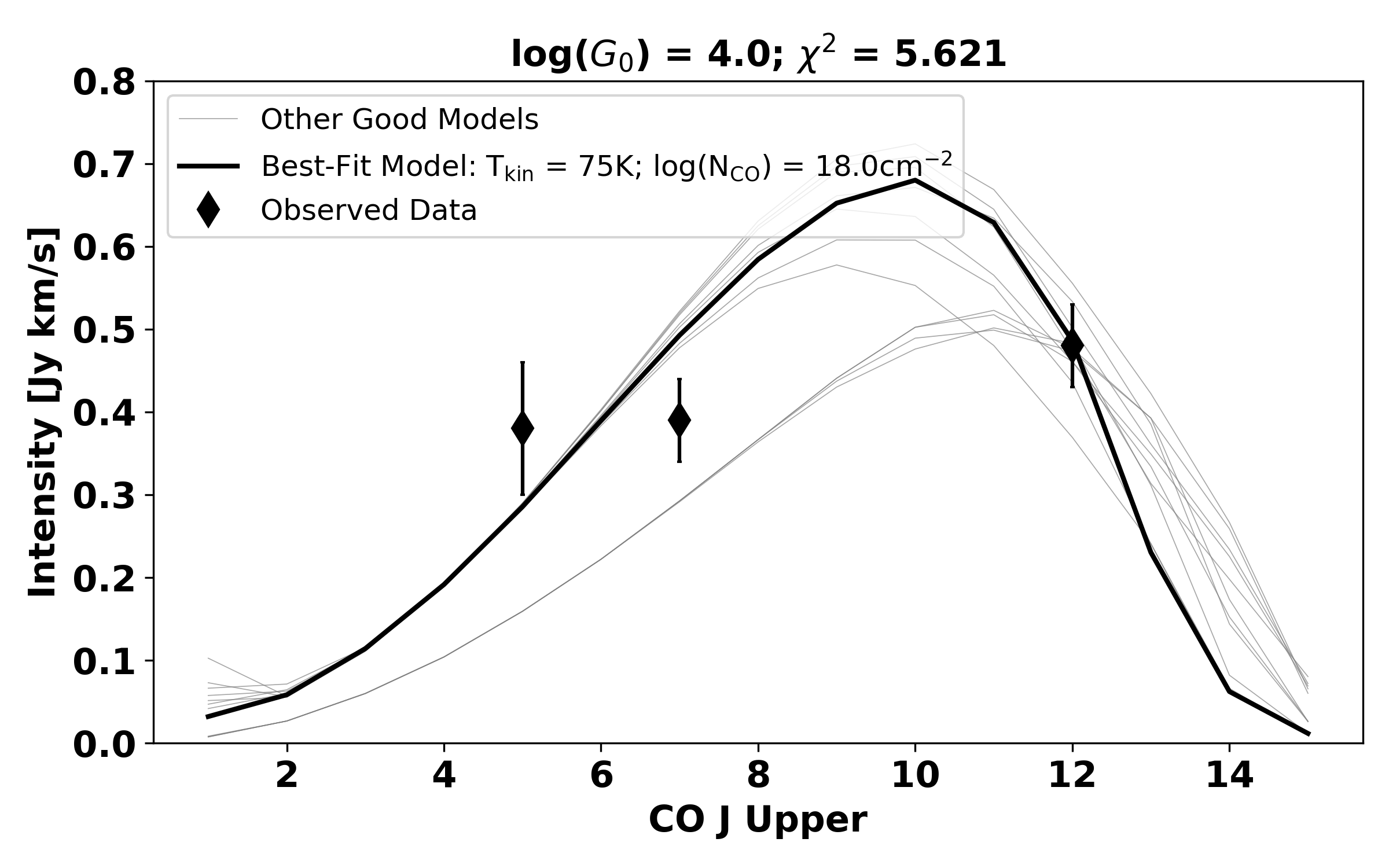}
    \includegraphics[width = 0.9\columnwidth]{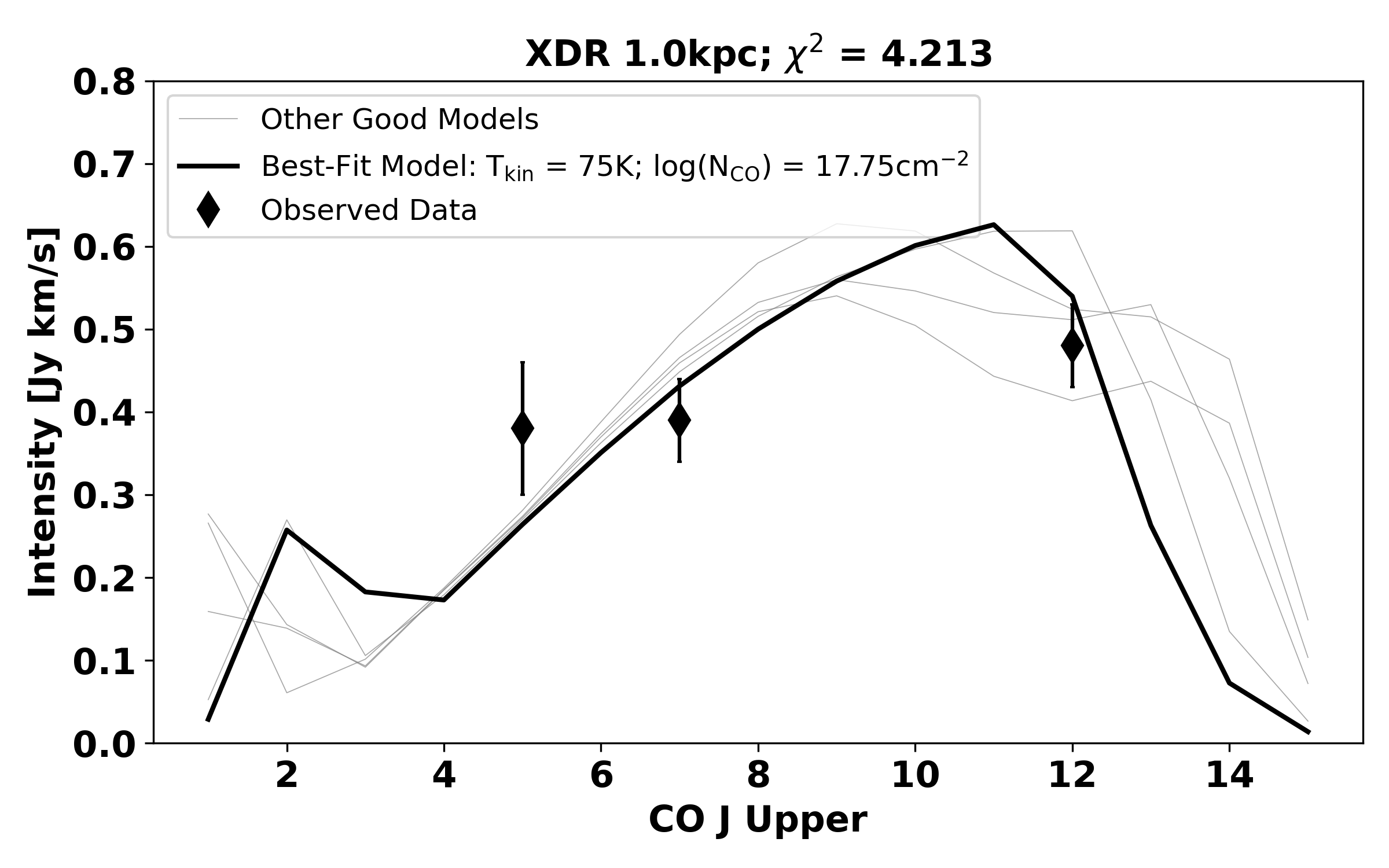}
    \includegraphics[width = 0.9\columnwidth]{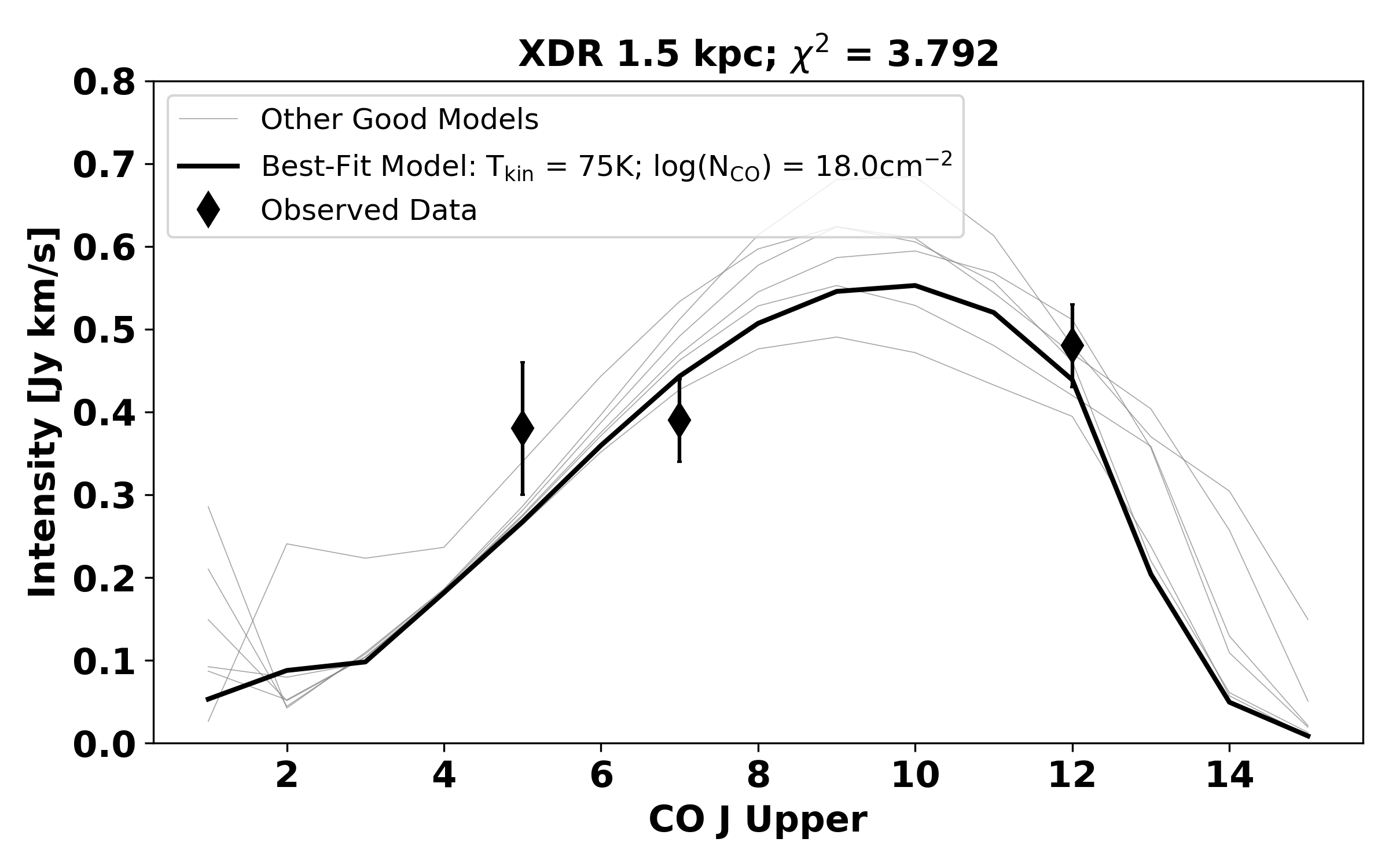}
    \caption{CO SLEDs showing different heating mechanisms including CMB heating, PDRs (log($G_0$) = 0.0, 1.0, 2.0, 3.0, 4.0), and AGN heating via X-rays at 1.0\,kpc and 1.5\,kpc. The solid black curve is the best-fit model and the lighter grey lines show the models within the lowest 1\% of the $\chi^{2}$ values for the different models. Note that the method we employ for isolating the best-fit models is a percentile cut, and it is for this reason that a different number of models are shown for different heating mechanisms.}
    \label{fig:co_sleds_1per}
\end{figure*}

\section{MOLPOP-CEP Two-Component Model 1$\sigma$ parameter range} 
This appendix provides the range for each parameter in a two-component model of the ISM where $\chi^{2} \leq 1\sigma$. 

\begin{table*}[h]
    \centering
    \caption{MOLPOP-CEP parameter $1\sigma$}
    \begin{tabular}{c c c c c c c}
        \hline \hline
        Model & $T_{\rm kin}$ & $n(\rm H_2)$ & $\Delta L$ & $T_{\rm kin, PDR}$ & $n(\rm H_{2, PDR})$ & $\Delta L_{PDR}$  \\

         & [K] & [cm$^{-3}$] & [cm] & [K] & [cm$^{-3}$] & [cm] \\
        \hline

        XDR\,1.0\,kpc + PDR & 50 & 4.25 - 5.5 & 16.0 - 19.0 & 50 - 150 & 3.0 - 4.0 & 16.0 - 19.0\\
        
        XDR\,1.5\,kpc + PDR & 50 & 4.75 - 5.5 & 16.25 - 18.75 & 50 - 200 & 3.5 - 4.25 & 16.0-16.5\\
        
       log($G_0$= 2.0) + PDR & 50 - 200 & 3.0 - 5.75 & 16.0 - 19.0 & 50 - 200 & 3.0 - 4.5 & 16.0 - 19.0 \\
        
       log($G_0$= 3.0) + PDR &  50 - 200 & 3.0 - 5.75 & 16.0 - 19.0 & 50 - 200 & 3.0 - 4.5 & 16.0 - 19.0 \\

       log($G_0$= 4.0) + PDR & 50 - 200 & 3.0 - 5.75 & 16.0-19.0 & 50-200 & 3.0 - 4.5 & 16.0 -19.0  \\
        \hline
    \end{tabular} \label{tab:molpop_1sig_parameterrange}
\end{table*}

\end{appendix}
\end{document}